\documentclass{elsarticle}
\usepackage{microtype}% improves general appearance of the text
\usepackage{amsmath,amssymb,amstext}% math support
\usepackage{newtxtext,newtxmath}% font support
\usepackage{enumitem}% customizing enumerated lists
\usepackage{subcaption}% subfigures
\graphicspath{{./}{./img/}}
\journal{arXiv}
\biboptions{sort&compress}
\providecommand{\Vector}[1]{\boldsymbol{#1}}% General vectors in bold italic
\providecommand{\unitVector}[1]{\boldsymbol{\mathbf{#1}}}% Unit vectors in bold roman
\providecommand{\Tensor}[1]{\boldsymbol{\mathsf{#1}}}% Tensor in sans-serif bold italic
\providecommand{\unitTensor}[1]{\boldsymbol{{\mathsf{#1}}}}% Identity tensor in sans-serif bold
\providecommand{\Matrix}[1]{\boldsymbol{\mathbf{#1}}}% Matrix in bold roman
\providecommand{\Unit}[1]{\,\mathrm{#1}}% Units in roman
\providecommand{\Set}[1]{\mathbb{#1}}% Special sets (e.g. real numbers) in blackboard bold
\providecommand{\Des}[1]{\mathrm{#1}}% description
\providecommand{\Order}{\operatorname{O}}% order operator
\providecommand{\Loptr}{\operatorname{\mathcal{L}}}% spatial operator
\providecommand{\Toptr}{\operatorname{\mathcal{LL}}}% temporal operator
\providecommand{\Soptr}{\operatorname{\mathcal{S}}}% solution operator
\begin{document}

\begin{frontmatter}

\title{An immersed boundary method for solving compressible flow with arbitrarily irregular and moving geometry}

\author[uw]{Huangrui Mo\corref{correspondingauthor}}
\ead{huangrui.mo@uwaterloo.ca}

\author[uw]{Fue-Sang Lien}
\ead{fslien@uwaterloo.ca}

\author[defence]{Fan Zhang}
\ead{fan.zhang@drdc-rddc.gc.ca}

\author[uw]{Duane S. Cronin}
\ead{duane.cronin@uwaterloo.ca}

\cortext[correspondingauthor]{Corresponding author}
\address[uw]{Department of Mechanical Engineering, University of Waterloo, 200 University Avenue West, Waterloo, ON N2L 3G1, Canada}
\address[defence]{Defence Research and Development Canada, P.O. Box 4000, Station Main, Medicine Hat, AB T1A 8K6, Canada}

\begin{abstract}
In this paper, a novel immersed boundary method is developed, validated, and applied. Through devising a second-order three-step flow reconstruction scheme, the proposed method is able to enforce the Dirichlet, Neumann, Robin, and Cauchy boundary conditions in a straightforward and consistent manner. Equipped with a fluid-solid coupling framework that integrates high-order temporal and spatial discretization schemes, numerical experiments concerning flow involving stationary and moving objects, convex and concave geometries, no-slip and slip wall boundary conditions, as well as subsonic and supersonic motions are conducted to validate the method. It is demonstrated that the proposed method can provide efficient, accurate, and robust boundary treatment for solving flow with arbitrarily irregular and moving geometries on Cartesian grids.
\end{abstract}


\end{frontmatter}

\section{Introduction}\label{sec:intro}

Being an important field of research in both fundamental sciences and engineering applications, fluid-solid interaction involving complex geometric settings has received substantial attention in the past and remains as an active field of research and development, owing to the ubiquitous presence of fluid-solid interaction phenomena and the great difficulty in tackling those problems via mathematical modeling. To solve flow involving irregular and moving geometries, one of the main challenges is related to enforcing boundary conditions.

In recent years, immersed-boundary-type methods \citep{peskin1972flow, peskin2002immersed, fedkiw1999non, fedkiw2002coupling} have gained increasing popularity in interface boundary treatment \citep{wang2004extended, mori2008implicit, liao2012simulations, kapahi2013three, kempe2015imposing, schwarz2016immersed}. As an attractive alternative to the Arbitrary Lagrangian--Eulerian method \citep{hirt1974arbitrary}, in which a body-conformal grid following the movement of phase interfaces is employed, immersed boundary methods are able to solve problems with complex interfaces on a generic Cartesian grid. Benefited from the use of a stationary Cartesian gird, grid generation is greatly simplified, and the per-grid-point computation exempts from operations associated with grid topology or transformations. Since the main data structures for numerical computation are simple arrays, the memory requirement is largely reduced, and high-order spatial discretization schemes are easy to implement \citep{mittal2005immersed}.

Extensions of the immersed boundary method, which was originally introduced by \citet{peskin1972flow}, have been continuously developed to improve the numerical properties of the method, particularly in aspects related to interface resolution, stability constraints, mass conservation, computational efficiency and robustness \citep{peskin2002immersed, mittal2005immersed, sotiropoulos2014immersed}. 

To simplify numerical discretization and relax stability constraints, \citet{yusof1997combined} and \citet{fadlun2000combined} proposed the direct forcing immersed boundary method, in which boundary forces are implicitly imposed via flow reconstruction. \citet{balaras2004modeling} later improved the reconstruction procedure of the direct forcing approach and applied to large-eddy simulations. Integrating ideas from the ghost fluid method \citep{fedkiw1999non, fedkiw2002coupling} and the direct forcing method \citep{fadlun2000combined}, \citet{tseng2003ghost} systematically developed a polynomial reconstruction based ghost-cell immersed boundary method to further increase implementation flexibility while maintaining sharp interfaces. \citet{kapahi2013three} proposed a least square interpolation approach and applied to solving impact problems. Employing the adaptive mesh refinement technique for resolving boundary layers, \citet{brehm2015locally} developed a locally stabilized immersed boundary method and applied to simulating the laminar to turbulent transition process on no-slip walls.

For a direct forcing immersed boundary method, its robustness highly depends on the numerical stability and stencil adaption capability of the employed interpolation method \citep{tseng2003ghost, gao2007improved, kapahi2013three}. Polynomial reconstruction based methods involve constructing linear systems on neighboring stencils of the interpolated node. When the stencil nodes are not well distributed in space, the resulting linear systems may suffer from numerical singularities \citep{tseng2003ghost, gao2007improved}. Additionally, a fixed minimum number of stencil nodes is needed to avoid under-determined linear systems. Therefore, special treatments are required when strongly concave or convex interfaces exist \citep{gao2007improved, kapahi2013three}. To enhance numerical stability and stencil adaption capability, the idea of using inverse distance weighting interpolation was firstly introduced by \citet{tseng2003ghost}, and a hybrid Taylor series expansion / inverse distance weighting approach was later developed by \citet{gao2007improved} for flow with no-slip walls.

In addition to numerical stability and stencil adaption capability, being able to enforce different types of boundary conditions in a straightforward and consistent manner is another vital factor in obtaining an efficient, accurate, and robust immersed boundary method, since a variety of boundary conditions are required to be repeatedly enforced on numerical boundaries. For instance, in solving Navier--Stokes equations, constant temperature at a wall and velocity at a no-slip wall have Dirichlet boundary conditions, pressure at a wall and temperature at an adiabatic wall have Neumann boundary conditions, and velocity at a slip wall has a type of Cauchy boundary conditions.

Excluding the Dirichlet boundary conditions in which boundary values are determined and known, the enforcement of other types of boundary conditions, particularly the Cauchy type of boundary conditions, for immersed boundaries demands considerable efforts \citep{crockett2011cartesian, kempe2015imposing, schwarz2016immersed}. \citet{kempe2015imposing} devised a numerical implementation of the slip-wall boundary conditions in the context of immersed boundary methods. However, the realization is not straightforward due to its complexity \citep{kempe2015imposing}. In addition, most direct forcing immersed boundary methods require constructing and solving a designated linear system for each type of boundary conditions. Therefore, to enforce a variety of boundary conditions in a straightforward and consistent manner is beneficial but can be challenging.

Employing the direct forcing approach, this paper develops a novel immersed boundary method via devising a second-order three-step flow reconstruction scheme. The developed method can enforce the Dirichlet, Neumann, Robin, and Cauchy boundary conditions in a straightforward and consistent manner, and is able to provide efficient, accurate, and robust boundary treatment for solving flow with arbitrarily irregular and moving geometries on Cartesian grids.

\section{Method development}\label{sec:method}

\subsection{A generalized framework}

\begin{figure}[!htbp]
    \centering
    \begin{subfigure}[b]{0.48\textwidth}
        \includegraphics[width=\textwidth]{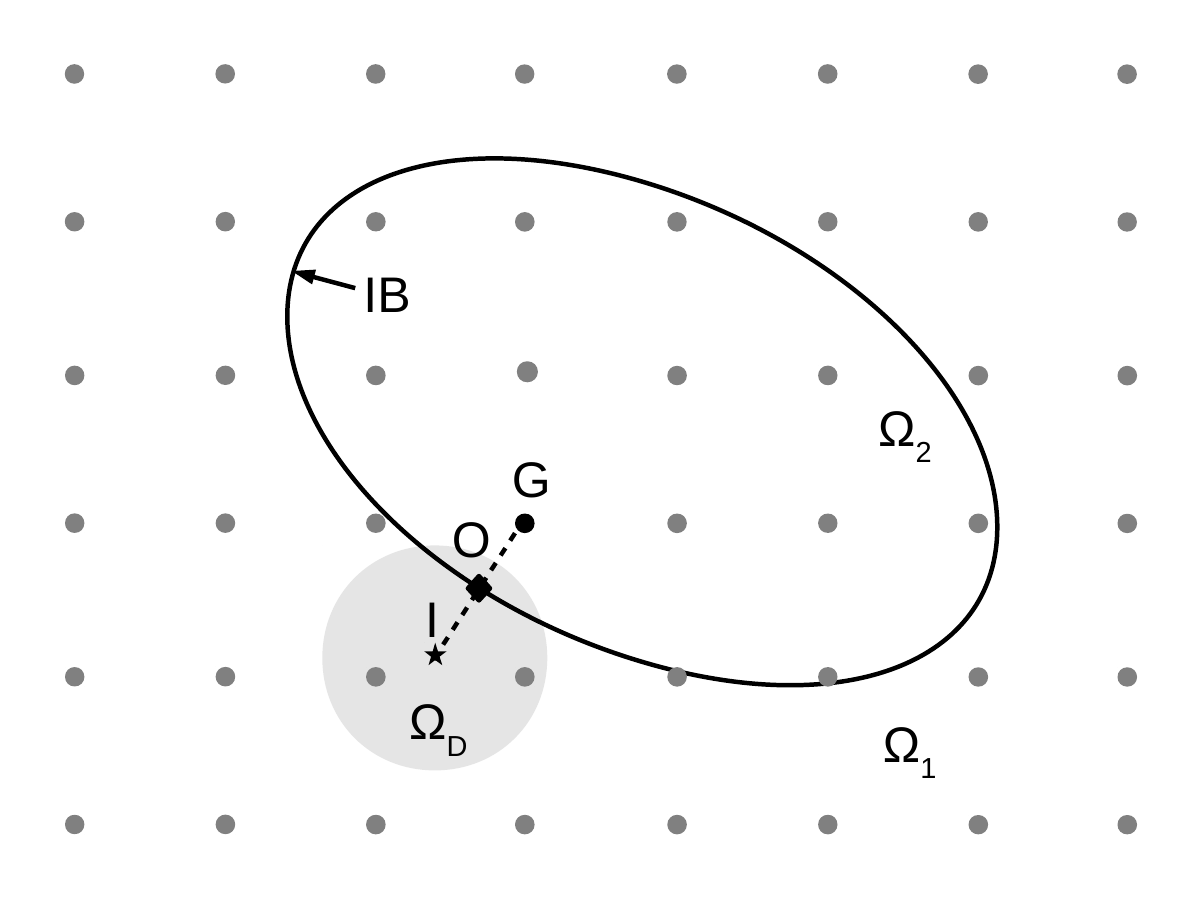}
        \caption{}
        \label{fig:gcibm_demo_2D}
    \end{subfigure}%
    ~
    \begin{subfigure}[b]{0.48\textwidth}
        \includegraphics[width=\textwidth]{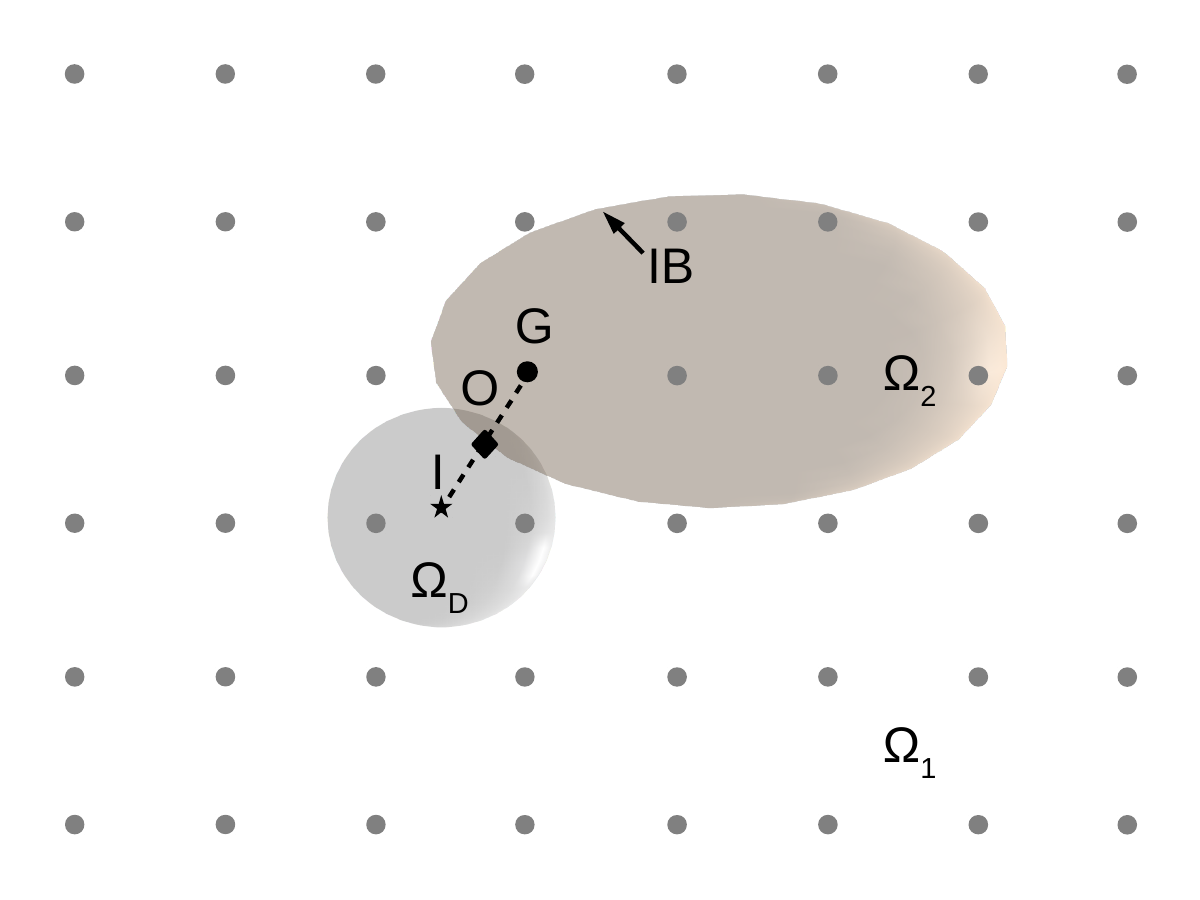}
        \caption{}
        \label{fig:gcibm_demo_3D}
    \end{subfigure}%
    \caption{Schematic diagrams of Cartesian-grid-based computational domains with immersed boundaries. (a) 2D space. (b) 3D Space. [Nomenclature: $G$, ghost node; $O$, boundary point; $I$, image point; $\Omega_{\Des{D}}$, domain of dependence; $\Omega_1$, fluid domain; $\Omega_2$, solid domain; $IB$, immersed boundary.]}
    \label{fig:gcibm_demo}
\end{figure}

Two- and three-dimensional Cartesian-grid-based computational domains with immersed boundaries are illustrated in Fig.~\ref{fig:gcibm_demo}, in which $G$ denotes a ghost node, a computational node that is outside the physical domain and locates at the numerical boundaries, $O$ denotes a boundary point with $\Vector{GO}$ as the outward normal vector, and $I$ is the image point of the ghost node $G$ reflected by the physical boundary.

In a direct forcing immersed boundary method, boundary treatment is mainly about constructing the flow at numerical boundaries. To obtain a proper ghost flow with the existence of physical boundaries effectively admitted, a two-step approach incorporating the method of images is commonly adopted:
\begin{subequations} \label{eq:gcibm}
    \begin{align}
        \psi_G &= 2\psi_O - \psi_I \label{eq:gcibma} \\
        \psi_I &= f(x_I, y_I, z_I) \label{eq:gcibmb}
    \end{align}
\end{subequations}
where $\psi$ denotes a generic flow variable, $f(x, y, z)$ is a local reconstruction function of $\psi$ at the image point $I$. 

As the local reconstruction function will largely determine the numerical properties of the resulting immersed boundary method, we develop a novel three-step flow reconstruction scheme to achieve efficient, accurate, and robust boundary treatment for arbitrarily irregular and moving boundaries. 

\subsection{A three-step flow reconstruction scheme}

In this study, the local reconstruction function is established on the physical boundary conditions at the boundary point $O$ and the known values of $\psi$ at nearby fluid nodes:
\begin{equation} \label{eq:gcibmb_another}
    \psi_I = f(\{\psi_N\}, \psi_O) 
\end{equation}
where $\psi_O$ is the value of $\psi$ at the boundary point $O$, at which the physical boundary conditions are enforced. Except for a Dirichlet boundary condition, the value of $\psi_O$ is unknown and implicitly subject to different types of mathematical constraints, which are the main challenges in developing a local reconstruction function. $\{\psi_N\}$ represents the values of $\psi$ at fluid nodes $\{N\}$ that satisfy
\begin{equation} \label{eq:domain}
    d_N = ||\Vector{x}_I - \Vector{x}_N|| \le R_I
\end{equation}
in which $\Vector{x}_I$ and $\Vector{x}_N$ are the position vectors of the point $I$ and $N$, respectively. $R_I$, referred to as the domain of dependence of the point $I$ and illustrated in Fig.~\ref{fig:gcibm_demo}, is the maximum distance from the point $I$ to nearby fluid nodes that are employed for flow reconstruction. 

The incorporation of physical boundary conditions in the local reconstruction function of $\psi_I$ gives
\begin{equation} \label{eq:limit}
    \begin{aligned}
        \lim_{||\Vector{x}_G - \Vector{x}_O|| \to 0}\psi_G
        &= 2\psi_O - \lim_{||\Vector{x}_G - \Vector{x}_O|| \to 0}\psi_I\\
        &= 2\psi_O - \lim_{||\Vector{x}_I - \Vector{x}_O|| \to 0}\psi_I\\
        &= 2\psi_O - \lim_{||\Vector{x}_I - \Vector{x}_O|| \to 0}f(\{\psi_N\}, \psi_O)\\
        &= \psi_O
    \end{aligned}
\end{equation}
Therefore, the constructed $\psi_G$ converges to the exact physical boundary conditions when the ghost node $G$ converges to the boundary point $O$.

For $\psi$ representing a generic field variable, to construct $\psi_I=f(\{\psi_N\}, \psi_O)$ regarding the Dirichlet, Neumann, Robin, and Cauchy boundary conditions in a straightforward and consistent manner, a three-step flow reconstruction scheme is proposed.

\paragraph{Prediction step}

Pre-estimate the value of $\psi_I$ by applying inverse distance weighting on the fluid nodes that locate in the domain of dependence of the image point $I$:
\begin{equation}\label{eq:prediciton}
    \psi_I^* = \frac{\sum w(d_N) \psi_N}{\sum w(d_N)},\ d_N = \max(d_N, d_{\Des{tiny}})\ \text{and}\ \ d_N \le R_I
\end{equation}
where $\psi_I^*$ denotes the predicted value of $\psi_I$, the weighting function $w(d)$ employs an inverse-power law $1/d^q$ with $q \in \Set{N}$, $d_N$ is the distance from the node $N$ to the image point $I$, $d_{\Des{tiny}} = \varepsilon \min(\Delta_x, \Delta_y, \Delta_z)$ is a positive real number to avoid a zero denominator, $\Delta_x, \Delta_y, \Delta_z$ are the mesh sizes in $x$, $y$, $z$ directions, respectively. The power $q=2$, the constant $\varepsilon=10^{-6}$, and the domain of dependence $R_I =2 \max(\Delta_x, \Delta_y, \Delta_z)$ are adopted in the test cases herein, which are shown to be well adequate for a variety of flow problems, as demonstrated in the numerical experiments.

\paragraph{Physical boundary condition enforcement step}

Determine the value of $\psi_O$ via the physical boundary conditions that $\psi$ needs to satisfy at the boundary point $O$ and the values of $\psi$ at interior physical domain.

\subparagraph{Dirichlet boundary condition}

If $\psi$ satisfies the Dirichlet boundary condition, the value of $\psi_O$ is exactly determined by the specified boundary condition:
\begin{equation} \label{eq:dirichlet}
    \psi_O = g
\end{equation}
where $g$ is a given value or function.

\subparagraph{Neumann boundary condition}

$\psi$ is required to satisfy
\begin{equation} \label{eq:neumann}
    \left. \frac{\partial \psi}{\partial n} \right|_O = \frac{\partial \psi_O}{\partial n}
\end{equation}
where ${\partial \psi_O}/{\partial n}$ is a given value or function.

In other words,
\begin{equation} \label{eq:reneumann}
    \lim_{l \to 0} \frac{\psi(\Vector{x}_O + l \unitVector{n}) - \psi(\Vector{x}_O)}{l} = \frac{\partial \psi_O}{\partial n}
\end{equation}
where $\Vector{x}_O$ and $\unitVector{n}$ are the position vector and the unit outward surface normal vector at the boundary point $O$, respectively.

Since point $I$ is on the normal direction of point $O$, it gives
\begin{equation} \label{eq:direction}
    \unitVector{n} = \frac{\Vector{x}_I - \Vector{x}_O}{||\Vector{x}_I - \Vector{x}_O||}
\end{equation}

Therefore,
\begin{equation} \label{eq:approxneumann}
    \frac{\psi_I - \psi_O}{||\Vector{x}_I - \Vector{x}_O||} - \left. \frac{\partial^2 \psi}{\partial n^2} \right|_O ||\Vector{x}_I - \Vector{x}_O|| + \Order(||\Vector{x}_I - \Vector{x}_O||^2)= \frac{\partial \psi_O}{\partial n}
\end{equation}

Due to Eq.~\eqref{eq:gcibma}, the second-order derivative term has
\begin{equation} \label{eq:secondderivative}
    \left. \frac{\partial^2 \psi}{\partial n^2} \right|_O = \frac{\psi_I + \psi_G - 2\psi_O}{2||\Vector{x}_I - \Vector{x}_O||^2} + \Order(||\Vector{x}_I - \Vector{x}_O||^2) = \Order(||\Vector{x}_I - \Vector{x}_O||^2)
\end{equation}

Hence, a second-order accurate approximation of $\psi_O$ can be given as
\begin{equation} \label{eq:resultneumann}
    \psi_O = \psi_I - ||\Vector{x}_I - \Vector{x}_O||\frac{\partial \psi_O}{\partial n}
\end{equation}

\subparagraph{Robin boundary condition}

A linear combination of the value of $\psi$ and its normal derivative on the boundary point $O$ is specified:
\begin{equation} \label{eq:robin}
    \alpha \psi_O + \beta \left. \frac{\partial \psi}{\partial n} \right|_O = g
\end{equation}
where $\alpha$ and $\beta$ are the linear combination coefficients, $g$ is a given value or function.

After approximating the normal derivative with second-order accuracy, it gives
\begin{equation} \label{eq:approxrobin}
    \alpha \psi_O + \beta \frac{\psi_I - \psi_O}{||\Vector{x}_I - \Vector{x}_O||} = g
\end{equation}

Then,
\begin{equation} \label{eq:resultrobin}
    \psi_O = \frac{\beta \psi_I - ||\Vector{x}_I - \Vector{x}_O|| g}{\beta - ||\Vector{x}_I - \Vector{x}_O|| \alpha}
\end{equation}

\subparagraph{Cauchy boundary condition}

For illustration purpose, $\psi$ is replaced by the velocity vector $\Vector{V}=(u,v,w)$ that satisfies the slip-wall boundary condition:
\begin{equation} \label{eq:slipwall}
    \begin{aligned}
        \left. (\Vector{V} \cdot \unitVector{n}) \right|_{\Vector{x}=\Vector{x}_O} &= \Vector{V}_{S} \cdot \unitVector{n} \\
        \left. \frac{\partial (\Vector{V} \cdot \hat{\Vector{t}})}{\partial n} \right|_{\Vector{x}=\Vector{x}_O} &= 0 \\
        \left. \frac{\partial (\Vector{V} \cdot \tilde{\Vector{t}})}{\partial n} \right|_{\Vector{x}=\Vector{x}_O} &= 0
    \end{aligned}
\end{equation}
where $\unitVector{n}$, $\hat{\unitVector{t}}$, and $\tilde{\unitVector{t}}$ are the unit normal vector, unit tangent vector, and unit bitangent vector at the boundary point $O$, respectively. $\Vector{V}_{S}$ is the velocity of the boundary surface at the point $O$.

After approximating the normal derivative with second-order accuracy, it gives
\begin{equation} \label{eq:velocity}
    \begin{aligned}
        u_O n_x + v_O n_y + w_O n_z &= u_{S} n_x + v_{S} n_y + w_{S} n_z \\
        u_O \hat{t}_x + v_O \hat{t}_y + w_O \hat{t}_z &= u_I \hat{t}_x + v_I \hat{t}_y + w_I \hat{t}_z \\
        u_O \tilde{t}_x + v_O \tilde{t}_y + w_O \tilde{t}_z &= u_I \tilde{t}_x + v_I \tilde{t}_y + w_I \tilde{t}_z
    \end{aligned}
\end{equation}

Using the orthogonality of the coefficient matrix, $\Vector{V}_O$ is determined as 
\begin{equation} \label{eq:determined}
    \begin{pmatrix}
        u_O \\
        v_O \\
        w_O
    \end{pmatrix}
    =
    \begin{bmatrix}
        n_x & n_y & n_z \\
        \hat{t}_x & \hat{t}_y & \hat{t}_z \\
        \tilde{t}_x & \tilde{t}_y & \tilde{t}_z
    \end{bmatrix}^{\Des{T}}
    \begin{pmatrix}
        u_{S} n_x + v_{S} n_y + w_{S} n_z \\
        u_I \hat{t}_x + v_I \hat{t}_y + w_I \hat{t}_z \\
        u_I \tilde{t}_x + v_I \tilde{t}_y + w_I \tilde{t}_z
    \end{pmatrix}
\end{equation}

The solution equations of $\psi_O$ for different types of boundary conditions now can be unified as
\begin{equation} \label{eq:unified}
    \psi_O = C \psi_I + \Des{R.R.H.S.}
\end{equation}
where the values of the coefficient $C$ and the rest right-hand side $\Des{R.R.H.S.}$ are summarized in Table~\ref{tab:bcmap}.
\begin{table}[!htbp]
    \centering
    \caption{Value map of $C$ and $\Des{R.R.H.S.}$ for different types of boundary conditions.}
    \label{tab:bcmap}
    \scriptsize
    \setlength{\tabcolsep}{3pt}
    \renewcommand{\arraystretch}{1.15}
    \begin{tabular}{lccc}
        \hline
        Type & Example form & $C$ & $\Des{R.R.H.S.}$\\
        \hline
        Dirichlet & $\psi_O = g$ & $0$ & $g$ \\[8pt]
        Neumann & $\left. \frac{\partial \psi}{\partial n} \right|_O = \frac{\partial \psi_O}{\partial n}$ & $1$ & $- ||\Vector{x}_I - \Vector{x}_O||\frac{\partial \psi_O}{\partial n}$ \\[8pt]
        Robin & $\alpha \psi_O + \beta \left. \frac{\partial \psi}{\partial n} \right|_O = g$ & $\frac{\beta}{\beta - ||\Vector{x}_I - \Vector{x}_O|| \alpha}$ & $\frac{- ||\Vector{x}_I - \Vector{x}_O|| g}{\beta - ||\Vector{x}_I - \Vector{x}_O|| \alpha}$ \\[8pt]
        Cauchy & $\begin{aligned} \left. (\Vector{V} \cdot \unitVector{n}) \right|_{\Vector{x}=\Vector{x}_O} &= \Vector{V}_{S} \cdot \unitVector{n} \\ \left. \frac{\partial (\Vector{V} \cdot \hat{\Vector{t}})}{\partial n} \right|_{\Vector{x}=\Vector{x}_O} &= 0 \\ \left. \frac{\partial (\Vector{V} \cdot \tilde{\Vector{t}})}{\partial n} \right|_{\Vector{x}=\Vector{x}_O} &= 0 \end{aligned}$ & $\begin{bmatrix} n_x & n_y & n_z \\ \hat{t}_x & \hat{t}_y & \hat{t}_z \\ \tilde{t}_x & \tilde{t}_y & \tilde{t}_z \end{bmatrix}^{\Des{T}} \begin{bmatrix} 0 & 0 & 0 \\ \hat{t}_x & \hat{t}_y & \hat{t}_z \\ \tilde{t}_x & \tilde{t}_y & \tilde{t}_z \end{bmatrix}$ & $\begin{bmatrix} n_x & n_y & n_z \\ \hat{t}_x & \hat{t}_y & \hat{t}_z \\ \tilde{t}_x & \tilde{t}_y & \tilde{t}_z \end{bmatrix}^{\Des{T}} \begin{bmatrix} n_x & n_y & n_z \\ 0 & 0 & 0 \\ 0 & 0 & 0 \end{bmatrix} \cdot \Vector{V}_S$\\
        \hline
    \end{tabular}
\end{table}

\paragraph{Correction step}

Solve the value of $\psi_I$ by adding the boundary point $O$ as a stencil node for the inverse distance weighting of $\psi_I$:
\begin{equation}\label{eq:correction}
    \psi_I = \frac{\sum w(d_N) \psi_N + w(d_O) \psi_O}{\sum w(d_N) + w(d_O)} = \frac{\psi_I^* + \frac{w(d_O)}{\sum w(d_N)}\psi_O}{1+\frac{w(d_O)}{\sum w(d_N)}}
\end{equation}
in which the repetition of calculations on fluid nodes can be avoided, since the sum of weights and sum of weighted values are already obtained in the prediction step.

Due to the unknown $\psi_I$ in Eq.~\eqref{eq:unified}, the solution equation of $\psi_O$ is coupled with the solution equation of $\psi_I$ in the correction step. To solve this problem, one method is a synchronous solving approach to solve $\psi_O$ and $\psi_I$ simultaneously:
\begin{equation} \label{eq:synchronous}
    \begin{cases}
        \psi_O = \ C \psi_I + \Des{R.R.H.S.} \\
        \psi_I = \ \frac{\psi_I^* + \frac{w(d_O)}{\sum w(d_N)}\psi_O}{1+\frac{w(d_O)}{\sum w(d_N)}}
    \end{cases}
\end{equation}

The other is an asynchronous solving approach: first, solve $\psi_O$ via approximating the unknown $\psi_I$ with the pre-estimated $\psi_I^*$; then, solve $\psi_I$ in the correction step.
\begin{equation} \label{eq:asynchronous}
    \begin{cases}
        \psi_O = C \psi_I^* + \Des{R.R.H.S.} \\
        \psi_I = \frac{\psi_I^* + \frac{w(d_O)}{\sum w(d_N)}\psi_O}{1+\frac{w(d_O)}{\sum w(d_N)}}
    \end{cases}
\end{equation}

The enforcement of the Dirichlet and trivial Neumann boundary conditions is equivalent in these two approaches. For the other types of boundary conditions, when the asynchronous solving approach is used, the physical boundary condition enforcement step and the correction step can be iteratively implemented. However, our numerical tests suggest that the effects of iterative implementation on the overall solution accuracy are insignificant.

The asynchronous solving approach without iterative implementation is adopted and examined herein, since the validity of the synchronous solving approach is established when the validity of the asynchronous solving approach is proved.

\section{Numerical implementation}\label{sec:implement}

\subsection{Fluid-solid coupling}

A partitioned fluid-solid interaction algorithm with second-order temporal accuracy, which is obtained via applying Strang splitting \citep{strang1968construction} to split physical processes, is employed to model the coupling between fluid and solid motions:
\begin{equation}
    \Vector{U}^{n+1} = \Soptr_{\Des{s}}(\frac{\Delta t}{2})\Soptr_{\Des{f}}(\frac{\Delta t}{2})\Soptr_{\Des{f}}(\frac{\Delta t}{2})\Soptr_{\Des{s}}(\frac{\Delta t}{2}) \Vector{U}^n
\end{equation}
where $\Vector{U}^{n}$ and $\Vector{U}^{n+1}$ denote the solution vectors of physical quantities at time $t^n$ and $t^{n+1}$, respectively; $\Soptr_{\Des{s}}$ and $\Soptr_{\Des{f}}$ represent the solution operators of solid dynamics and fluid dynamics, respectively.

\subsection{Governing equations and discretization}

\subsubsection{Fluid dynamics}

The motion of fluids is described by the conservative form of the three-dimensional Navier--Stokes equations in Cartesian coordinates: 
\begin{equation}
    \frac{\partial \Vector{U}}{\partial t}+\frac{\partial \Vector{F}_i}{\partial x_i} = \frac{\partial \Vector{F}^{\Des{v}}_i}{\partial x_i}+\Vector{\Phi}
\end{equation}

The vectors of conservative variables $\Vector{U}$, convective fluxes $\Vector{F}_i$, diffusive fluxes $\Vector{F}^{\Des{v}}_i$, and source terms $\Vector{\Phi}$ are as follows:
\begin{equation}
    \Vector{U} =
    \begin{pmatrix}
        \rho\\
        \rho V_j\\
        \rho e_{\Des{T}}
    \end{pmatrix}
    ,\,\,
    \Vector{F}_i =
    \begin{pmatrix}
        \rho V_i\\
        \rho V_i V_j + p \delta_{ij}\\
        (\rho e_{\Des{T}}+p) V_i
    \end{pmatrix}
    ,\,\,
    \Vector{F}^{\Des{v}}_i =
    \begin{pmatrix}
        0\\
        \tau_{ij}\\
        k \frac{\partial T}{\partial x_i} + \tau_{il} V_l
    \end{pmatrix}
    ,\,\,
    \Vector{\Phi} =
    \begin{pmatrix}
        0\\
        f^{\Des{b}}_j\\
        f^{\Des{b}}_l V_l
    \end{pmatrix}
\end{equation}
where $\rho$ is the density, $\Vector{V}$ is the velocity, $e_{\Des{T}} = e +\Vector{V}\cdot\Vector{V}/2$ is the specific total energy, $e$ is the specific internal energy, $p$ is the thermodynamic pressure, $\Tensor{\tau}$ is the viscous stress tensor, $T$ is the temperature, $k$ is the thermal conductivity, $\Vector{f}^{\Des{b}}$ represents external body forces such as gravity, $i$ is a free index, $j$ is an enumerator, $l$ is a dummy index. 

Currently, the closure of the system is through supplying the Newtonian fluid relation with the Stokes hypothesis
\begin{equation}
    \tau_{ij} = \mu\left(\frac{\partial V_i}{\partial x_j} + \frac{\partial V_j}{\partial x_i} - \frac{2}{3}(\nabla\cdot\Vector{V}) \delta_{ij}\right)
\end{equation}
and the perfect gas law
\begin{equation}
    \begin{gathered}
        p = \rho R T \\
        e = C_v T
    \end{gathered}
\end{equation}
where $\mu$ is the dynamic viscosity and is determined by the Sutherland viscosity law, $R$ is the specific gas constant, and $C_v$ is the specific heat capacity at constant volume.

The temporal integration is achieved via the third-order SSP Runge--Kutta method \citep{shu1988efficient, gottlieb2001strong}:
\begin{equation}
    \begin{aligned}
        &\Vector{U}^{(1)} = \Toptr\Vector{U}^n\\
        &\Vector{U}^{(2)} = 3/4\Vector{U}^n+1/4\Toptr\Vector{U}^{(1)}\\
        &\Vector{U}^{n+1} = 1/3\Vector{U}^n+2/3\Toptr\Vector{U}^{(2)}\\
        &\Toptr = (\Matrix{I} + \Delta t \Loptr)
    \end{aligned}
\end{equation}
where $\Matrix{I}$ is the identity matrix, operator $\Loptr = \Loptr_x + \Loptr_y + \Loptr_z$, $\Loptr_x$, $\Loptr_y$, and $\Loptr_z$ represent the spatial operators of $x$, $y$, and $z$ dimension, respectively.

For systems of conservation laws in multidimensional space, the discretization of spatial operators can be conducted using dimension-by-dimension \citep{shu1988efficient} or dimensional-splitting \citep{strang1968construction} approximation. While the former preserves temporal accuracy, the latter has a much less severe stability constraint. To guarantee discrete mass conservation, conservative discretization is adopted for all the spatial derivatives. Taking the $x$ dimension as an example, the flux derivative at a node $i$ is approximated as
\begin{equation}
    \left. \frac{\partial \Vector{F}}{\partial x} \right|_i = \frac{1}{\Delta x}\left[\hat{\Vector{F}}_{i+\frac{1}{2}}-\hat{\Vector{F}}_{i-\frac{1}{2}}\right]
\end{equation}
where $\Vector{F}$ represents either the convective flux vector or the diffusive flux vector, $\hat{\Vector{F}}_{i+{1}/{2}}$ is a numerical flux at the interface between the discretization interval $\Omega_i=[x_{i-{1}/{2}},x_{i+{1}/{2}}]$ and $\Omega_{i+1}=[x_{i+{1}/{2}},x_{i+{3}/{2}}]$.

In convective flux discretization, local characteristic splitting is employed to locally decompose the vector system into a set of scalar conservation laws. Then, scalar flux splitting is conducted to ensure upwinding property
\begin{equation}
    f(u) = f^+(u) + f^-(u), \,\, \frac{\mathrm{d} f^+(u)}{\mathrm{d} u} \ge 0, \,\, \frac{\mathrm{d} f^-(u)}{\mathrm{d} u} \le 0
\end{equation}
where $f$ is a scalar characteristic flux.

Since both the forward flux and backward flux are discretized in conservative form, the discretization of a scalar flux derivative has the form
\begin{equation}
    \left. \frac{\partial f}{\partial x} \right|_i = \frac{1}{\Delta x} \left[ \hat{f}_{i+\frac{1}{2}} - \hat{f}_{i-\frac{1}{2}} \right], \,\, \hat{f}_{i+\frac{1}{2}} = \hat{f}_{i+\frac{1}{2}}^+ + \hat{f}_{i+\frac{1}{2}}^-, \,\, \hat{f}_{i-\frac{1}{2}} = \hat{f}_{i-\frac{1}{2}}^+ + \hat{f}_{i-\frac{1}{2}}^-
\end{equation}

The fifth-order WENO scheme \citep{jiang1996efficient} is then applied for the reconstruction of numerical fluxes. Due to the symmetry between the numerical flux $\hat{f}_{i+{1}/{2}}^+$ and $\hat{f}_{i-{1}/{2}}^-$, only the reconstruction of the former is presented herein. To obtain the latter, one replaces all $+$ and $-$ signs in the superscript and subscript of each variable in the equations by the corresponding opposite signs $-$ and $+$, respectively.
\begin{equation}
    \hat{f}_{i+\frac{1}{2}}^{+} = \sum_{n=0}^{N} \omega_{n}^+ q_{n}^+(f_{i+n-N}^+, \dotsc, f_{i+n}^+), \,\, N=(r-1)=2
\end{equation}
where
\begin{equation}
    \begin{gathered}
        q_0^+(f_{i-2}^+, \dotsc, f_{i}^+) = (2f_{i-2}^+ - 7f_{i-1}^+ + 11f_{i}^+) / 6\\
        q_1^+(f_{i-1}^+, \dotsc, f_{i+1}^+) = (-f_{i-1}^+ + 5f_{i}^+ + 2f_{i+1}^+) / 6\\
        q_2^+(f_{i}^+, \dotsc, f_{i+2}^+) = (2f_{i}^+ + 5f_{i+1}^+ - f_{i+2}^+) / 6\\
        \omega_{n}^+ = \frac{\alpha_n^+}{\alpha_0^+ + \dotsb + \alpha_N^+}, \,\, \alpha_n^+ = \frac{C_n}{(\varepsilon + IS_n^+)^2}, \,\, \varepsilon = 10^{-6}\\
        C_0 = \frac{1}{10}, \,\, C_1 = \frac{6}{10}, \,\, C_2 = \frac{3}{10}\\
        IS_{0}^+ = \frac{13}{12}(f_{i-2}^+ - 2f_{i-1}^+ + f_{i}^+)^2 + \frac{1}{4}(f_{i-2}^+ - 4f_{i-1}^+ + 3f_{i}^+)^2\\
        IS_{1}^+ = \frac{13}{12}(f_{i-1}^+ - 2f_{i}^+ + f_{i+1}^+)^2 + \frac{1}{4}(f_{i-1}^+ - f_{i+1}^+)^2\\
        IS_{2}^+ = \frac{13}{12}(f_{i}^+ - 2f_{i+1}^+ + f_{i+2}^+)^2 + \frac{1}{4}(3f_{i}^+ - 4f_{i+1}^+ + f_{i+2}^+)^2
    \end{gathered}
\end{equation}
where $r$ is the number of candidate stencils, $q_n$ are the $r$-th order approximations of $\hat{f}_{i+1/2}$ on the candidate stencils $S_n=(x_{i+n-N},\dotsc,x_{i+n})$, $\omega_{n}$ are the actual weights of $q_n$, which are determined by the smoothness of solution in the candidate stencils $S_n$, as measured by $IS_n$, and $C_n$ are optimal weights to ensure that the convex combination of $q_n$ converges to a $(2r-1)$-th order approximation of $\hat{f}_{i+1/2}$ on the undivided stencil $S=(x_{i-N},\dotsc,x_{i+N})$ in smooth regions.
\begin{figure}[!htbp]
    \centering
    \includegraphics[trim = 0mm 25mm 0mm 18mm, clip, width=0.48\textwidth]{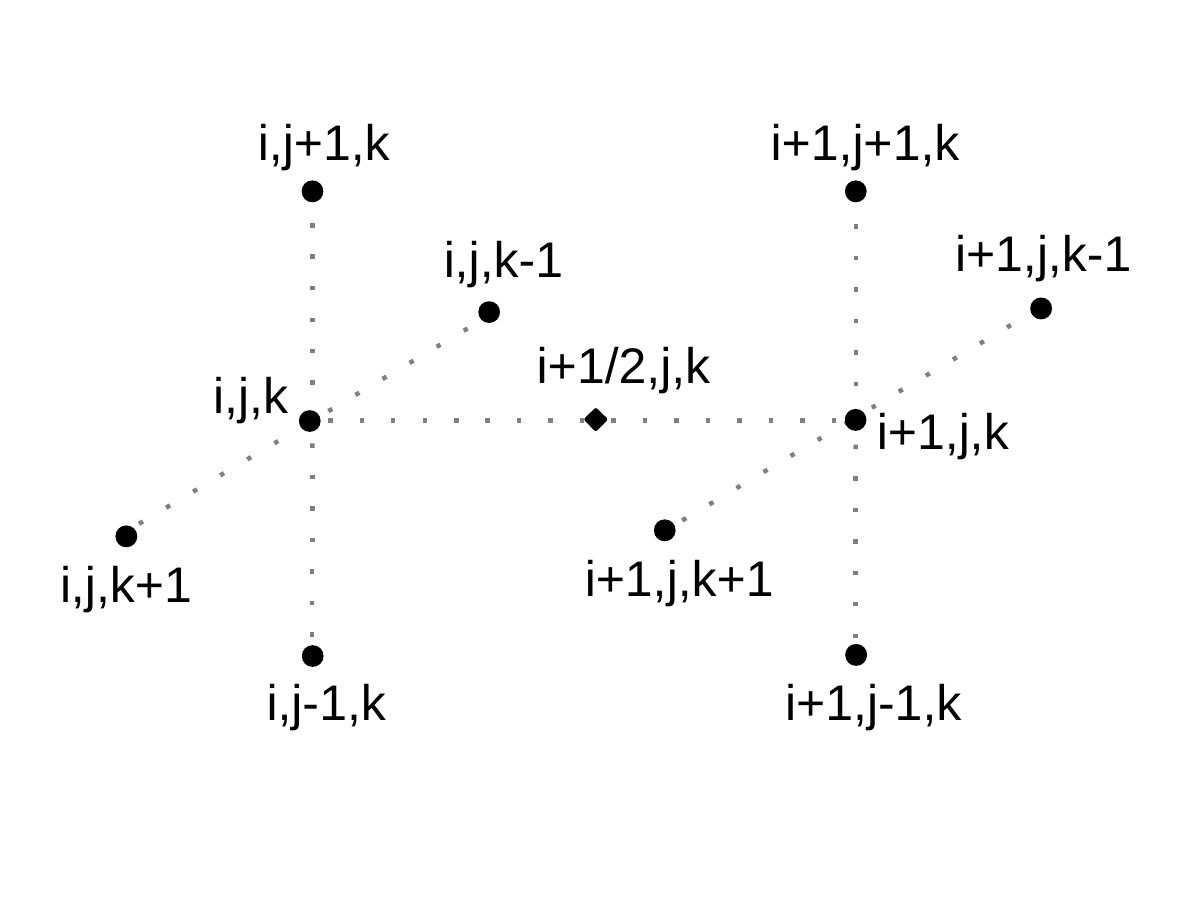}
    \caption{A schematic diagram illustrating interfacial diffusive flux reconstruction.}
    \label{fig:diffusive_flux_discretize}
\end{figure}

The second-order central difference discretization is employed for diffusive fluxes. A conservative discretization of the diffusive fluxes involves consecutive differentiation, which may lead to an even-odd decoupling issue \citep{pirozzoli2011numerical, brehm2015locally}. Therefore, the reconstruction function should be carefully devised. In this work, the interfacial flux $\hat{\Vector{F}}^{\Des{v}}_{i+{1}/{2}}$ is reconstructed on the discretized space $[i, i+1] \times [j-1, j+1] \times [k-1, k+1]$, as illustrated in Fig.~\ref{fig:diffusive_flux_discretize}. Let $\phi$ denote a physical quantity in $\Vector{F}^{\Des{v}}$, in order to avoid even-odd decoupling resulting from applying consecutive derivative discretization, the following reconstructions can be adopted:
\begin{equation}
    \begin{aligned}
        \phi_{i+\frac{1}{2}, j, k} &= \frac{\phi_{i, j, k} + \phi_{i+1, j, k}}{2} \\
        \left.\frac{\partial \phi}{\partial x}\right|_{i+\frac{1}{2}, j, k} &= \frac{\phi_{i+1, j, k} - \phi_{i, j, k}}{\Delta x} \\
        \left.\frac{\partial \phi}{\partial y}\right|_{i+\frac{1}{2}, j, k} &= \frac{\phi_{i, j+1, k} + \phi_{i+1, j+1, k} - \phi_{i, j-1, k} - \phi_{i+1, j-1, k}}{4\Delta y} \\
        \left.\frac{\partial \phi}{\partial z}\right|_{i+\frac{1}{2}, j, k} &= \frac{\phi_{i, j, k+1} + \phi_{i+1, j, k+1} - \phi_{i, j, k-1} - \phi_{i+1, j, k-1}}{4\Delta z}
    \end{aligned}
\end{equation}

\subsubsection{Solid dynamics}

The motion of solids is governed by the equation system comprising the Newton's second law of translational motion and the Euler equations of rotational motion:
\begin{equation}
    \frac{\mathrm{d} \Vector{U}}{\mathrm{d} t} = \Vector{\Phi}
    ,\,\,
    \Vector{U} =
    \begin{pmatrix}
        \Vector{V}\\
        \Vector{x}_{\Des{c}}\\
        \Matrix{I}_{\Des{c}} \Vector{\omega}\\
        \Vector{\theta}
    \end{pmatrix}
    ,\,\,
    \Vector{\Phi} =
    \begin{pmatrix}
        \frac{1}{m}\int\limits_{\partial\Omega} \unitVector{n} \cdot (-p \unitTensor{I} + \Tensor{\tau}) \, \mathrm{d}S + \Vector{g} \\
        \Vector{V}\\
        \int\limits_{\partial\Omega} (\Vector{x} - \Vector{x}_{\Des{c}}) \times [\unitVector{n} \cdot (-p \unitTensor{I} + \Tensor{\tau})] \, \mathrm{d}S\\
        \Vector{\omega}
    \end{pmatrix}
\end{equation}
where $\Vector{x}$ is the position vector of spatial points, $\Omega$ is the spatial domain occupied by a solid, $\Vector{x}_{\Des{c}}$ is the position vector of the solid centroid, $\Vector{\theta}$ is the orientation (vector of Euler angles) of the solid, $\Vector{V}$ and $\Vector{\omega}$ are the translational and angular velocities of the solid, respectively, $m$ is the mass of the solid, $\Matrix{I}_{\Des{c}}$ is the moment of inertia matrix, $\unitVector{n}$ is the unit outward surface normal vector, $p$ and $\Tensor{\tau}$ are the pressure and viscous stress tensor field exerted on the solid surface via fluid, respectively, and $\Vector{g}$ is the body force per unit mass, such as gravitational acceleration, exerted by external fields.

The time integration of the ordinary differential equation system is via a second-order Runge--Kutta scheme:
\begin{equation}
    \begin{aligned}
        &\Vector{k}_1 = \Vector{\Phi}(t^n, \Vector{U}^n)\\
        &\Vector{k}_2 = \Vector{\Phi}(t^n + \Delta t, \Vector{U}^n + \Delta t \Vector{k}_1)\\
        &\Vector{U}^{n+1} = \Vector{U}^n + \Delta t(\Vector{k}_1 + \Vector{k}_2) / 2
    \end{aligned}
\end{equation}

\subsection{Interface description and evolution}

Interface description and evolution are conducted using the front-tracking method \citep{tryggvason2001front}. Therefore, interfaces are  explicitly tracked via individual Lagrangian grids such as triangulated facets, and the evolution of interfaces is governed by the laws of solid motion.

\section{Numerical experiments} \label{sec:validity}

\subsection{Supersonic flow over a wedge}

\begin{figure}[!htbp]
    \centering
    \begin{subfigure}[b]{0.4\textwidth}
        \includegraphics[width=\textwidth]{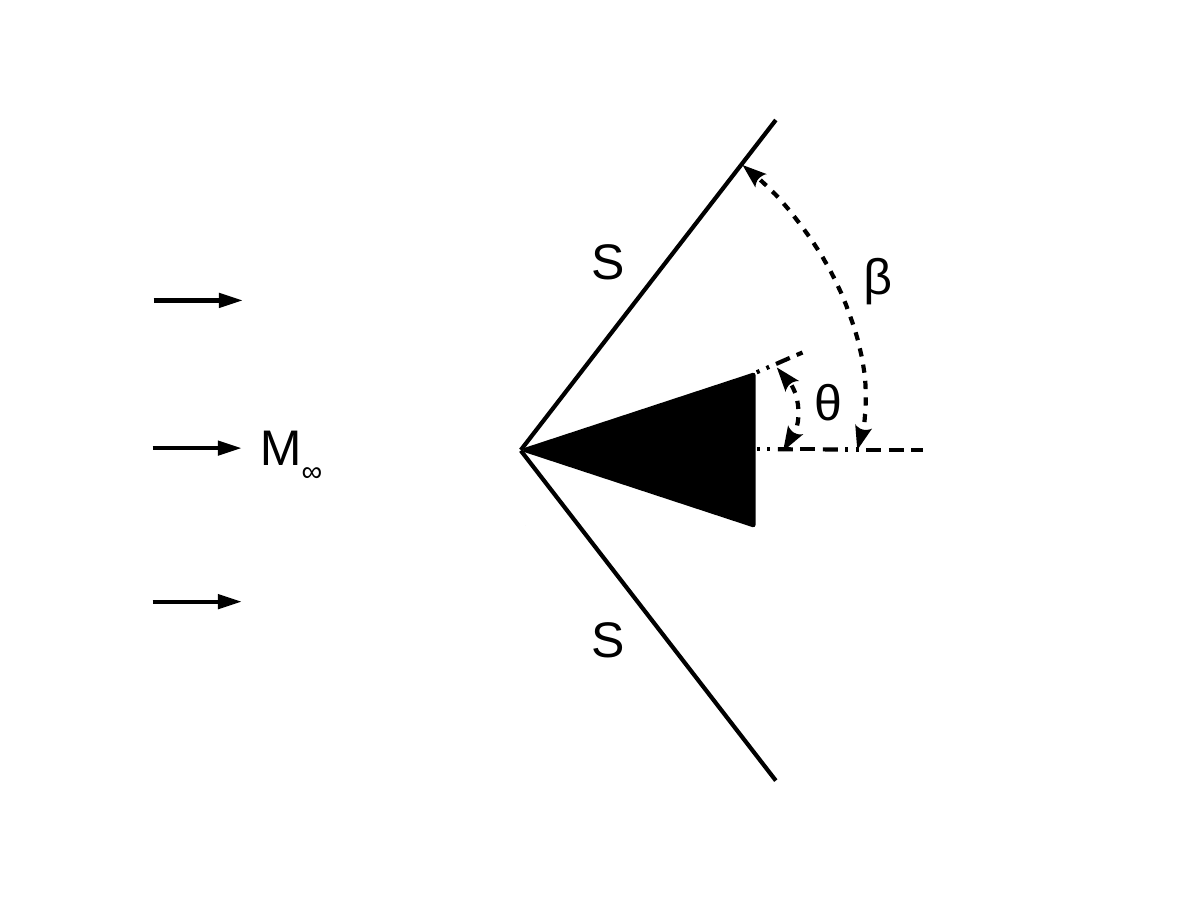}
        \caption{}
        \label{fig:flow_wedge_demo_a}
    \end{subfigure}%
    ~
    \begin{subfigure}[b]{0.4\textwidth}
        \includegraphics[width=\textwidth]{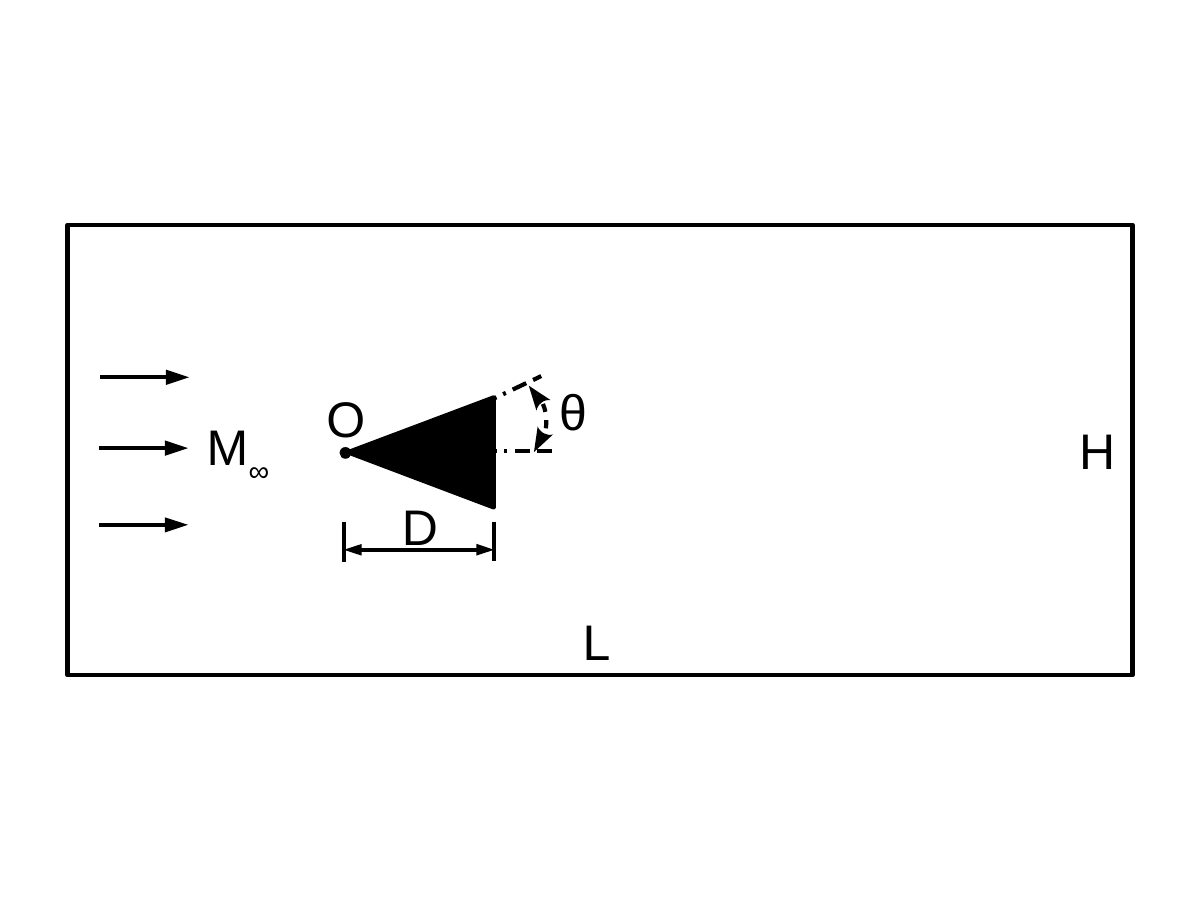}
        \caption{}
        \label{fig:flow_wedge_demo_b}
    \end{subfigure}%
    \caption{Schematic diagrams for the supersonic flow over a wedge problem. (a) Oblique shock relation. (b) Computational configuration. [Nomenclature: $M_{\infty}$, Mach number of the approaching flow; $S$, oblique shock; $\theta$, deflection angle; $\beta$, shock angle; $D$, length of wedge; $O$, the front vertex of wedge; $L$, domain length; $H$, domain height.] Schematic diagrams adapted from \citet{anderson2010fundamentals}.}
    \label{fig:flow_wedge_demo}
\end{figure}

As illustrated in Fig.~\ref{fig:flow_wedge_demo_a}, when a supersonic flow with Mach number $M_{\infty}$ passes over a wedge with an adequate deflection angle $\theta$, stationary oblique shock waves with a shock angle $\beta$ can be created at the nose of the wedge. The $M_{\infty}-\theta-\beta$ relation can be analytically obtained via a control volume analysis based on conservation laws and has the following form \citep{anderson2010fundamentals}:
\begin{equation}
    \tan \theta = \frac{2}{\tan \beta} \frac{M_{\infty}^2 \sin^2 \beta - 1}{M_{\infty}^2 (\gamma + \cos(2\beta)) + 2}
\end{equation}
where $\gamma$ is the heat capacity ratio.

To validate the proposed method, this supersonic flow over a wedge problem is solved. As illustrated in Fig.~\ref{fig:flow_wedge_demo_b}, in a $L \times H = [-0.5D, 9.5D] \times [-2.5D, 2.5D]$ domain with an initial flow state of $(\rho_0, u_0, v_0, p_0)=(1.4 \Unit{kg/m^3}, 40 \Unit{m/s}, 0, 400 \Unit{Pa})$, in which the speed of sound is $a_0=20 \Unit{m/s}$, a stationary wedge with $D=1 \Unit{m}$ is positioned at $O(0,0)$ and has the slip-wall boundary condition. The right domain boundary has the outflow condition, and the top and bottom boundaries are treated as slip walls. The left domain boundary has an inflow condition $(\rho_{\infty}, u_{\infty}, v_{\infty}, p_{\infty})=(\rho_0, M_{\infty} a_0, 0, p_0)$, and the inviscid flow is solved to $t=4L/(M_{\infty}a_0)$ for obtaining a steady state.

In order to simplify the discussion, notation $M_{\infty}(\hat{M})-\theta(\hat{\theta})-\beta_e(\hat{\beta}_e)-\beta_n(\hat{\beta}_n)-G(\hat{G})$ is used to denote a case with Mach number $\hat{M}$, deflection angle $\hat{\theta}$, analytical shock angle $\hat{\beta}_e$, and numerical shock angle $\hat{\beta}_n$ solved on grid $\hat{G}$. It is worth mentioning that the measurement of numerical shock angles is through manually picking up two points at the center of the computed oblique shock line and then computing the slope angle of that line. Therefore, the presented numerical shock angles herein are subject to sampling errors. To better facilitate the comparison of numerical and analytical solutions, straight lines with slope angles being equal to the analytical shock angles are visualized in the figures of results.

\begin{figure}[!htbp]
    \centering
    \begin{subfigure}[b]{0.48\textwidth}
        \includegraphics[width=\textwidth]{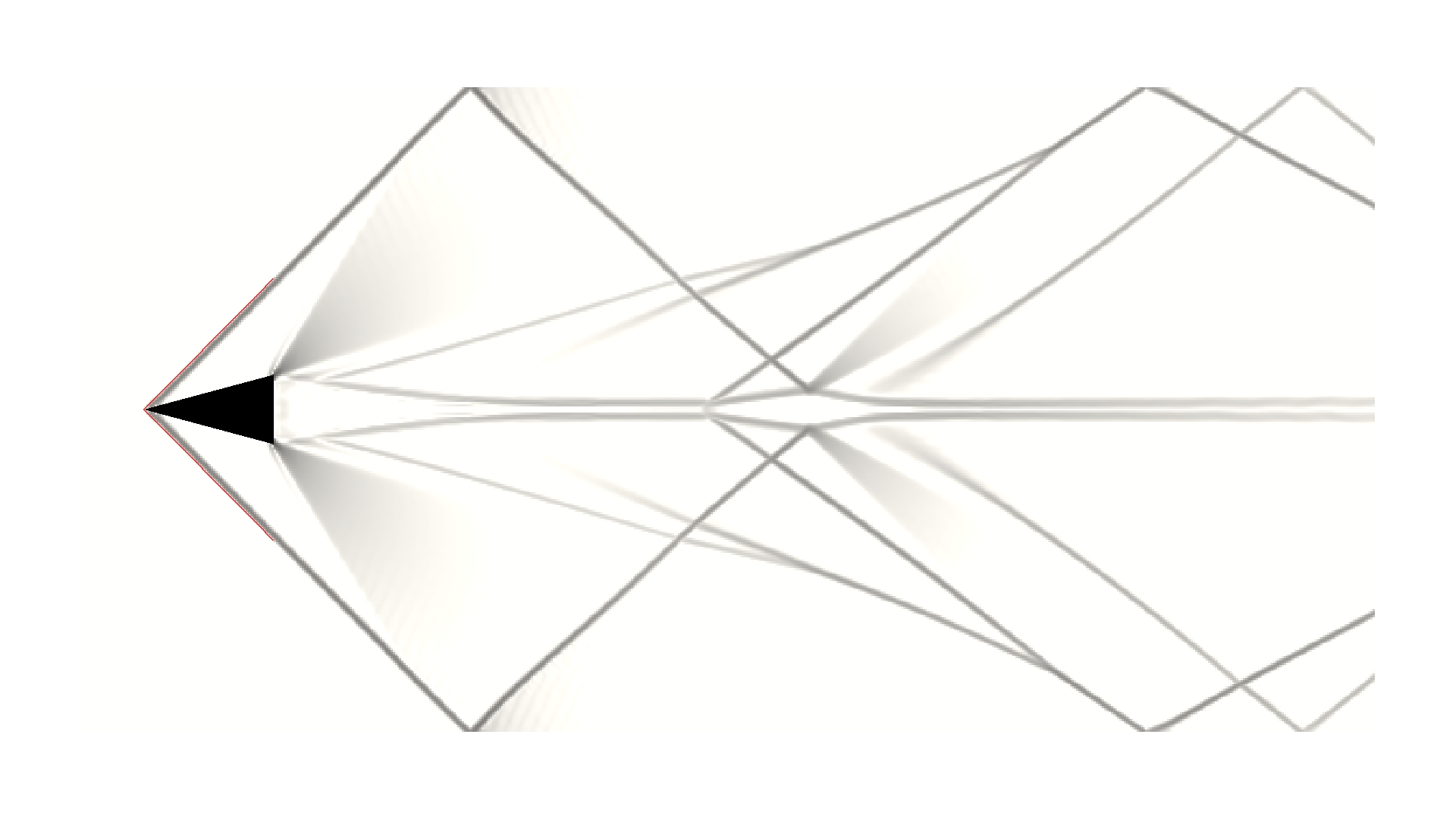}
        \caption{}
        \label{fig:1_wedge_nomv_deg15_mach2_m0600_lined}
    \end{subfigure}%
    ~
    \begin{subfigure}[b]{0.48\textwidth}
        \includegraphics[width=\textwidth]{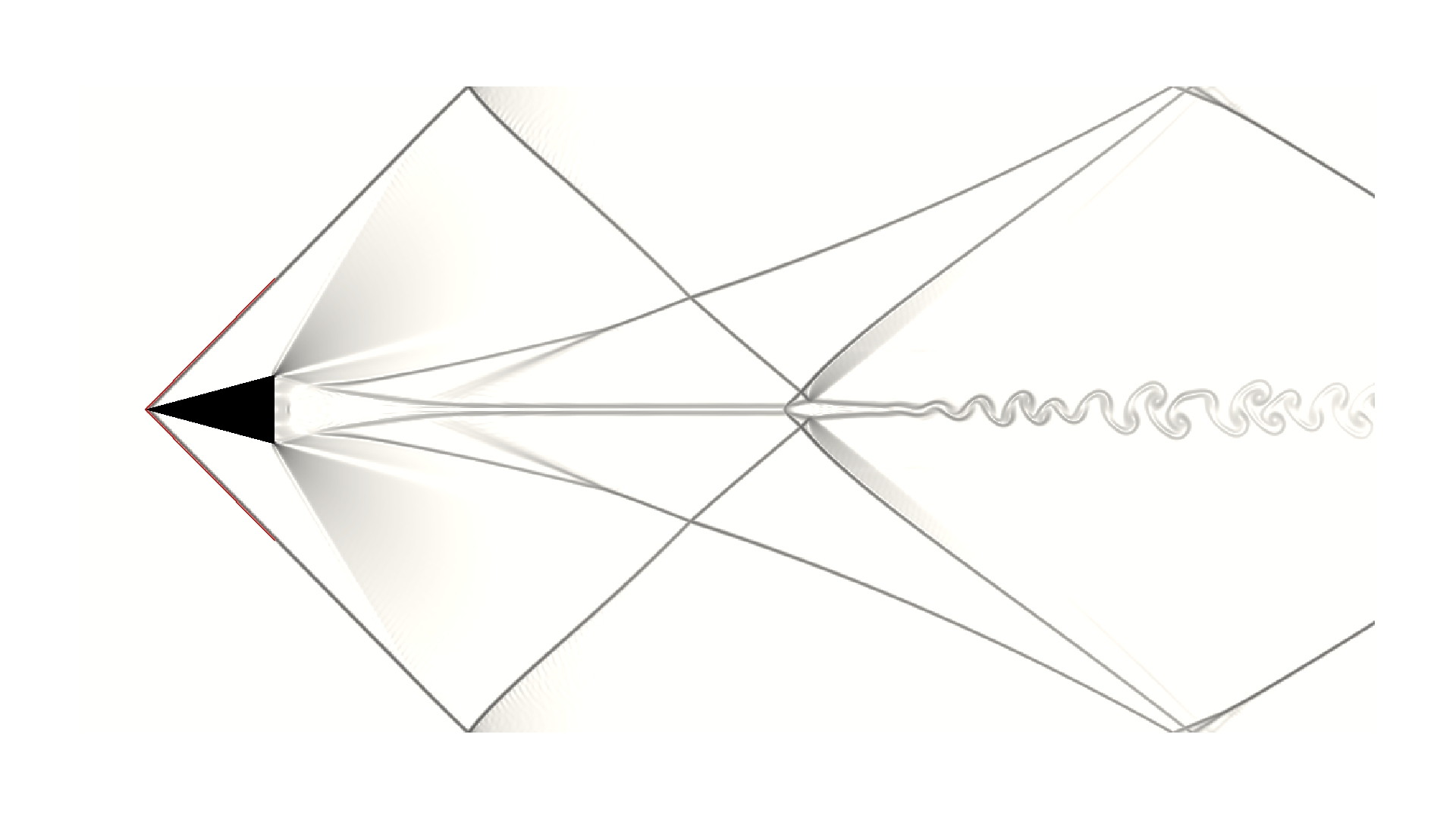}
        \caption{}
        \label{fig:1_wedge_nomv_deg15_mach2_m1200_lined}
    \end{subfigure}%
    \\
    \begin{subfigure}[b]{0.48\textwidth}
        \includegraphics[width=\textwidth]{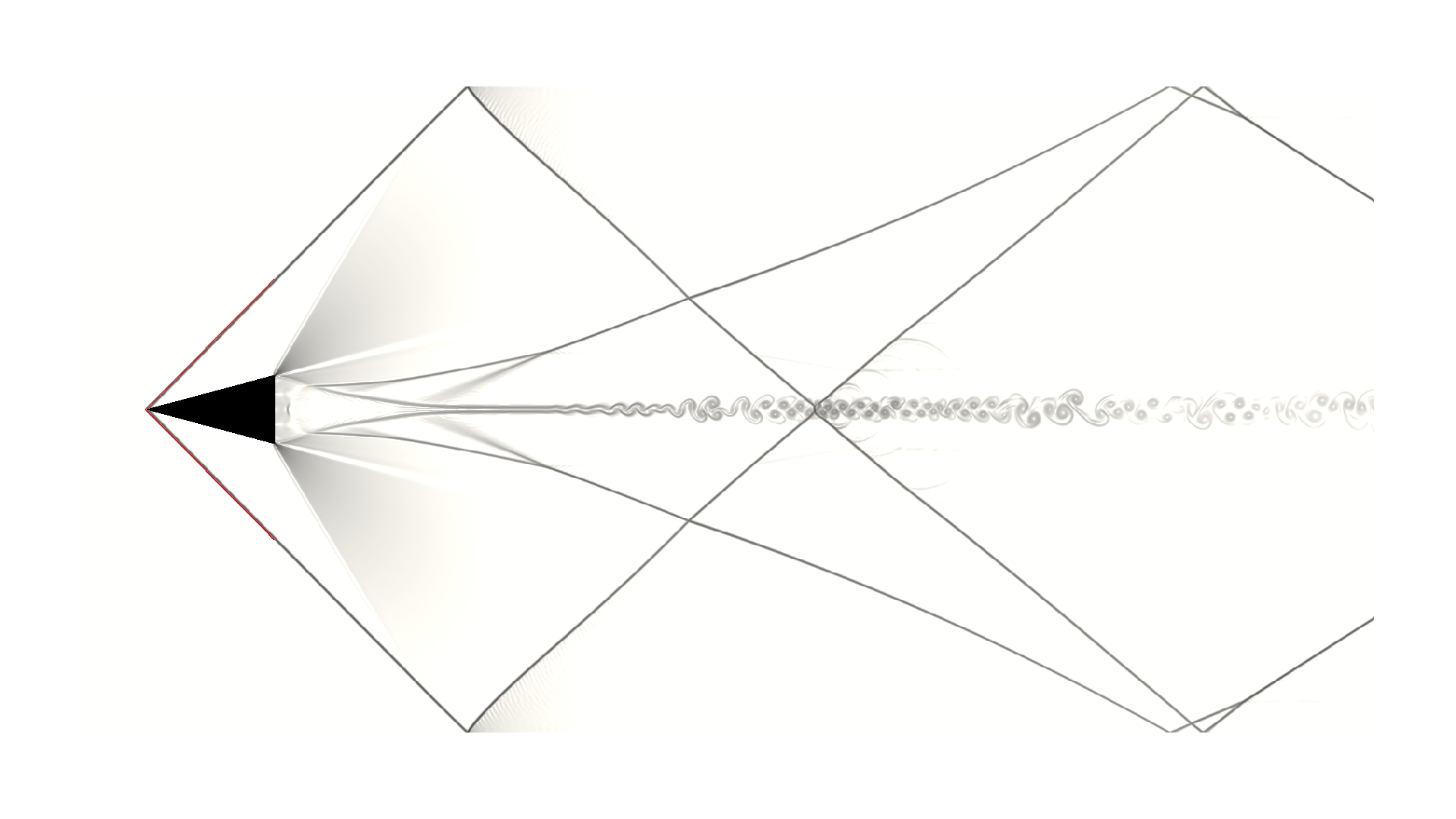}
        \caption{}
        \label{fig:1_wedge_nomv_deg15_mach2_m1800_lined}
    \end{subfigure}%
    ~
    \begin{subfigure}[b]{0.48\textwidth}
        \includegraphics[width=\textwidth]{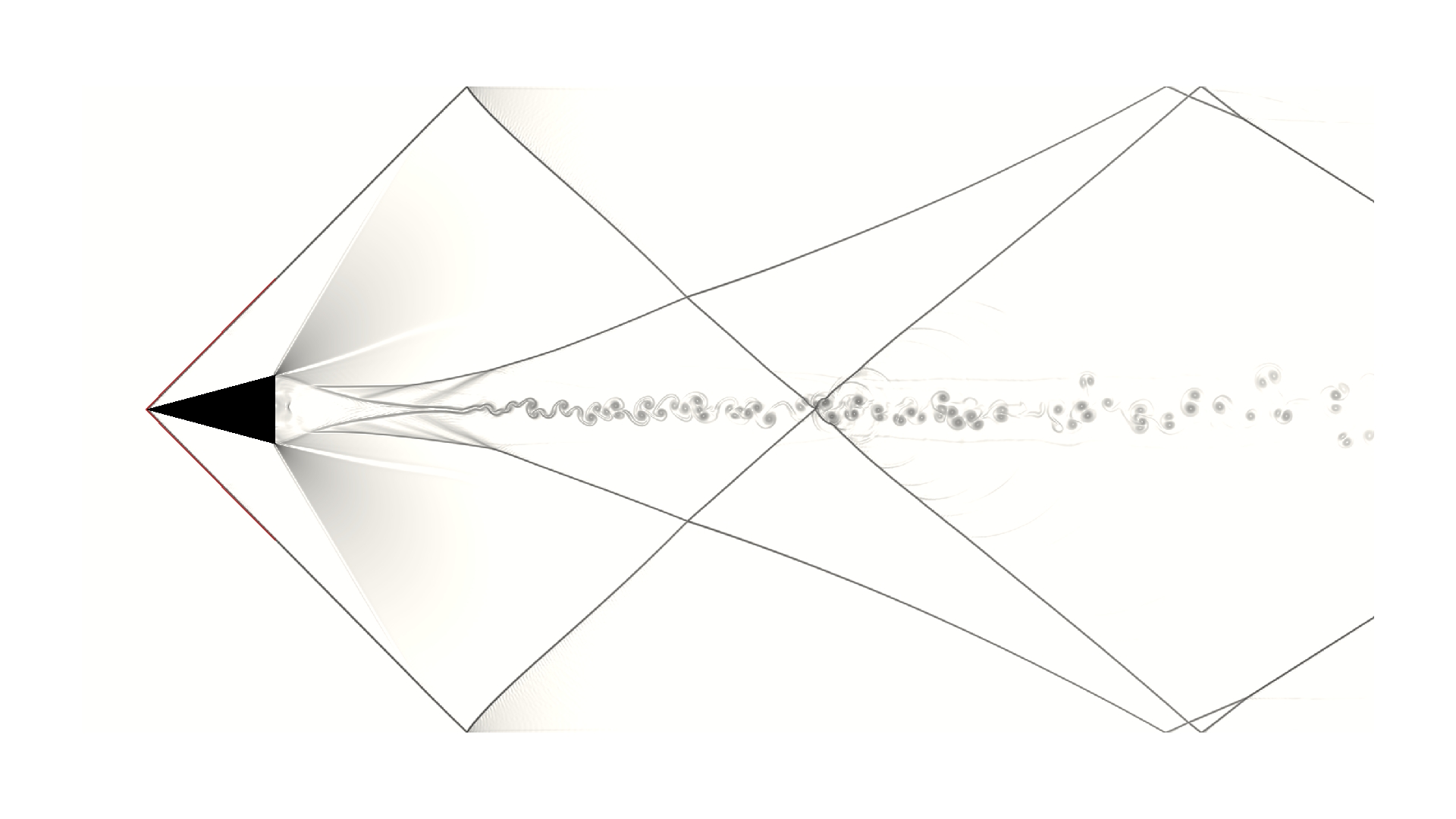}
        \caption{}
        \label{fig:1_wedge_nomv_deg15_mach2_m2400_lined}
    \end{subfigure}%
    \caption{Grid sensitivity study of supersonic flow over a wedge for case $M_{\infty}(2)-\theta(15^{\circ})-\beta_e(45.344^{\circ})$. Red lines represent the analytical solutions (closely overlapped with numerical solutions). (a) $\beta_n(44.927^{\circ})-G(600\times300)$. (b) $\beta_n(45.198^{\circ})-G(1200\times600)$. (c) $\beta_n(45.726^{\circ})-G(1800\times900)$. (d) $\beta_n(45.352^{\circ})-G(2400\times1200)$.}
    \label{fig:1_wedge_nomv_deg15_mach2_mesh}
\end{figure}

A grid sensitivity study is conducted for case $M_{\infty}(2)-\theta(15^{\circ})-\beta_e(45.344^{\circ})$ on a series of successively refined grids, and the results on four chosen grids are shown in Fig.~\ref{fig:1_wedge_nomv_deg15_mach2_mesh}. As the grid resolution increases, the sharpness of the computed shocks improves accordingly, and the perturbations in the wakes are less dissipated. The generation of oscillating wakes in the inviscid flow may be partially due to the numerical viscosity in the numerical schemes. Nonetheless, the predicted oblique shock angles are in excellent agreement with the analytical solution over a wide range of grid resolution, and Prandtl-Meyer expansion waves are also physically resolved at each rear corner of the wedge. Since good numerical accuracy and shock sharpness can be achieved on the $1200\times600$ grid, this grid is employed for the investigation of other cases.

\begin{figure}[!htbp]
    \centering
    \begin{subfigure}[b]{0.65\textwidth}
        \includegraphics[width=\textwidth]{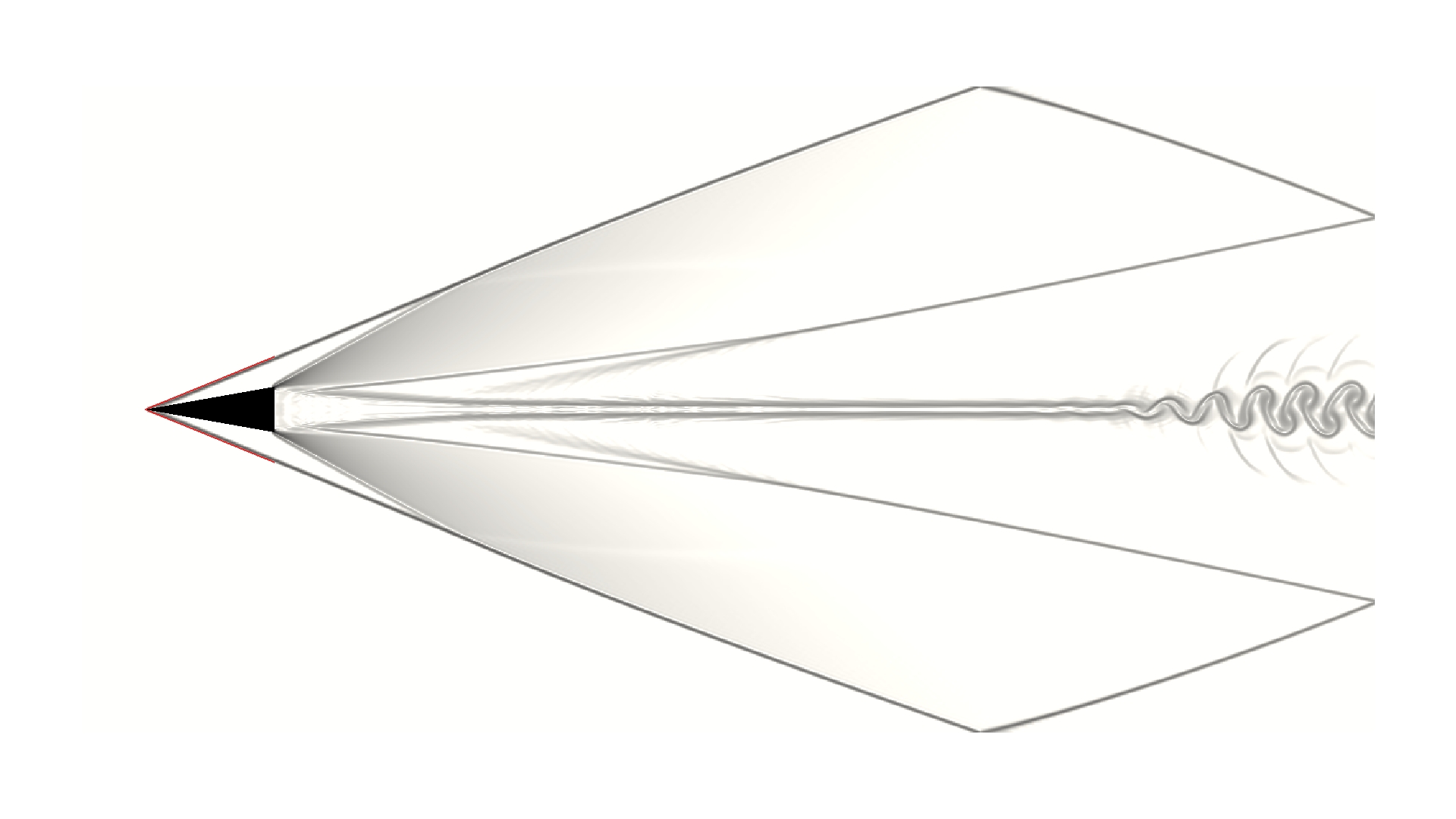}
        \caption{}
        \label{fig:1_wedge_nomv_deg10_mach4_m1200_lined}
    \end{subfigure}%
    \\
    \begin{subfigure}[b]{0.65\textwidth}
        \includegraphics[width=\textwidth]{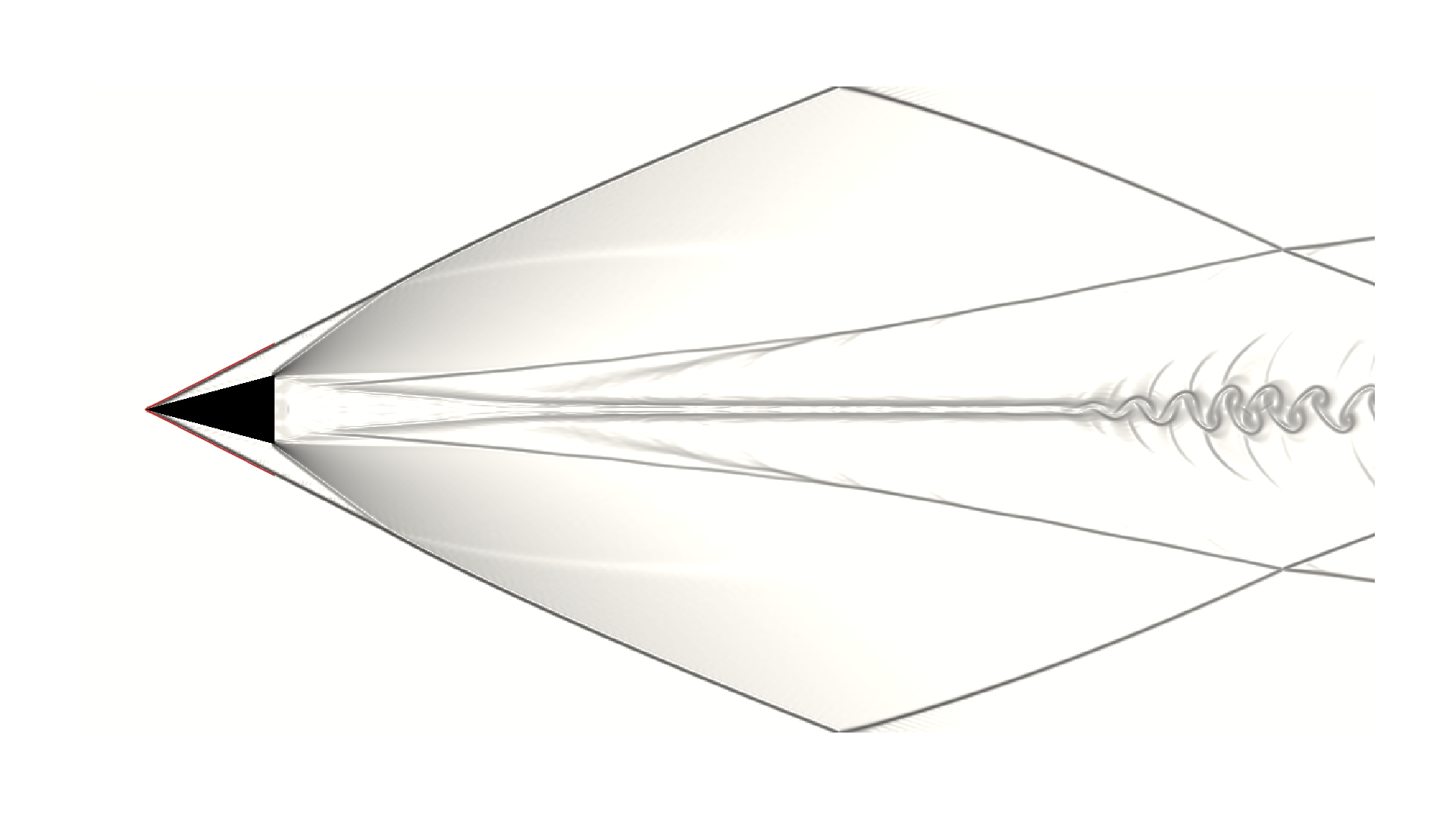}
        \caption{}
        \label{fig:1_wedge_nomv_deg15_mach4_m1200_lined}
    \end{subfigure}%
    \\
    \begin{subfigure}[b]{0.65\textwidth}
        \includegraphics[width=\textwidth]{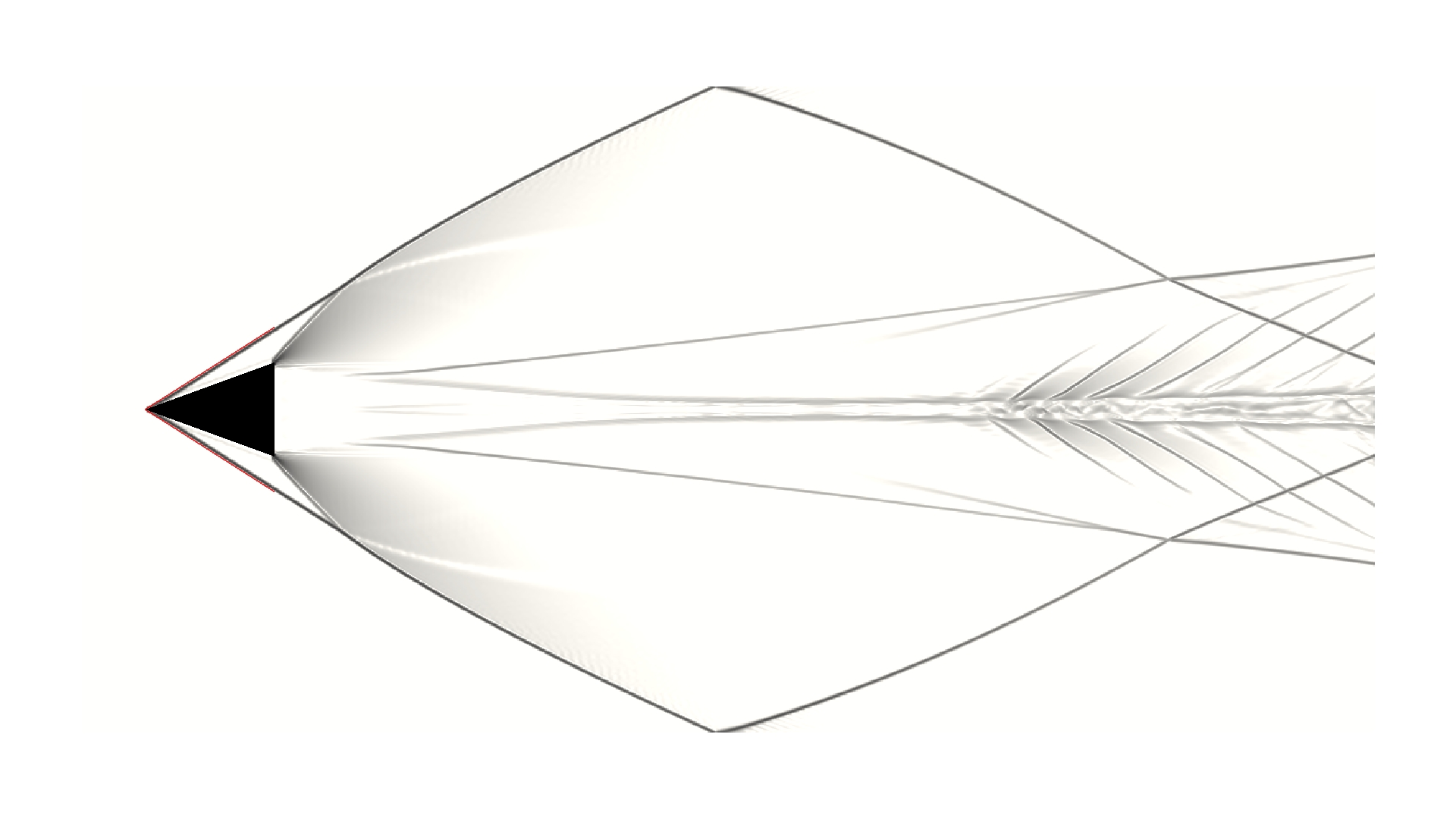}
        \caption{}
        \label{fig:1_wedge_nomv_deg20_mach4_m1200_lined}
    \end{subfigure}%
    \caption{Oblique shock relation of supersonic flow over a wedge for case $M_{\infty}(4)-G(1200\times600)$ over different deflection angles. Red lines represent the analytical solutions (closely overlapped with numerical solutions). (a) $\theta(10^{\circ})-\beta_e(22.234^{\circ})-\beta_n(22.227^{\circ})$. (b) $\theta(15^{\circ})-\beta_e(27.063^{\circ})-\beta_n(27.325^{\circ})$. (c) $\theta(20^{\circ})-\beta_e(32.464^{\circ})-\beta_n(32.293^{\circ})$.}
    \label{fig:1_wedge_nomv_deg}
\end{figure}

\begin{figure}[!htbp]
    \centering
    \begin{subfigure}[b]{0.65\textwidth}
        \includegraphics[width=\textwidth]{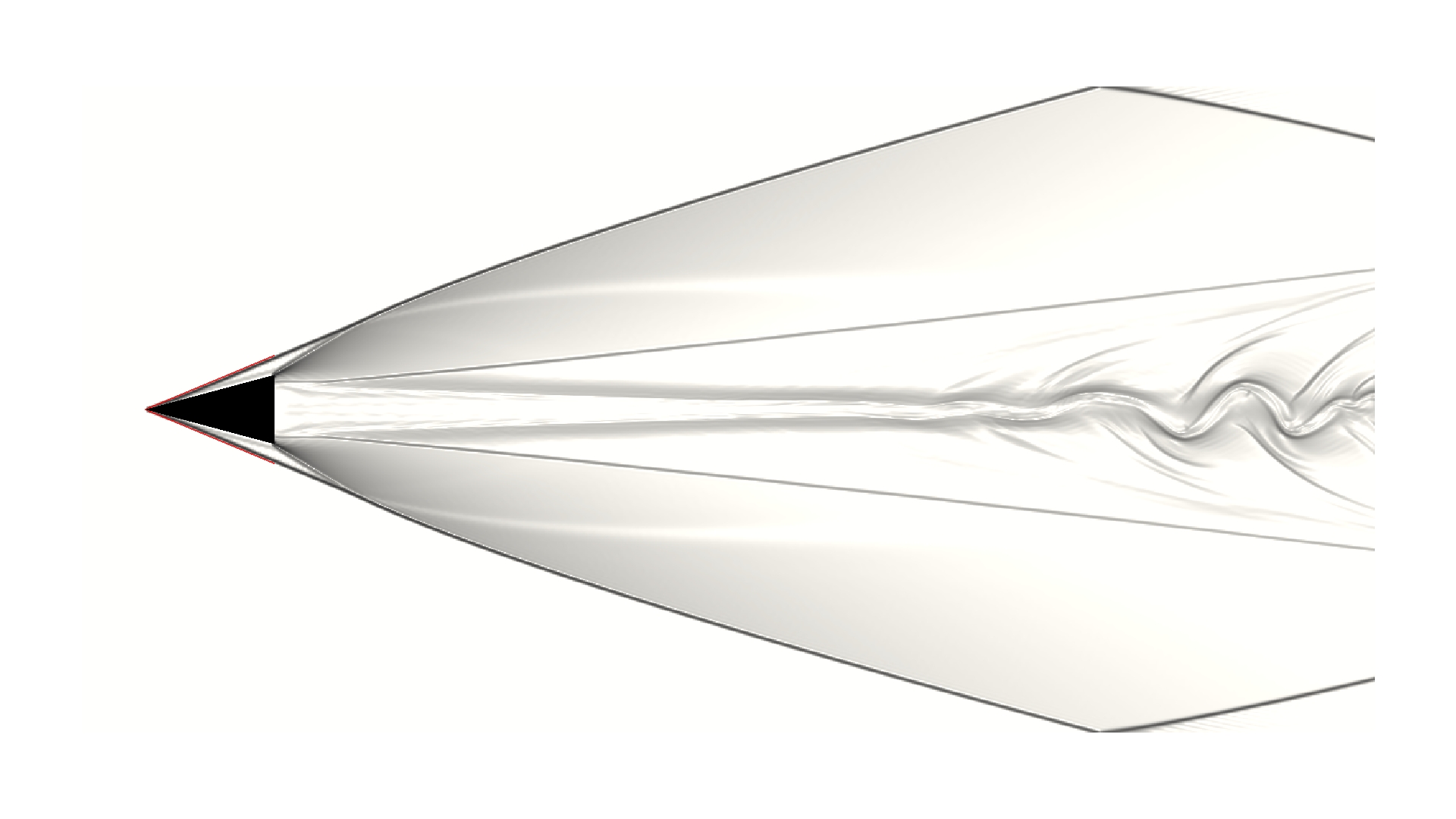}
        \caption{}
        \label{fig:1_wedge_nomv_deg15_mach6_m1200_lined}
    \end{subfigure}%
    \\
    \begin{subfigure}[b]{0.65\textwidth}
        \includegraphics[width=\textwidth]{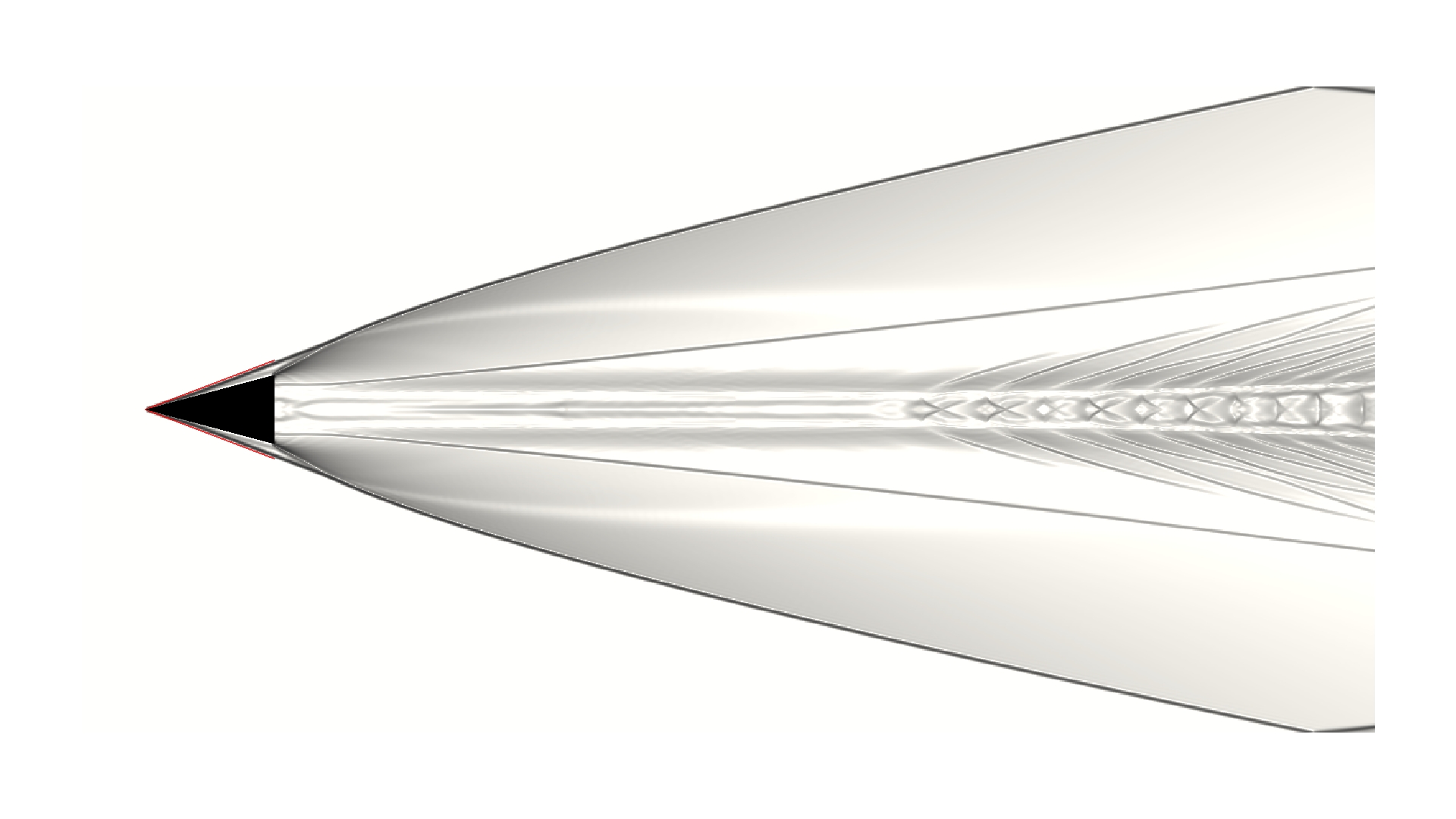}
        \caption{}
        \label{fig:1_wedge_nomv_deg15_mach8_m1200_lined}
    \end{subfigure}%
    \\
    \begin{subfigure}[b]{0.65\textwidth}
        \includegraphics[width=\textwidth]{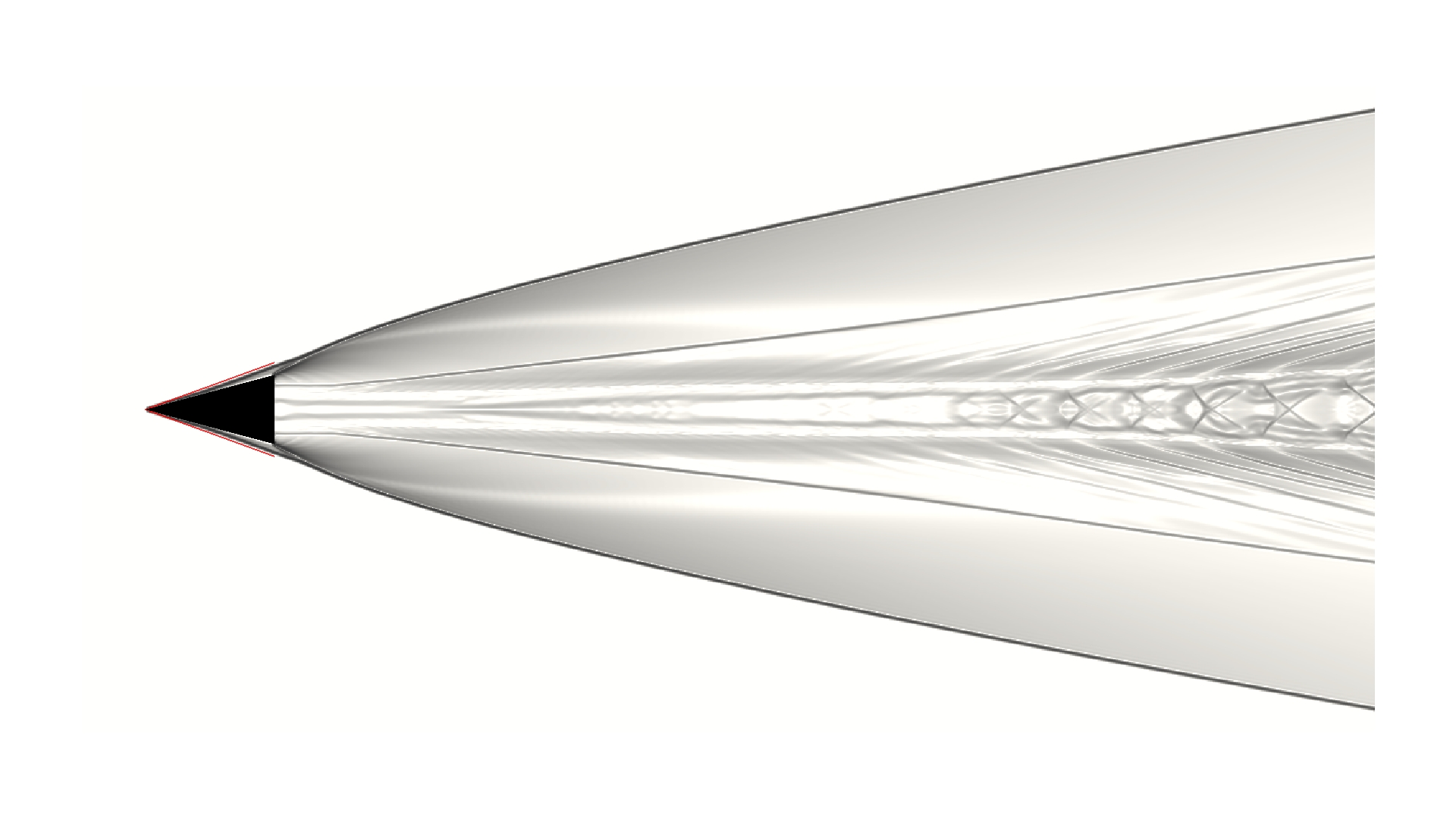}
        \caption{}
        \label{fig:1_wedge_nomv_deg15_mach10_m1200_lined}
    \end{subfigure}%
    \caption{Oblique shock relation of supersonic flow over a wedge for case $\theta(15^{\circ})-G(1200\times600)$ over different Mach numbers. Red lines represent the analytical solutions (closely overlapped with numerical solutions). (a) $M_{\infty}(6)-\beta_e(22.672^{\circ})-\beta_n(22.585^{\circ})$. (b) $M_{\infty}(8)-\beta_e(20.860^{\circ})-\beta_n(20.595^{\circ})$. (c) $M_{\infty}(10)-\beta_e(19.942^{\circ})-\beta_n(18.930^{\circ})$.}
    \label{fig:1_wedge_nomv_mach}
\end{figure}

The prediction capability on oblique shock relation over different deflection angles and Mach numbers are illustrated in Fig.~\ref{fig:1_wedge_nomv_deg} and Fig.~\ref{fig:1_wedge_nomv_mach}, respectively. Under the same Mach number, a higher deflection angle leads to a higher shock angle. For a fixed deflection angle, as the Mach number increases, the shock angle decreases. Due to the finite length of the wedge, oblique shocks with low shock angles may strongly interfere with the Prandtl-Meyer expansion waves and incline toward the rear corners of the wedge. For the oblique shock angles, the excellent agreement between the numerical and analytical solutions is consistently presented among all the cases, demonstrating the high validity of the proposed method.

\subsection{Supersonic translating wedge}

\begin{figure}[!htbp]
    \centering
    \begin{subfigure}[b]{0.4\textwidth}
        \includegraphics[width=\textwidth]{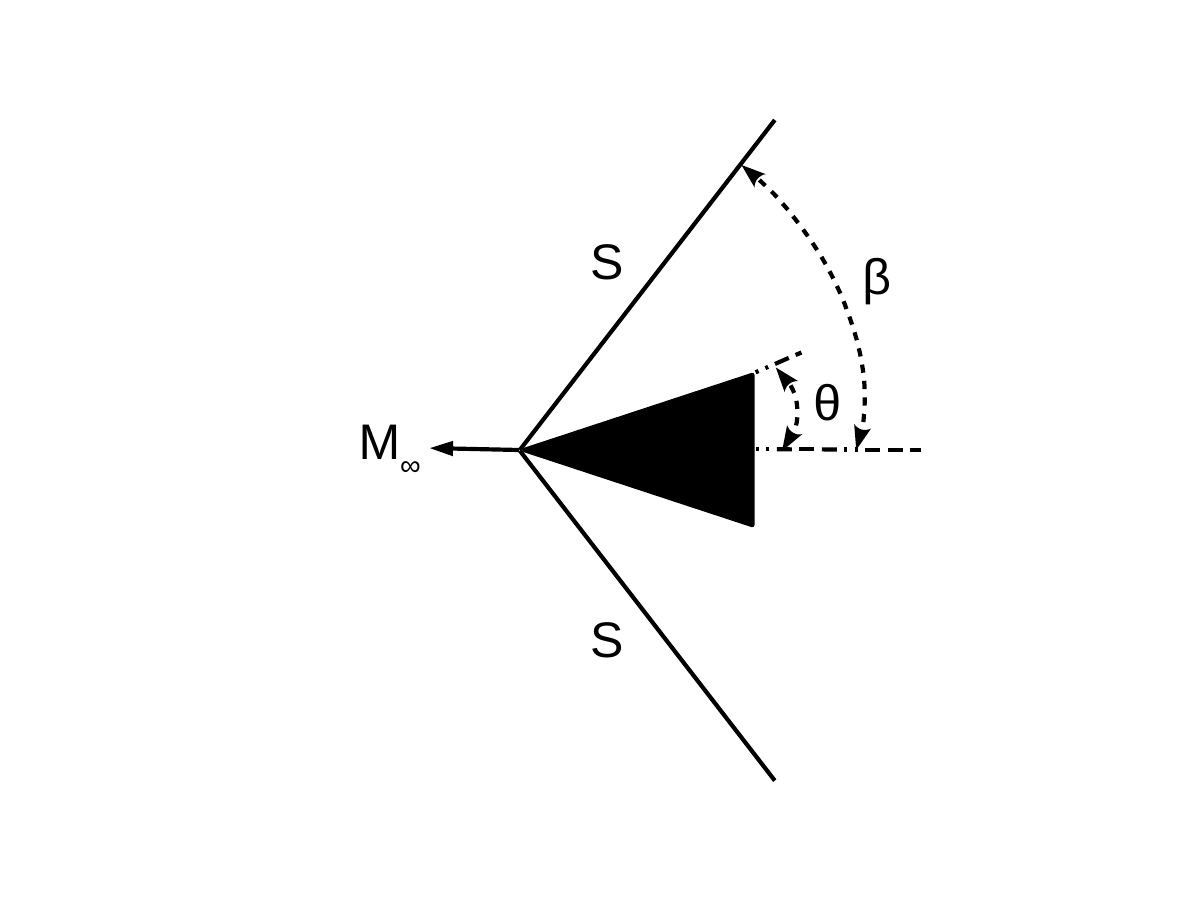}
        \caption{}
        \label{fig:flow_wedge_demo_c}
    \end{subfigure}%
    ~
    \begin{subfigure}[b]{0.4\textwidth}
        \includegraphics[width=\textwidth]{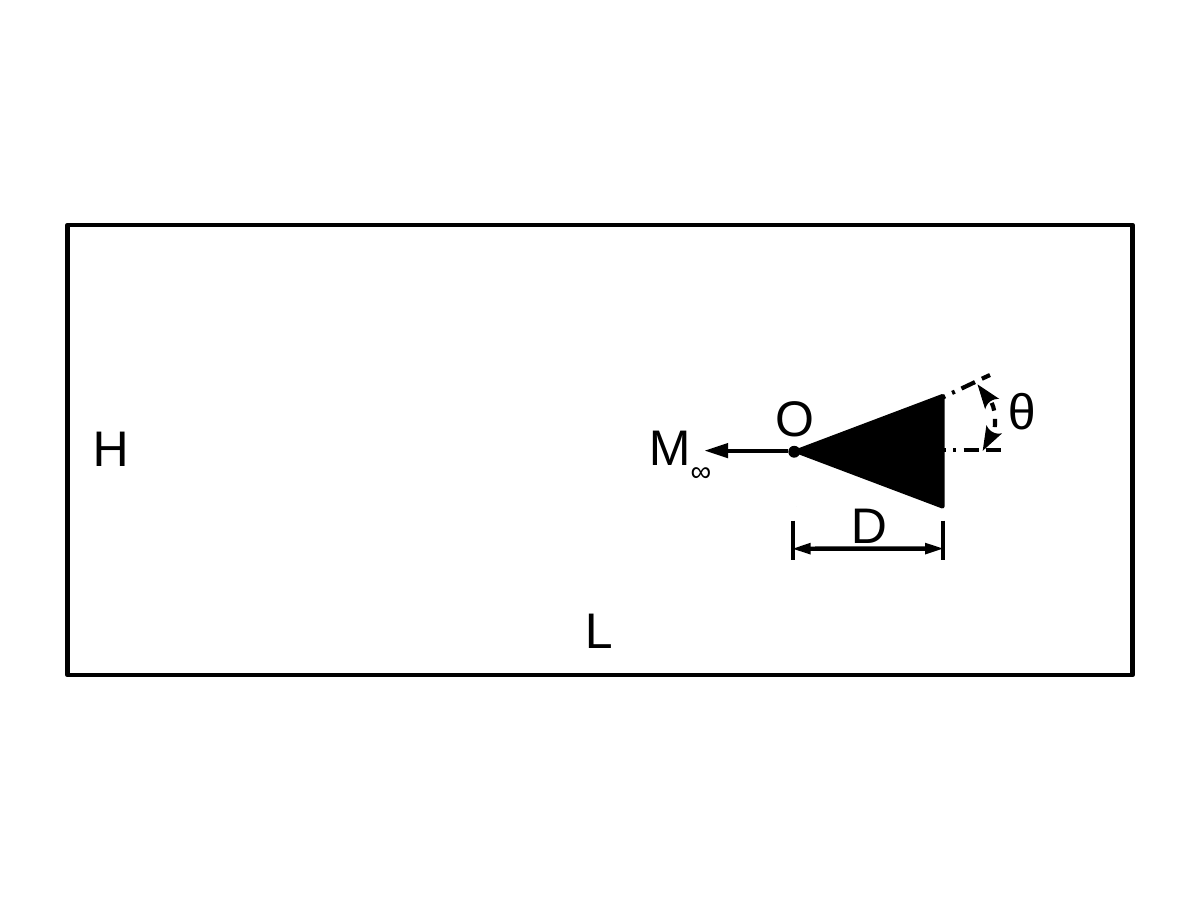}
        \caption{}
        \label{fig:flow_wedge_demo_d}
    \end{subfigure}%
    \caption{Schematic diagrams for the supersonic translating wedge problem. (a) Oblique shock relation. (b) Computational configuration. [Nomenclature: $M_{\infty}$, Mach number of the moving wedge; $S$, oblique shock; $\theta$, deflection angle; $\beta$, shock angle; $D$, length of wedge; $O$, the front vertex of wedge; $L$, domain length; $H$, domain height.] Schematic diagrams adapted from \citet{anderson2010fundamentals}.}
    \label{fig:wedge_flow_demo}
\end{figure}

In order to test the proposed method for solving flow involving moving geometries, the supersonic translating wedge problem is solved, which is a Galilean transformation of the supersonic flow over a wedge problem. As illustrated in Fig.~\ref{fig:wedge_flow_demo}, in a $L \times H = [-0.5D, 9.5D] \times [-2.5D, 2.5D]$ domain with an initial flow state of $(\rho_0, u_0, v_0, p_0)=(1.4 \Unit{kg/m^3}, 0, 0, 400 \Unit{Pa})$, in which the speed of sound is $a_0=20 \Unit{m/s}$, a wedge with $D=1 \Unit{m}$ and $M_{\infty}=2$ is initially positioned at $O(8D,0)$ and has the slip-wall boundary condition. The outflow condition is enforced at the left and right domain boundaries, while the slip-wall condition is imposed at the top and bottom boundaries. To limit the required size of the computational domain, the evolution is solved to $t=0.8L/(M_{\infty}a_0)$. Although the most parts of the flow region are still in an unsteady state, and the transient perturbations generated from the sudden motion of the wedge in the initial stationary flow are not advected out of the domain, the oblique shocks are well-developed within the given computational time.

\begin{figure}[!htbp]
    \centering
    \begin{subfigure}[b]{0.48\textwidth}
        \includegraphics[width=\textwidth]{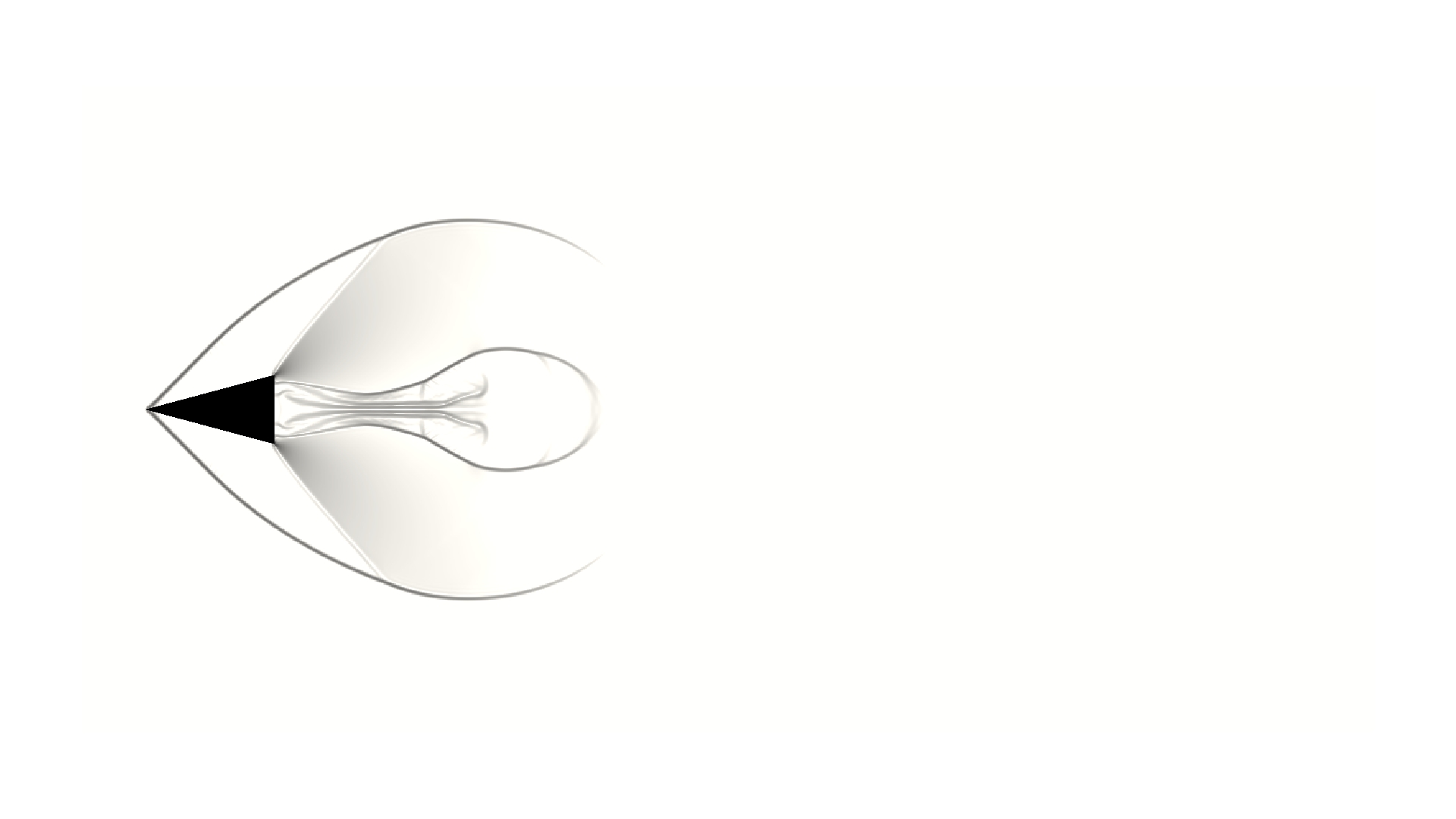}
        \caption{$0.05 \Unit{s}$}
        \label{fig:1_wedge_nomv_deg15_mach2_m1200_t050ms}
    \end{subfigure}%
    ~
    \begin{subfigure}[b]{0.48\textwidth}
        \includegraphics[width=\textwidth]{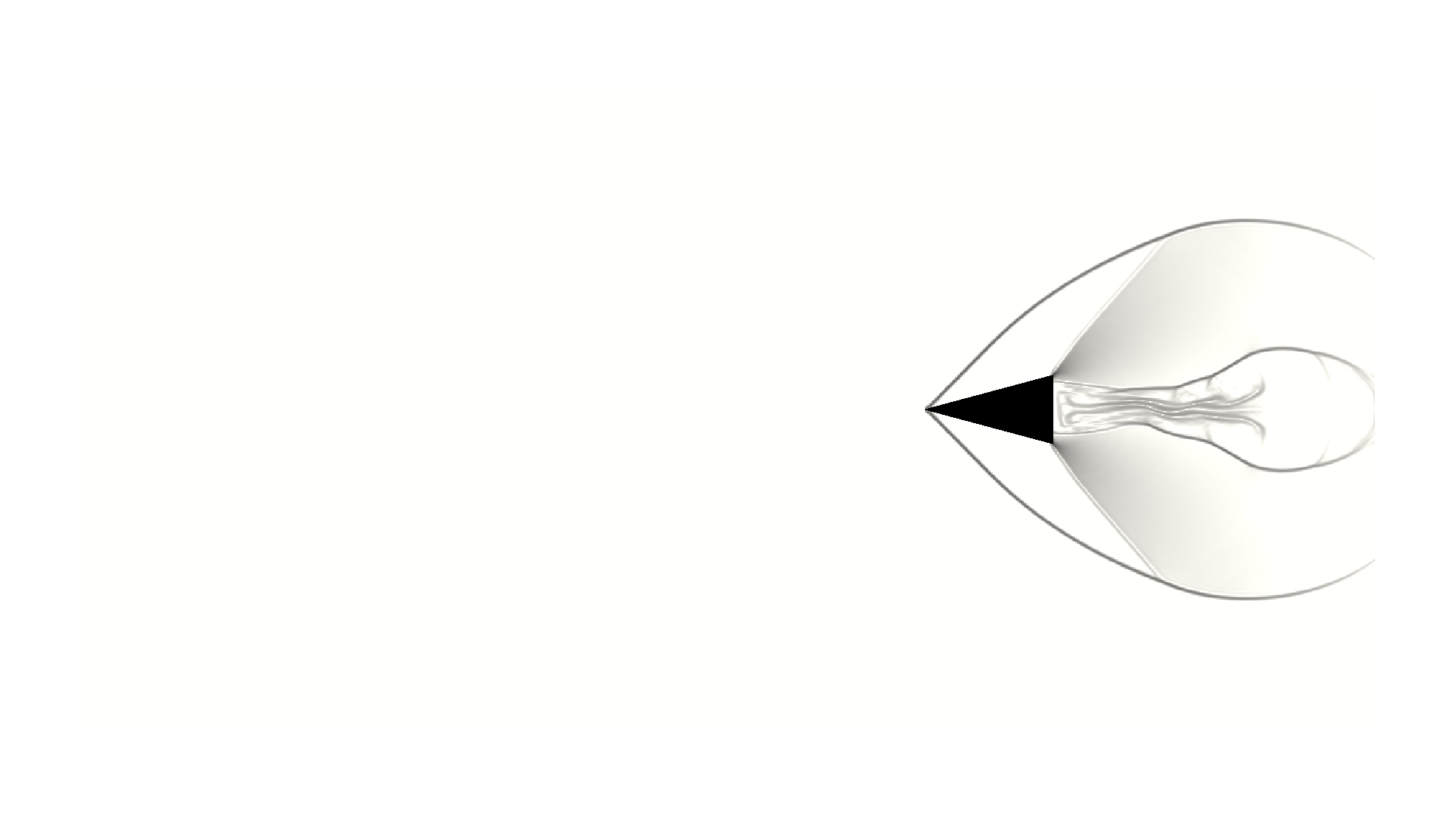}
        \caption{$0.05 \Unit{s}$}
        \label{fig:1_wedge_mv_deg15_mach2_m1200_t050ms}
    \end{subfigure}%
    \\
    \begin{subfigure}[b]{0.48\textwidth}
        \includegraphics[width=\textwidth]{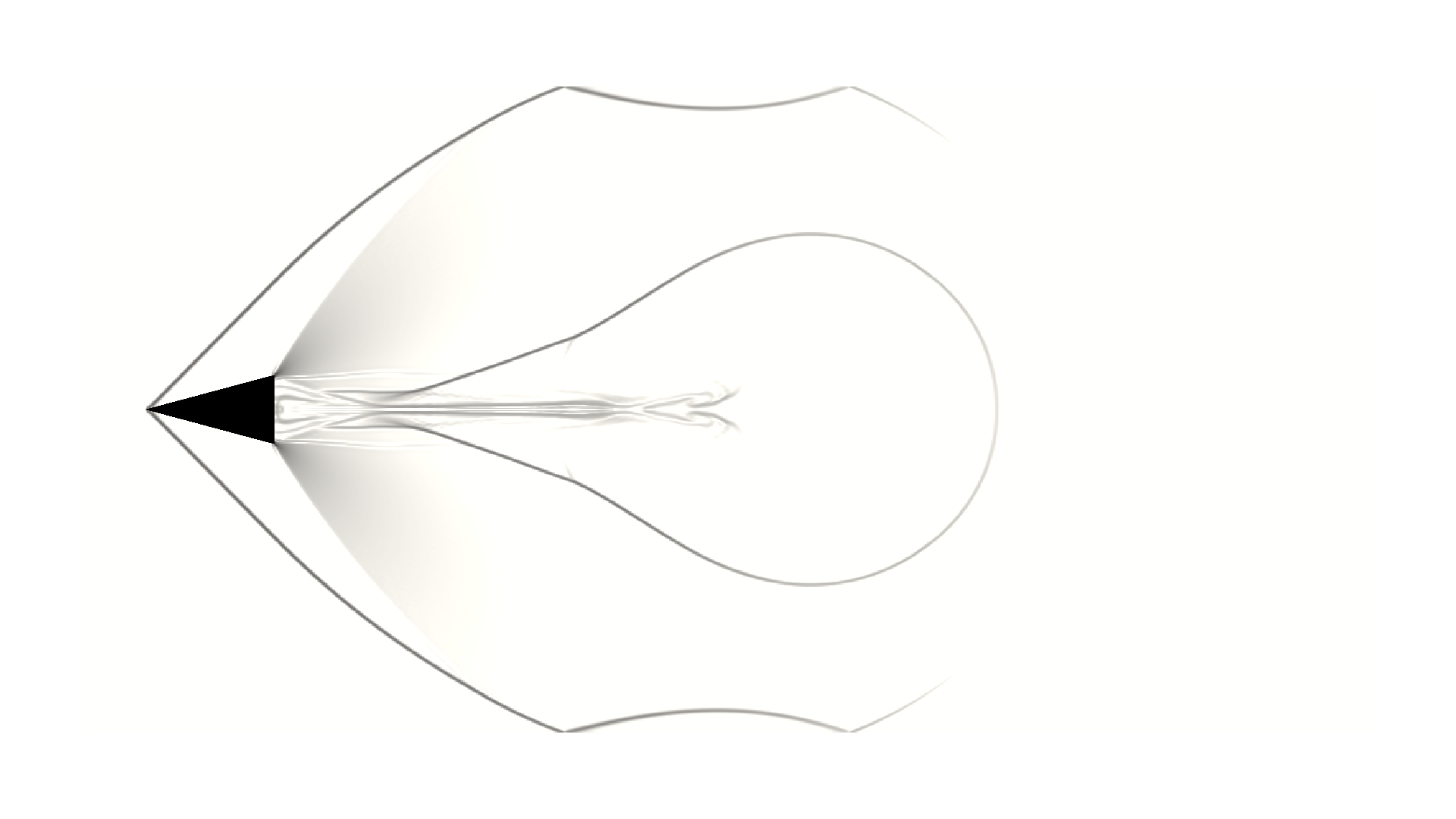}
        \caption{$0.10 \Unit{s}$}
        \label{fig:1_wedge_nomv_deg15_mach2_m1200_t100ms}
    \end{subfigure}%
    ~
    \begin{subfigure}[b]{0.48\textwidth}
        \includegraphics[width=\textwidth]{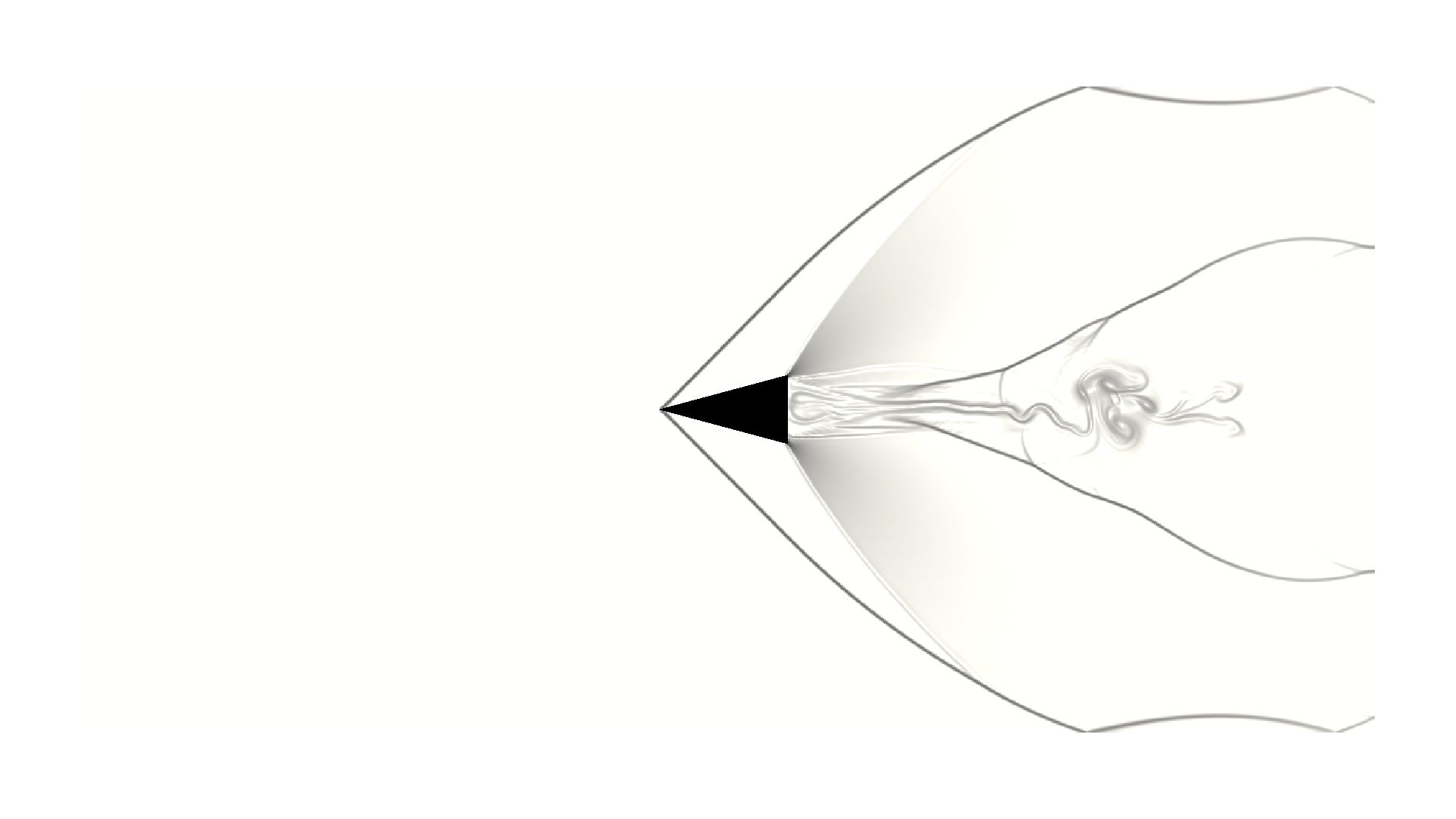}
        \caption{$0.10 \Unit{s}$}
        \label{fig:1_wedge_mv_deg15_mach2_m1200_t100ms}
    \end{subfigure}%
    \\
    \begin{subfigure}[b]{0.48\textwidth}
        \includegraphics[width=\textwidth]{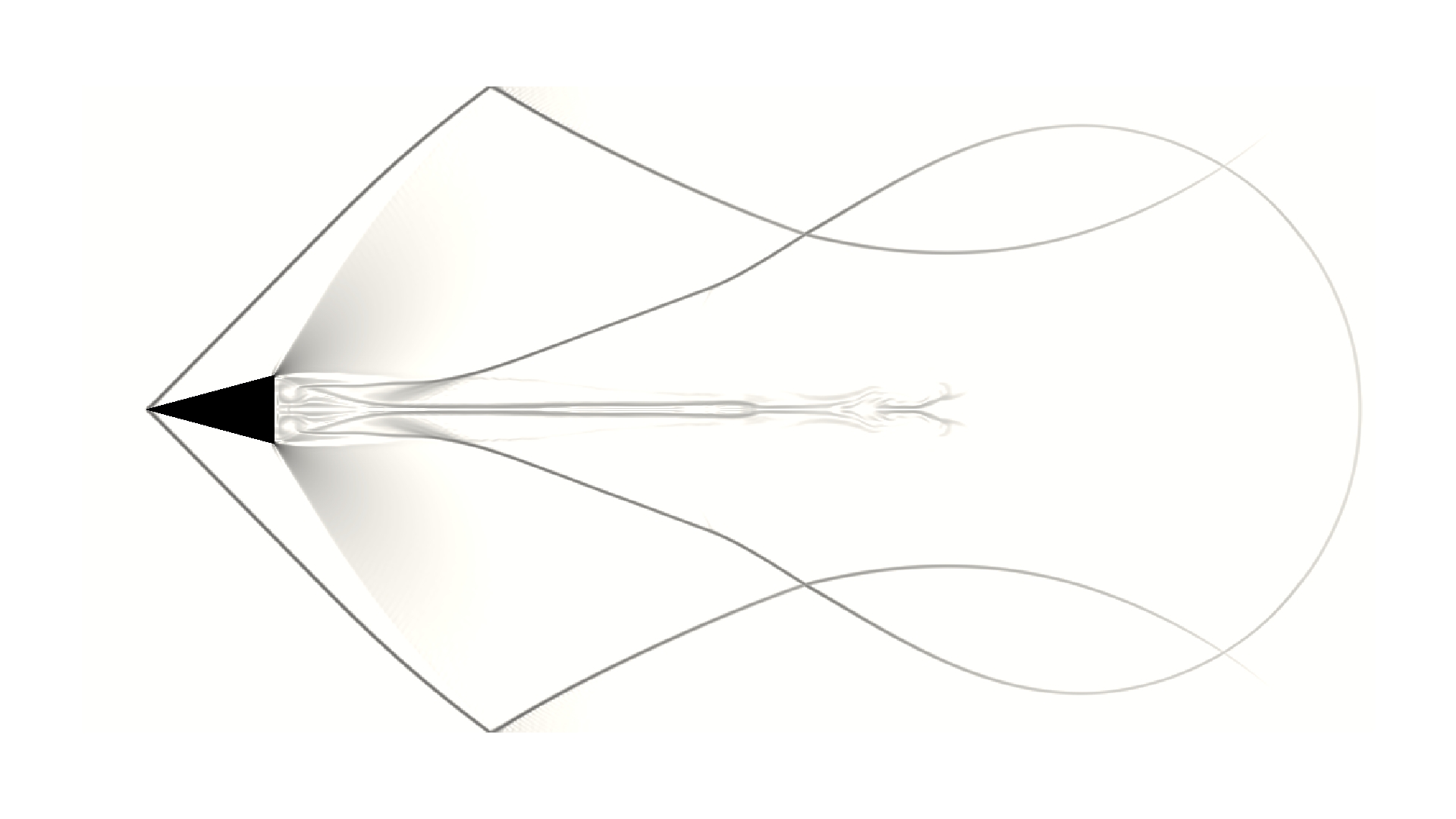}
        \caption{$0.15 \Unit{s}$}
        \label{fig:1_wedge_nomv_deg15_mach2_m1200_t150ms}
    \end{subfigure}%
    ~
    \begin{subfigure}[b]{0.48\textwidth}
        \includegraphics[width=\textwidth]{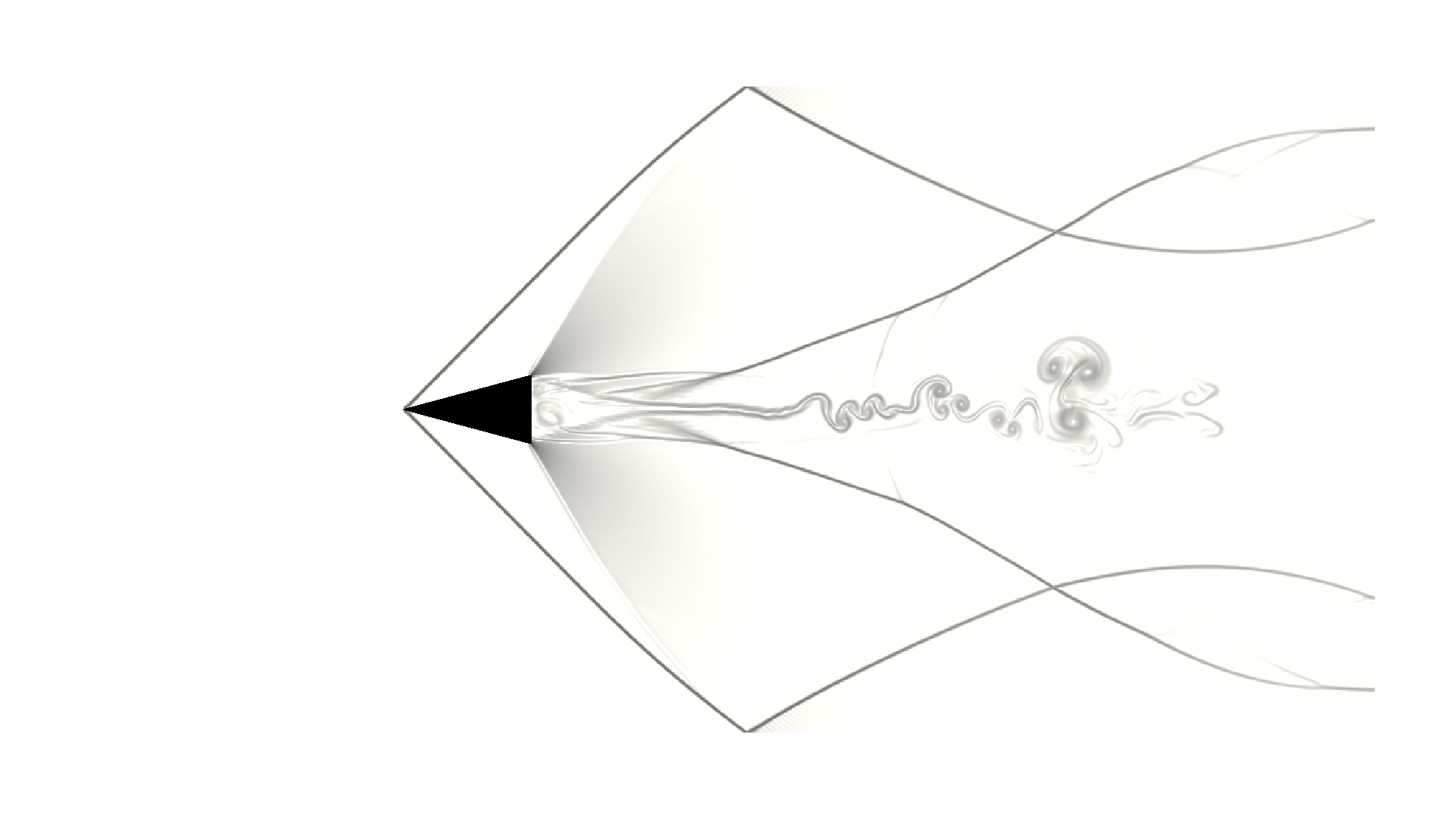}
        \caption{$0.15 \Unit{s}$}
        \label{fig:1_wedge_mv_deg15_mach2_m1200_t150ms}
    \end{subfigure}%
    \\
    \begin{subfigure}[b]{0.48\textwidth}
        \includegraphics[width=\textwidth]{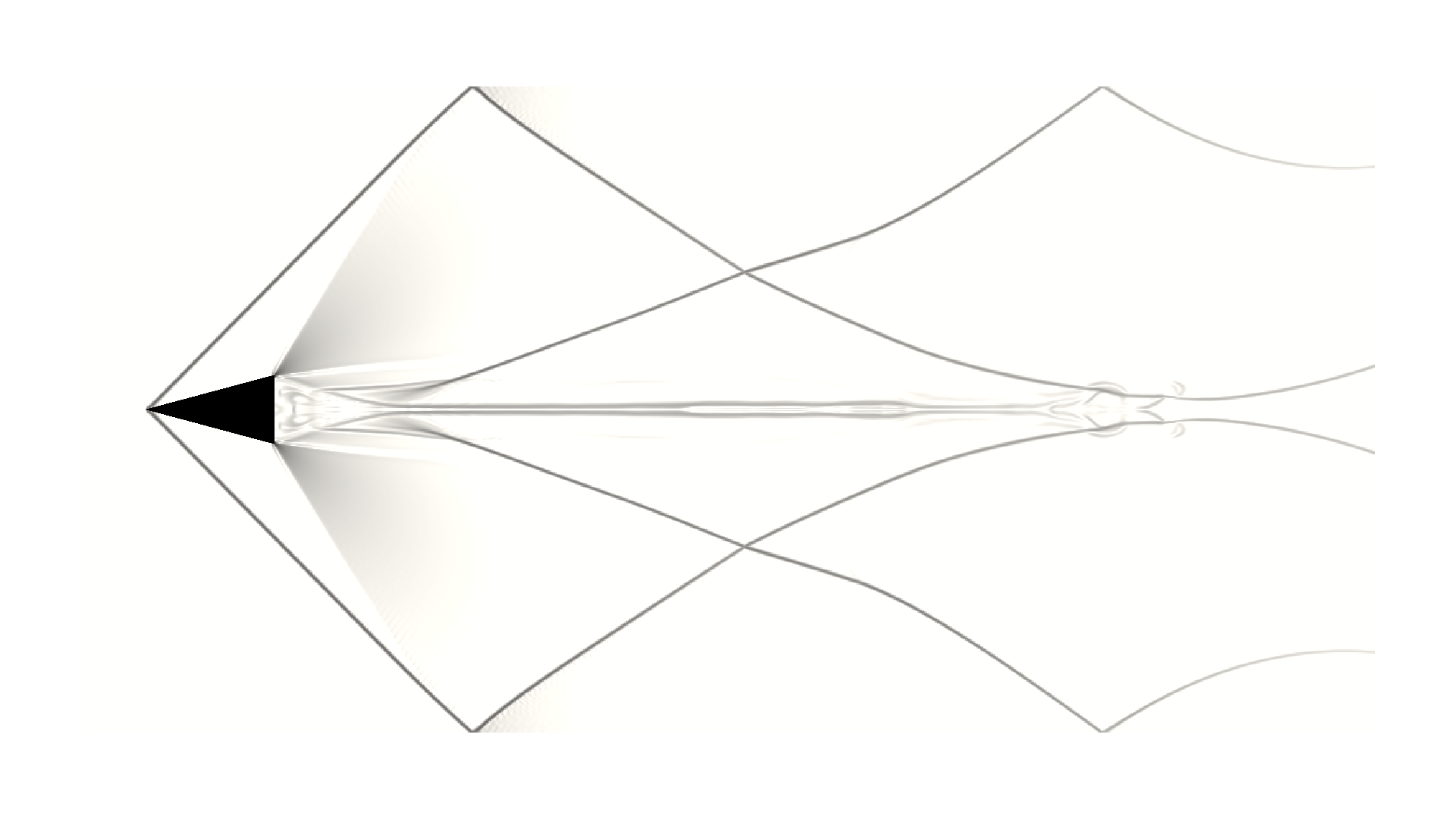}
        \caption{$0.20 \Unit{s}$}
        \label{fig:1_wedge_nomv_deg15_mach2_m1200_t200ms}
    \end{subfigure}%
    ~
    \begin{subfigure}[b]{0.48\textwidth}
        \includegraphics[width=\textwidth]{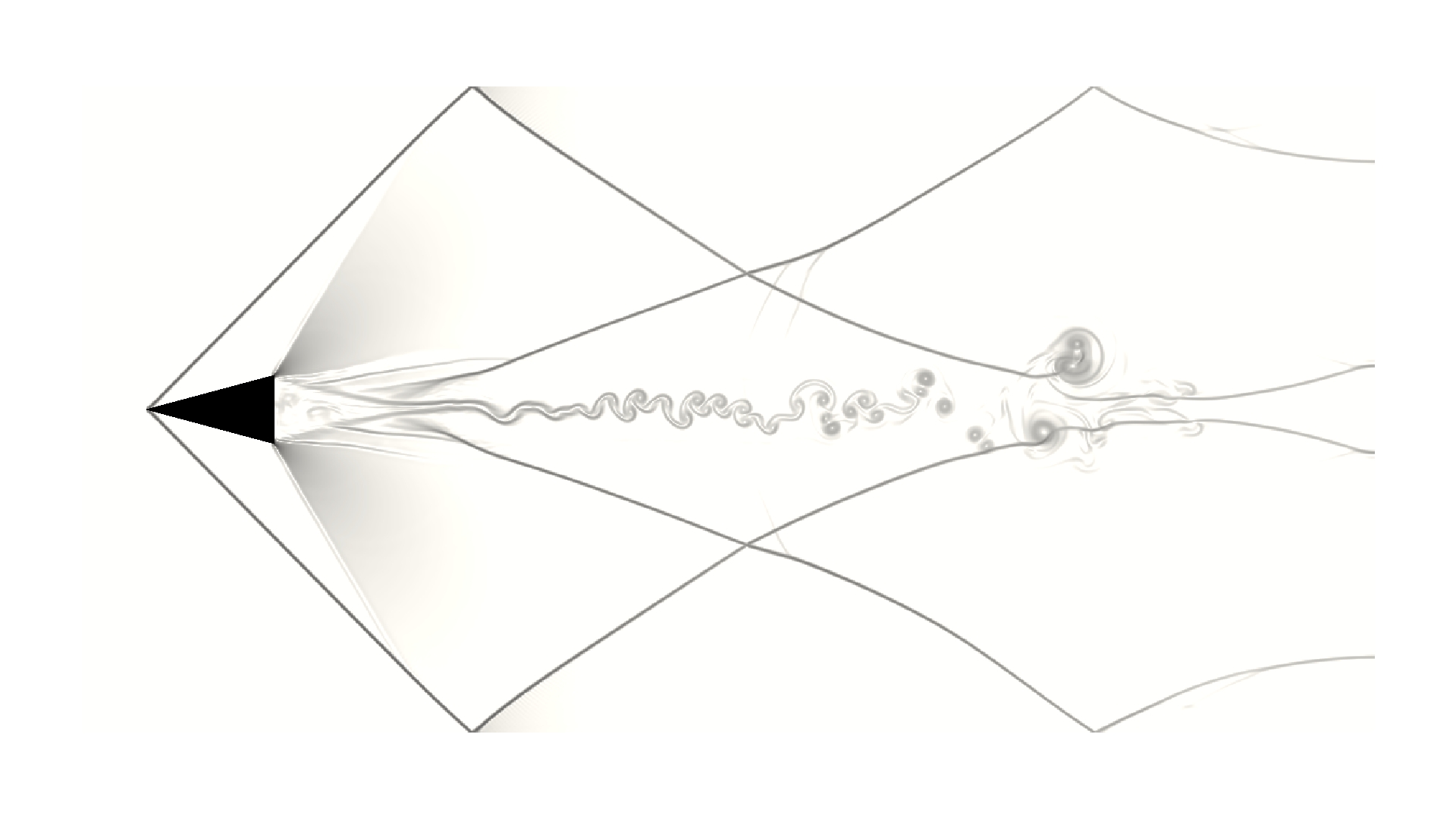}
        \caption{$0.20 \Unit{s}$}
        \label{fig:1_wedge_mv_deg15_mach2_m1200_t200ms}
    \end{subfigure}%
    \caption{Time evolution of solution for case $M_{\infty}(2)-\theta(15^{\circ})-G(1200\times600)$. (a), (c), (e), (g) The supersonic flow over a wedge problem. (b), (d), (f), (h) The supersonic translating wedge problem.}
    \label{fig:1_wedge_nomv_vs_mv_deg15_mach2_m1200}
\end{figure}

\begin{figure}[!htbp]
    \centering
    \begin{subfigure}[b]{0.65\textwidth}
        \includegraphics[width=\textwidth]{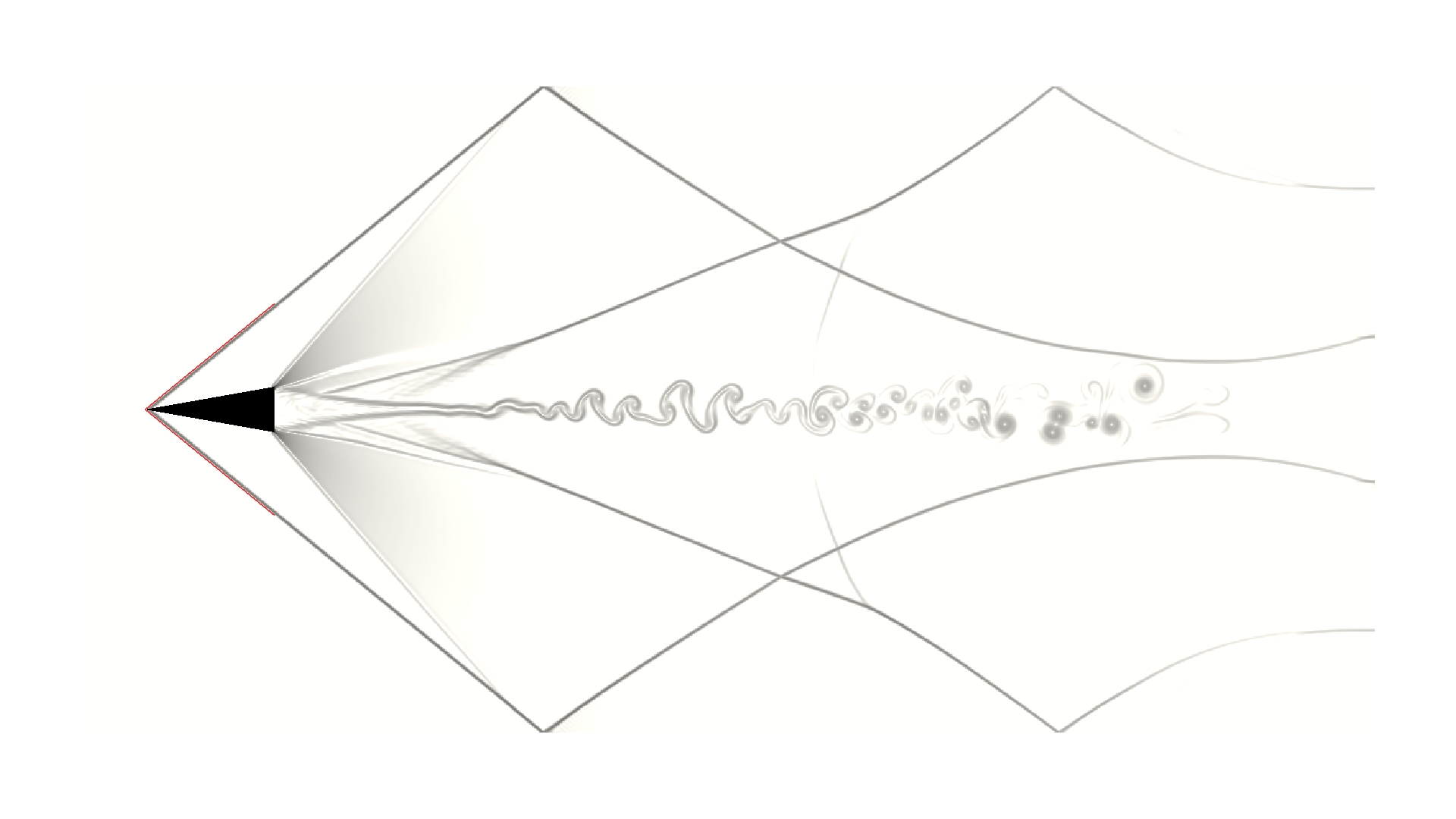}
        \caption{}
        \label{fig:1_wedge_mv_deg10_mach2_m1200_lined}
    \end{subfigure}%
    \\
    \begin{subfigure}[b]{0.65\textwidth}
        \includegraphics[width=\textwidth]{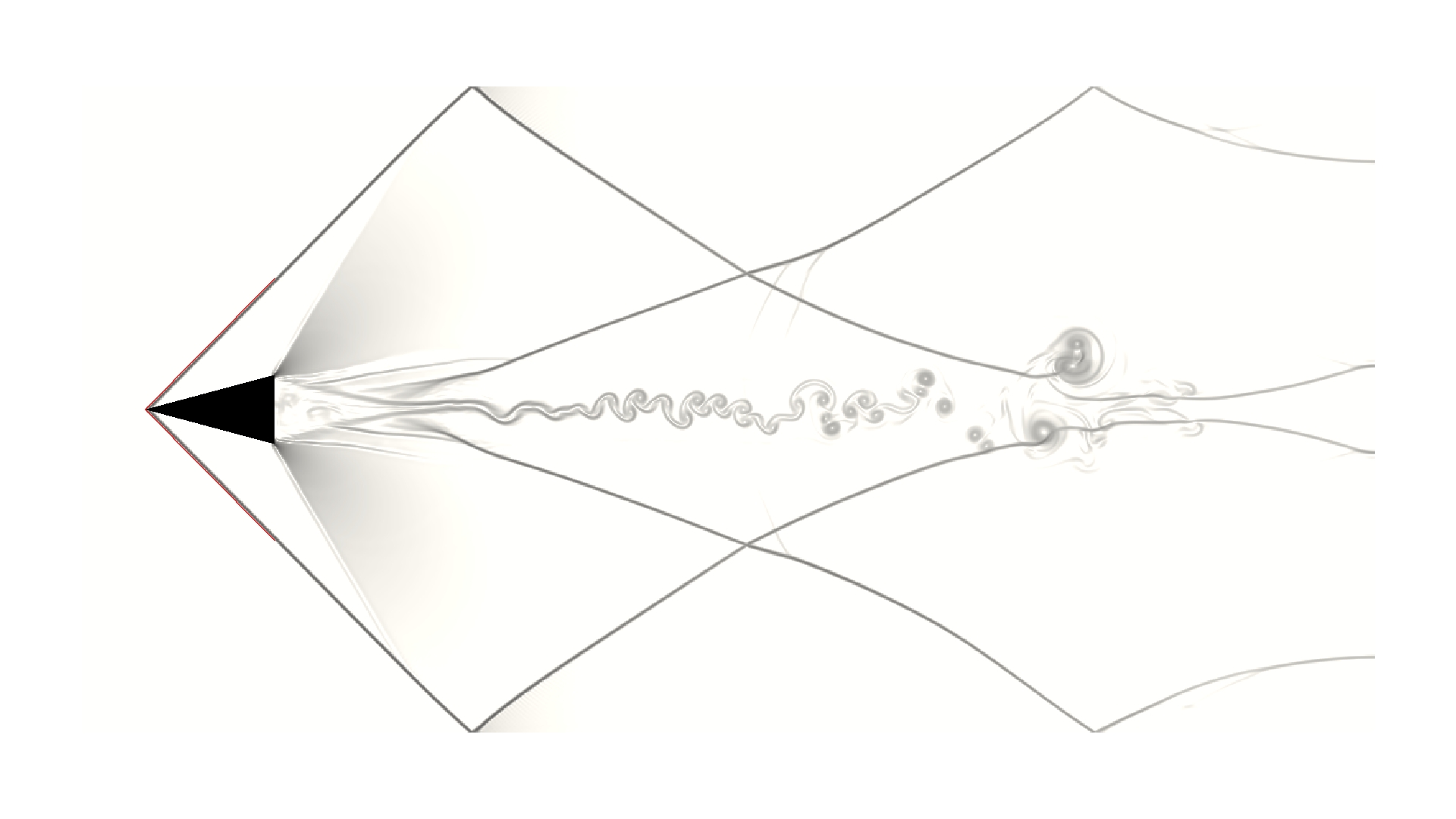}
        \caption{}
        \label{fig:1_wedge_mv_deg15_mach2_m1200_lined}
    \end{subfigure}%
    \\
    \begin{subfigure}[b]{0.65\textwidth}
        \includegraphics[width=\textwidth]{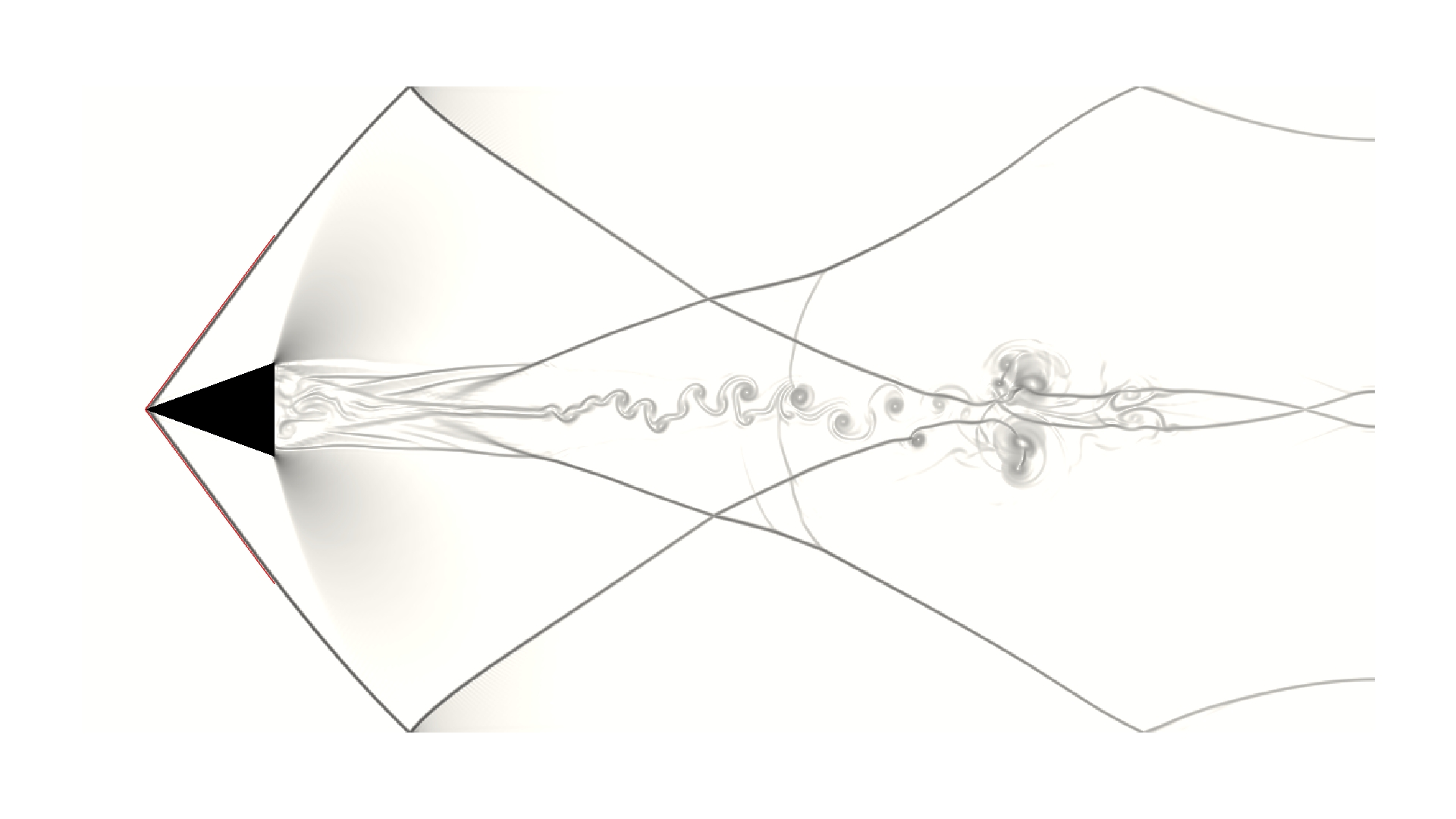}
        \caption{}
        \label{fig:1_wedge_mv_deg20_mach2_m1200_lined}
    \end{subfigure}%
    \caption{Oblique shock relation of supersonic translating wedge for case $M_{\infty}(2)-G(1200\times600)$ over different deflection angles. Red lines represent the analytical solutions (closely overlapped with numerical solutions). (a) $\theta(10^{\circ})-\beta_e(39.314^{\circ})-\beta_n(39.313^{\circ})$. (b) $\theta(15^{\circ})-\beta_e(45.344^{\circ})-\beta_n(45.034^{\circ})$. (c) $\theta(20^{\circ})-\beta_e(53.423^{\circ})-\beta_n(53.425^{\circ})$.}
    \label{fig:1_wedge_mv_deg}
\end{figure}

The time evolution of both the supersonic flow over a wedge and the supersonic translating wedge under the condition $M_{\infty}(2)-\theta(15^{\circ})-G(1200\times600)$ is captured in Fig.~\ref{fig:1_wedge_nomv_vs_mv_deg15_mach2_m1200}, in which the dynamic process of oblique shock formation at the wedge nose and Prandtl-Meyer expansion wave generation at the rear corners is clearly depicted. The predicted wakes are different for this Galilean transformation pair, which is mainly due to extra interpolation being required when computational nodes within the translating wedge are moved out of the wedge body into the fluid domain at the next time moment. The introduced perturbations caused by interpolation errors can be significantly magnified under the supersonic flow condition, resulting in a disturbed wake. Nevertheless, main flow features exempted from the extra interpolation errors, such as the oblique shocks, reflected shocks, and Prandtl-Meyer expansion waves, are well agreed in this transformation pair, which validates the correctness of the method in solving transient flows. To further verify the solution of the supersonic translating wedge problem, a comparison of the numerical and analytical solutions for different deflection angles is shown in Fig.~\ref{fig:1_wedge_mv_deg}. The excellent agreement in numerical and analytical solutions demonstrates the validity and accuracy of the method in solving flow with moving geometries.

\subsection{Shock diffraction over a cylinder} \label{case:1_cyn}

A Mach $2.81$ planar shock interacting with a stationary circular cylinder is studied to further evaluate the validity of the developed method. This classical shock diffraction problem has been widely investigated in the literature, including both experimental observations \citep{bryson1961diffraction, bazhenova1984unsteady, kaca1988interferometric} and numerical studies \citep{kaca1988interferometric,ripley2006numerical, sambasivan2009ghostb}.
\begin{figure}[!htbp]
    \centering
    \begin{subfigure}[b]{0.4\textwidth}
        \includegraphics[width=\textwidth]{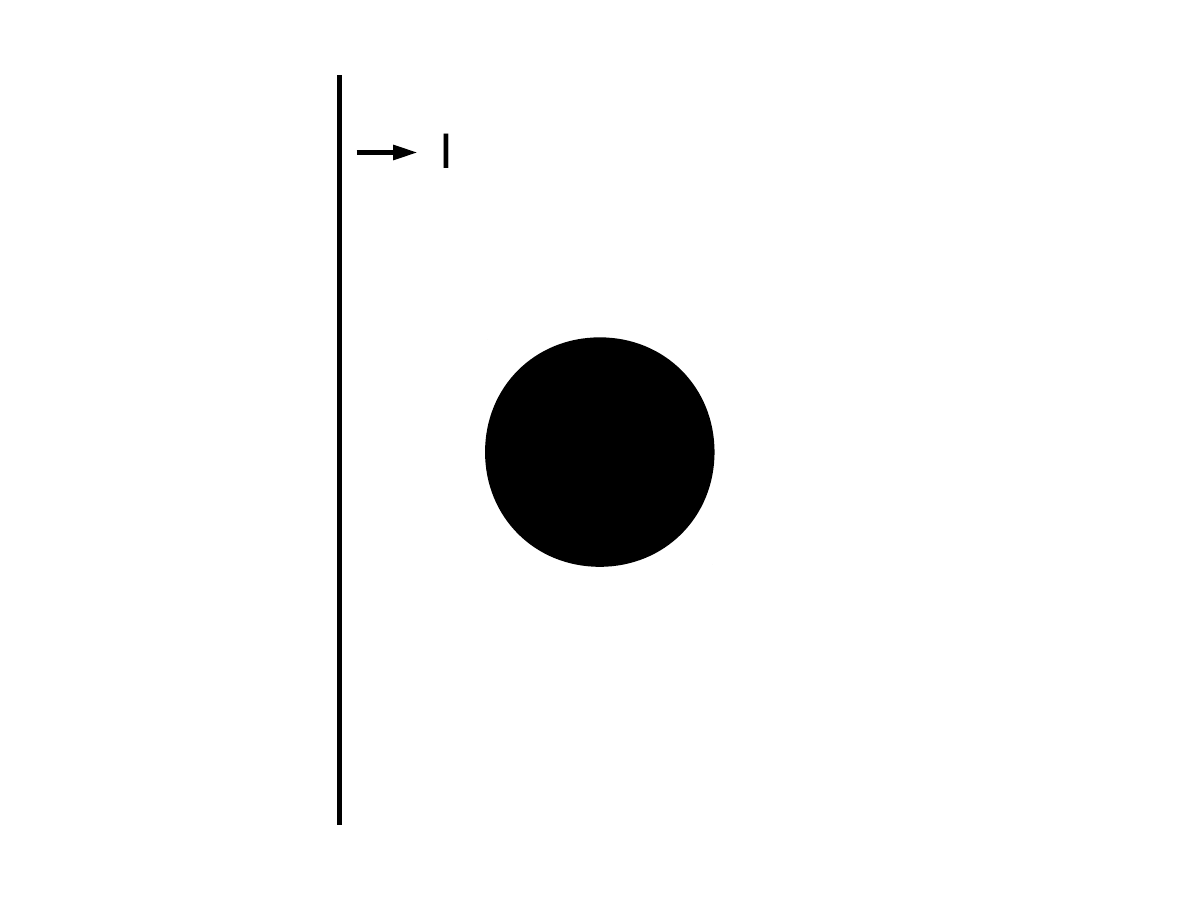}
        \caption{}
        \label{fig:shock_cylinder_demo_a}
    \end{subfigure}%
    ~
    \begin{subfigure}[b]{0.4\textwidth}
        \includegraphics[width=\textwidth]{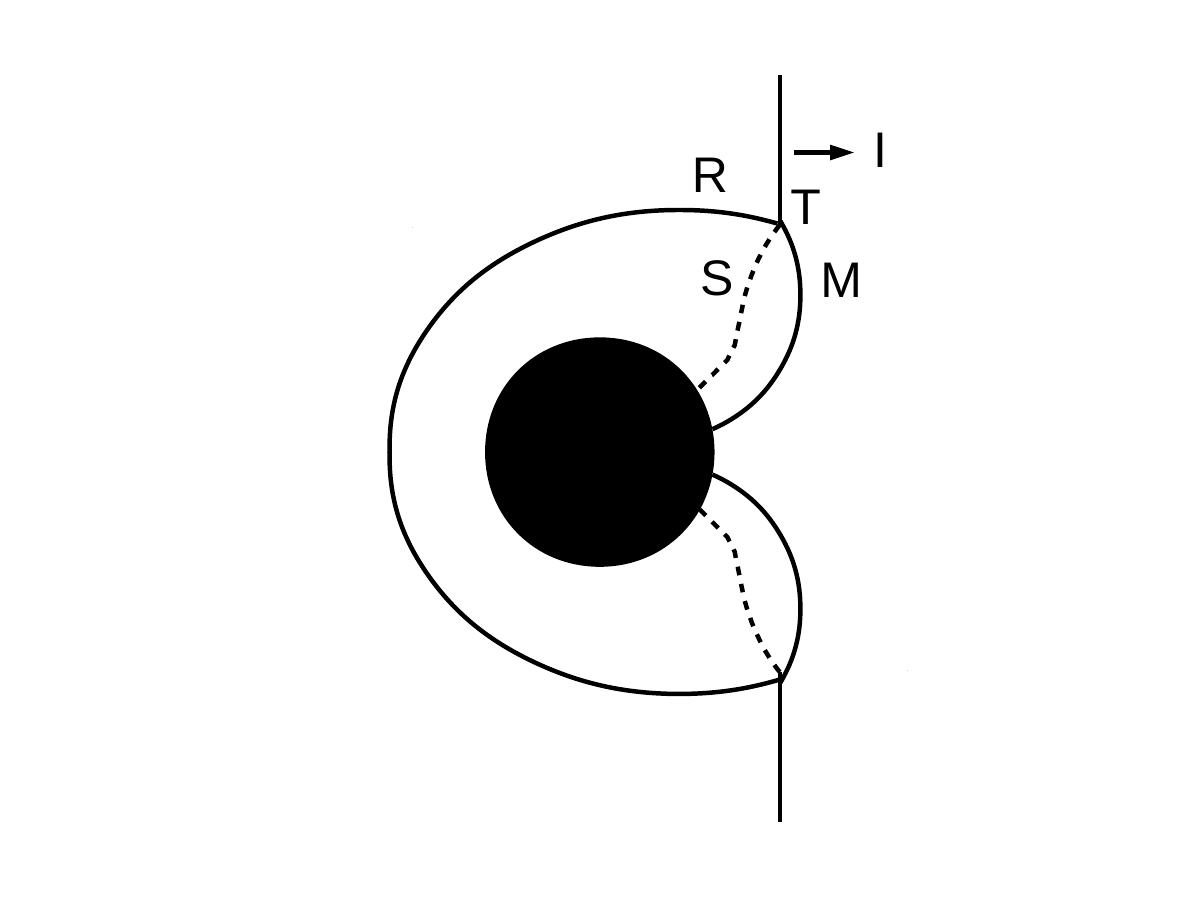}
        \caption{}
        \label{fig:shock_cylinder_demo_b}
    \end{subfigure}%
    \caption{Schematic diagrams for a planar shock interacting with a stationary circular cylinder. (a) Initial state. (b) Well-developed diffraction. [Nomenclature: $I$, incident shock; $R$, reflected shock; $M$, Mach stem (diffracted shock); $S$, slip line (contact discontinuity); $T$, triple point.] Schematic diagrams adapted from \citet{kaca1988interferometric}.}
    \label{fig:shock_cylinder_demo}
\end{figure}

As a time-dependent process, the interaction between the incident shock and cylinder encompasses complex compressible flow features such as shocks and contact discontinuities. As illustrated in Fig.~\ref{fig:shock_cylinder_demo}, the incident shock initially propagates freely toward the cylinder. Once colliding with the cylinder, the shock reflects as well as diffracts over the convex solid surface with the formation of a curved Mach stem and a slip line at each side of the plane of symmetry. During the evolution, triple points are produced through the intersection of the incident shock, reflected shock, and diffracted shock. At the later stage of evolution, the two diffracting Mach stems collide and form a shock-induced wake at the rear of the cylinder.

In the numerical configuration, a circular cylinder with diameter $D = 1 \Unit{m}$ is positioned at the center of a $6D \times 6D$ square domain while an initial shock is positioned $0.5D$ upstream of the cylinder. This computational configuration is similar to \citet{ripley2006numerical} except that a full domain size without symmetric boundary assumption is used herein. The flow is assumed to be inviscid, and the slip-wall boundary condition is enforced at the cylinder surface. The evolution process is solved to $t=1.0 \Unit{s}$.
\begin{figure}[!htbp]
    \centering
    \begin{subfigure}[b]{0.24\textwidth}
        \includegraphics[trim = 75mm 0mm 75mm 0mm, clip, width=\textwidth]{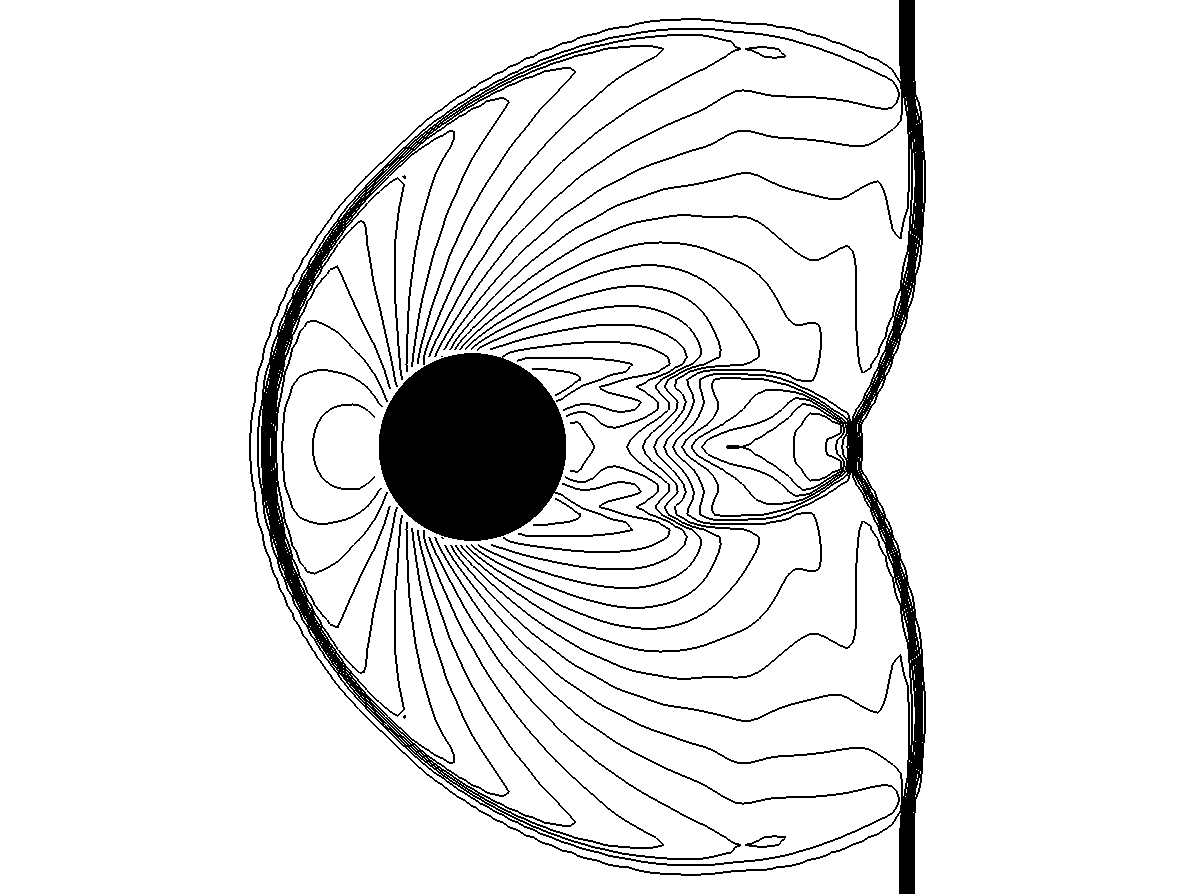}
        \caption{}
        \label{fig:1_cyn_nomv_invis_m0150_density_contour}
    \end{subfigure}%
    ~
    \begin{subfigure}[b]{0.24\textwidth}
        \includegraphics[trim = 75mm 0mm 75mm 0mm, clip, width=\textwidth]{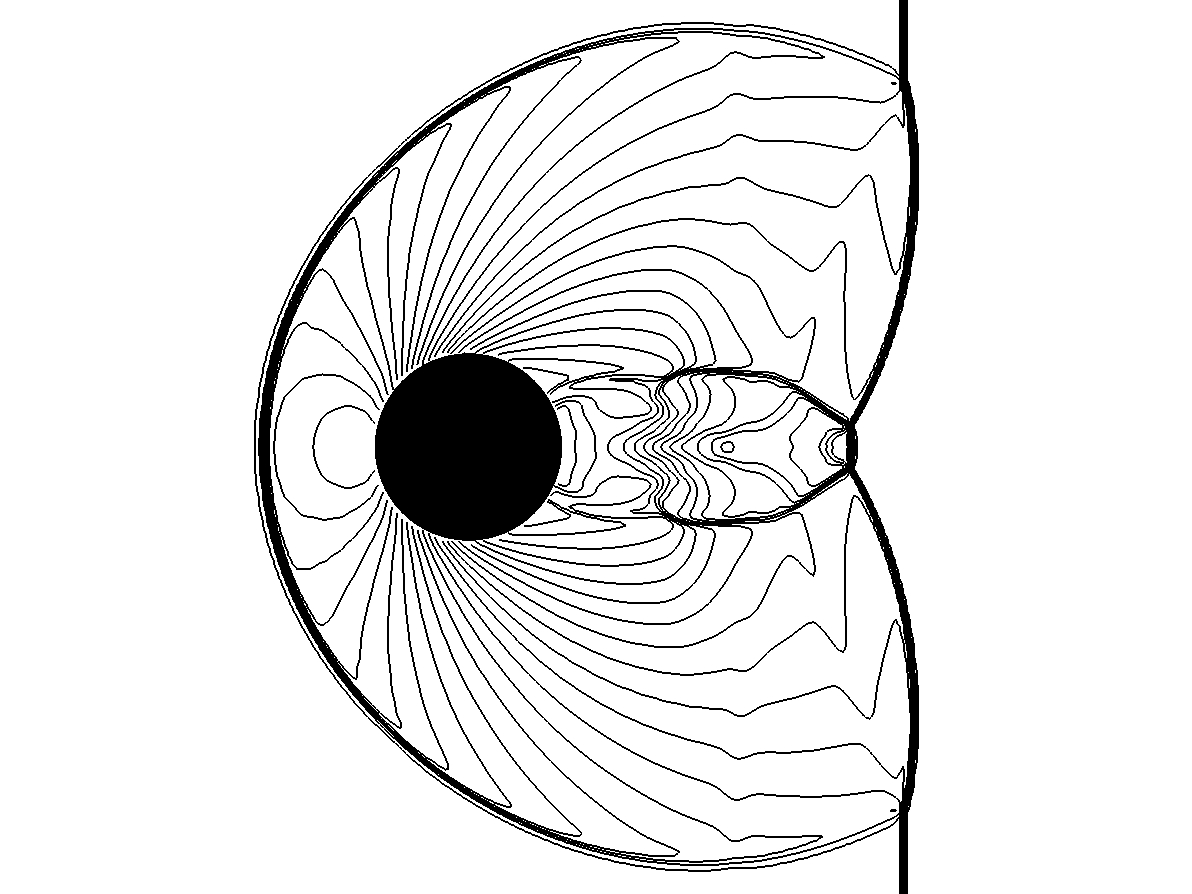}
        \caption{}
        \label{fig:1_cyn_nomv_invis_m0300_density_contour}
    \end{subfigure}%
    ~
    \begin{subfigure}[b]{0.24\textwidth}
        \includegraphics[trim = 75mm 0mm 75mm 0mm, clip, width=\textwidth]{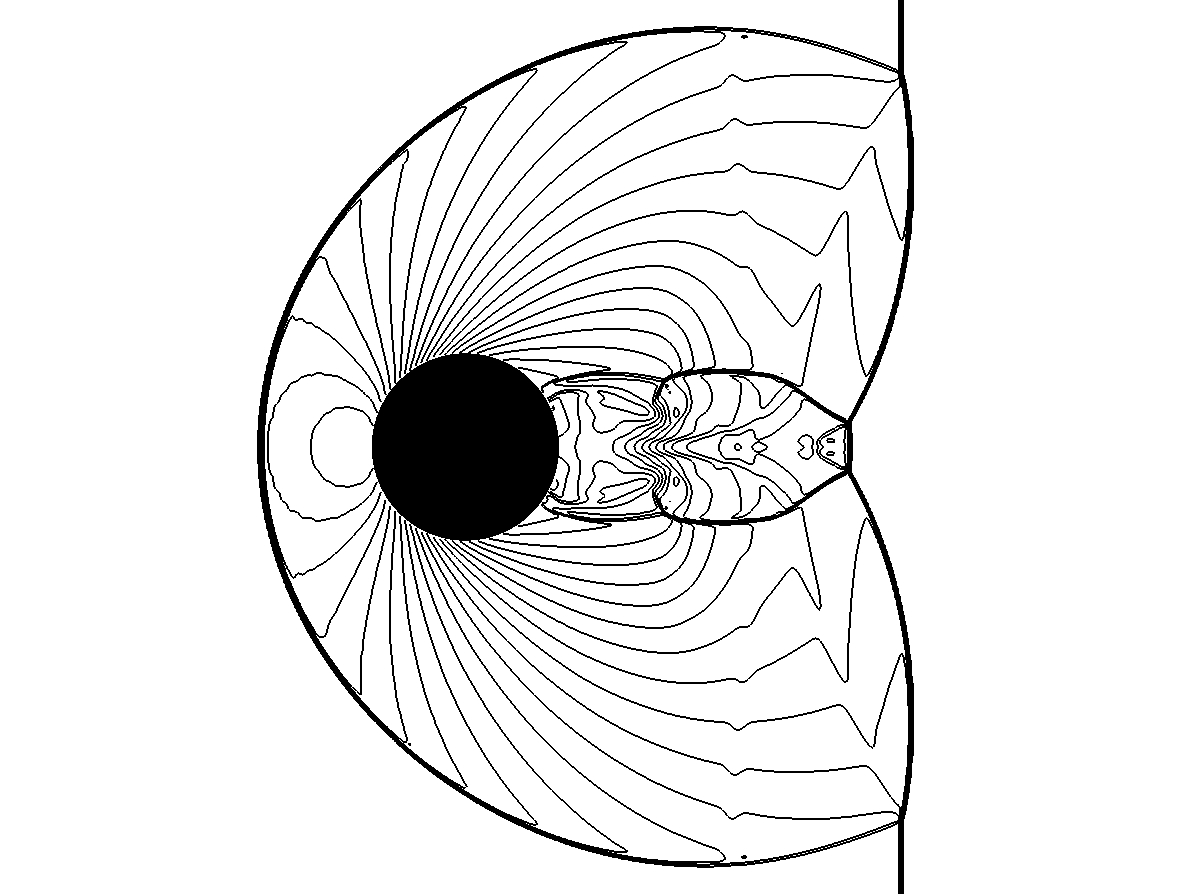}
        \caption{}
        \label{fig:1_cyn_nomv_invis_m0600_density_contour}
    \end{subfigure}%
    ~
    \begin{subfigure}[b]{0.24\textwidth}
        \includegraphics[trim = 75mm 0mm 75mm 0mm, clip, width=\textwidth]{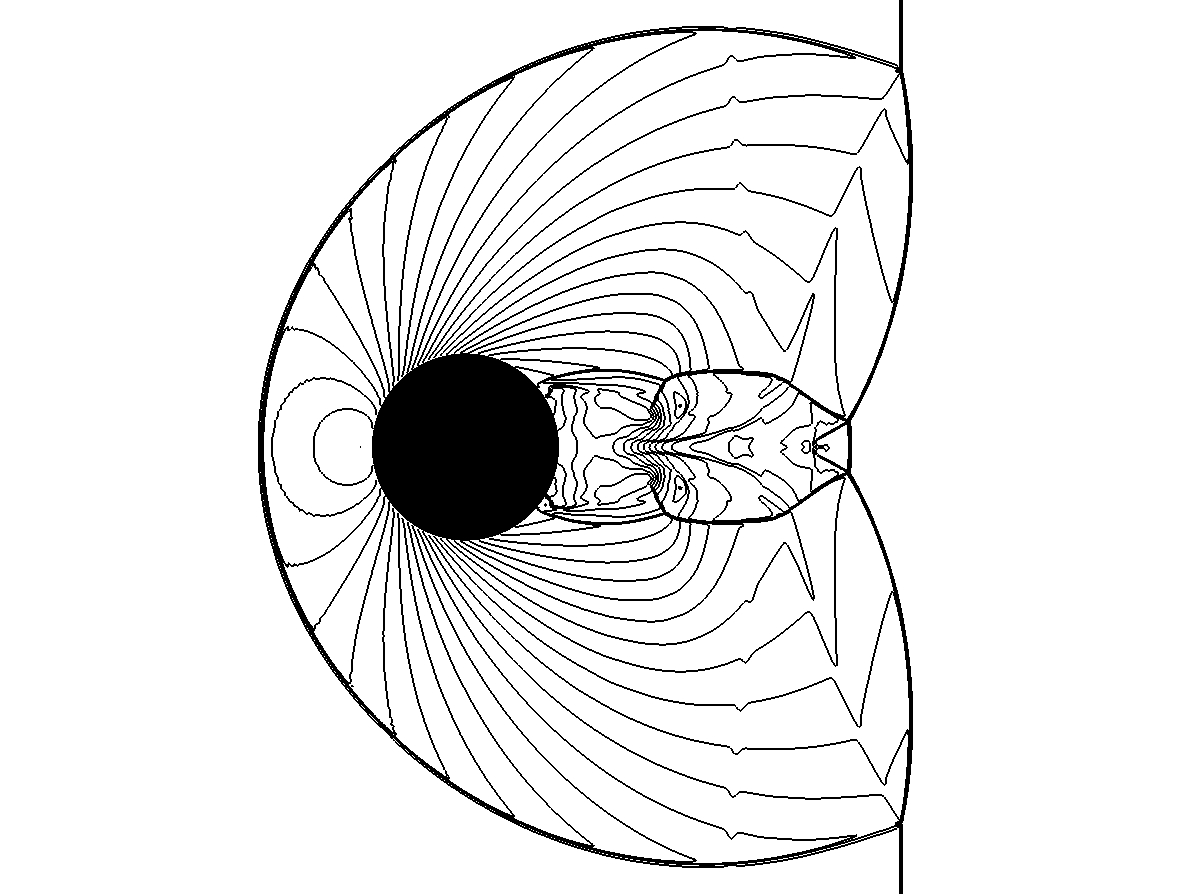}
        \caption{}
        \label{fig:1_cyn_nomv_invis_m1200_density_contour}
    \end{subfigure}%
    \caption{Density contour of shock diffraction over a cylinder solved on different grid sizes. (a) $150\times150$. (b) $300\times300$. (c) $600\times600$. (d) $1200\times1200$.}
    \label{fig:1_cyn_nomv_invis_density_contour}
\end{figure}
\begin{figure}[!htbp]
    \centering
    \begin{subfigure}[b]{0.24\textwidth}
        \includegraphics[trim = 75mm 0mm 75mm 0mm, clip, width=\textwidth]{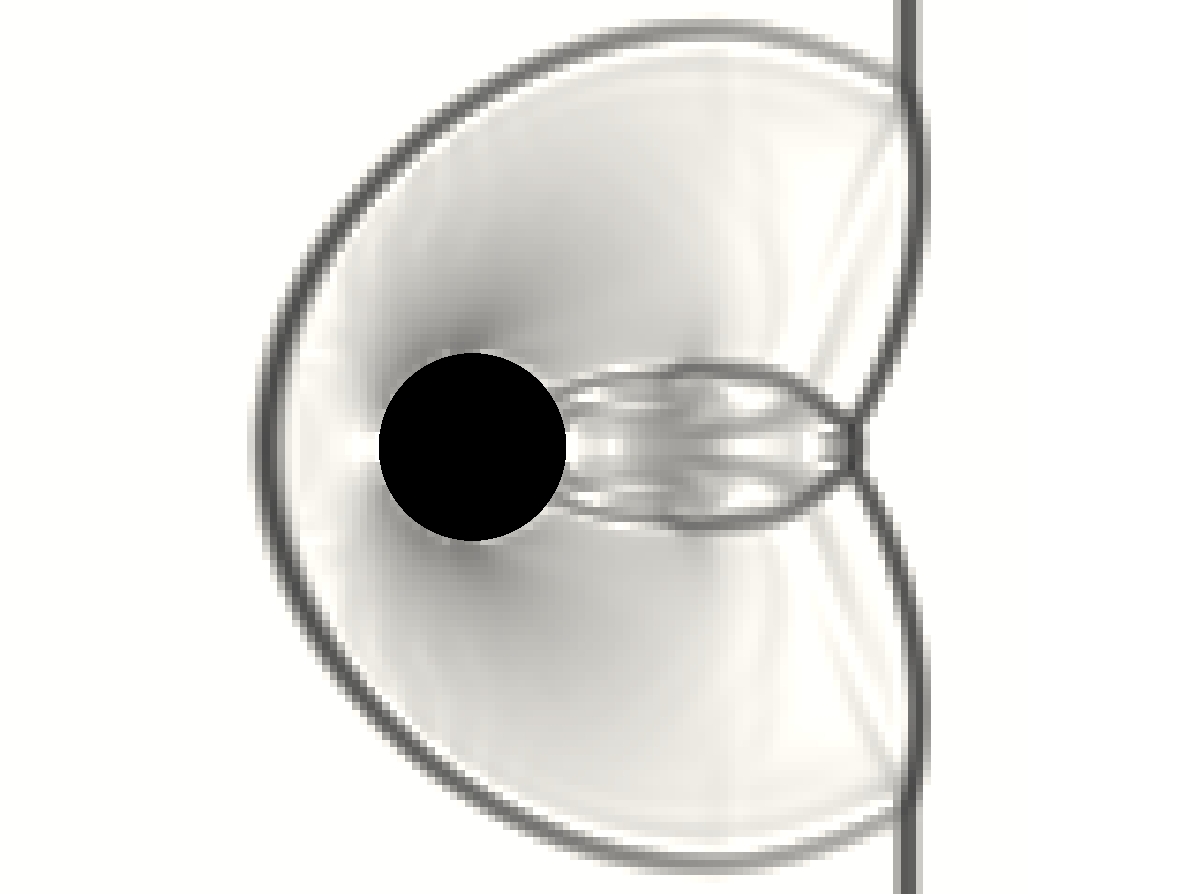}
        \caption{}
        \label{fig:1_cyn_nomv_invis_m0150_schlieren}
    \end{subfigure}%
    ~
    \begin{subfigure}[b]{0.24\textwidth}
        \includegraphics[trim = 75mm 0mm 75mm 0mm, clip, width=\textwidth]{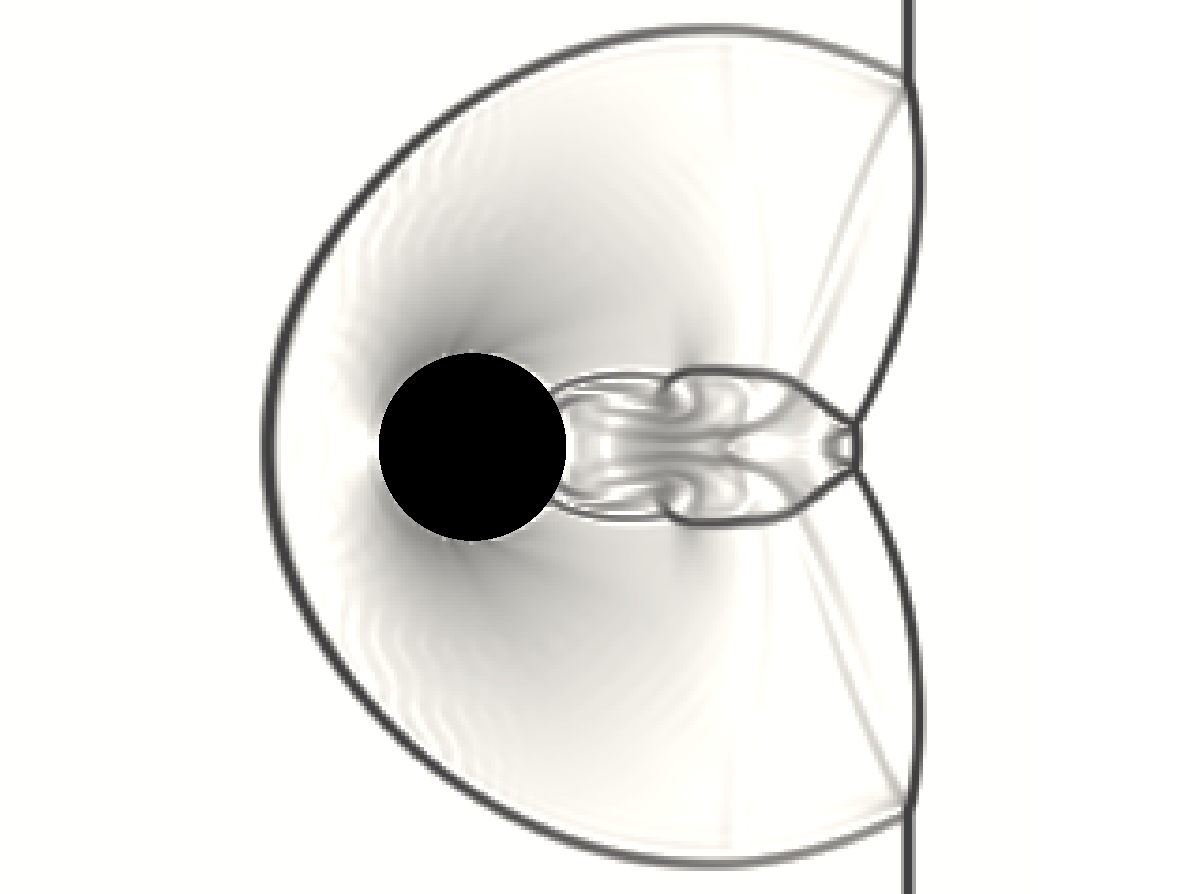}
        \caption{}
        \label{fig:1_cyn_nomv_invis_m0300_schlieren}
    \end{subfigure}%
    ~
    \begin{subfigure}[b]{0.24\textwidth}
        \includegraphics[trim = 75mm 0mm 75mm 0mm, clip, width=\textwidth]{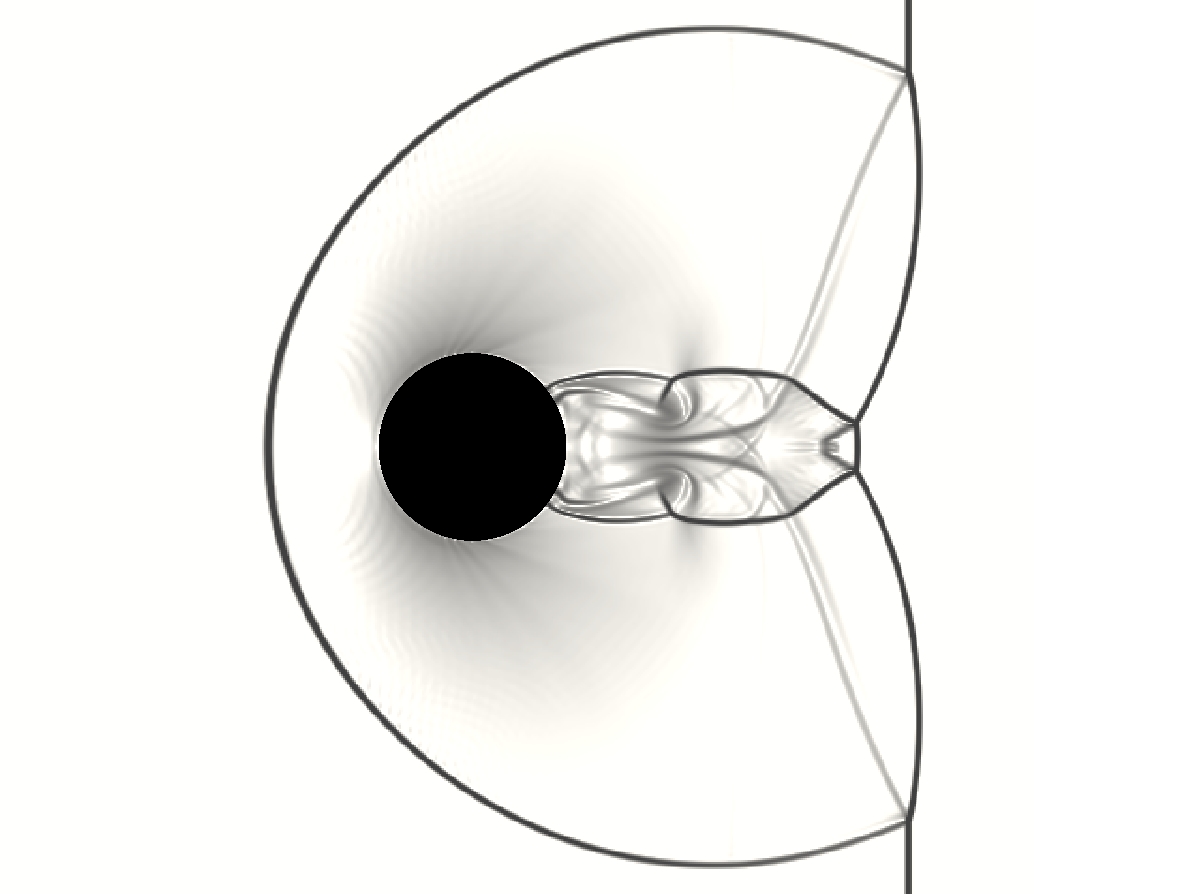}
        \caption{}
        \label{fig:1_cyn_nomv_invis_m0600_schlieren}
    \end{subfigure}%
    ~
    \begin{subfigure}[b]{0.24\textwidth}
        \includegraphics[trim = 75mm 0mm 75mm 0mm, clip, width=\textwidth]{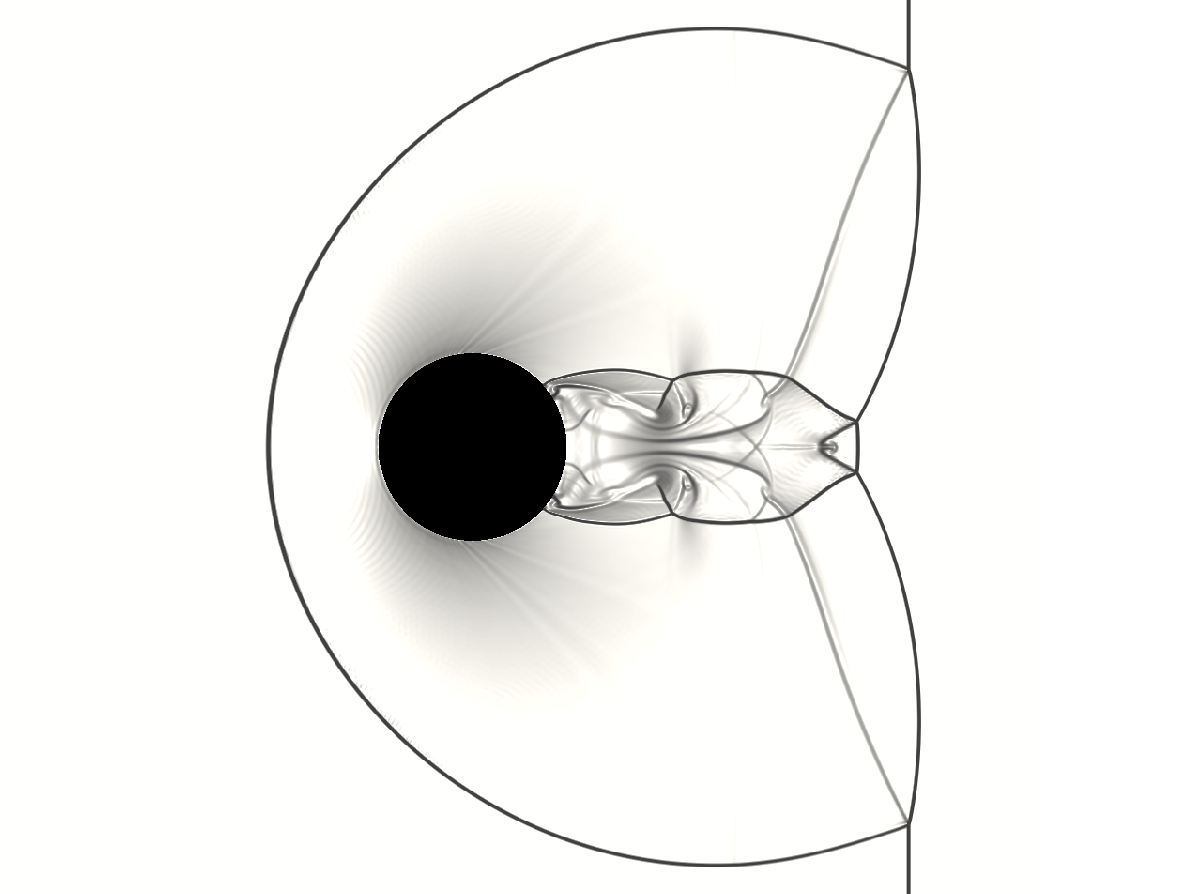}
        \caption{}
        \label{fig:1_cyn_nomv_invis_m1200_schlieren}
    \end{subfigure}%
    \caption{Numerical Schlieren of shock diffraction over a cylinder solved on different grid sizes. (a) $150\times150$. (b) $300\times300$. (c) $600\times600$. (d) $1200\times1200$.}
    \label{fig:1_cyn_nomv_invis_schlieren}
\end{figure}

The computed density contours over a series of grids are shown in Fig.~\ref{fig:1_cyn_nomv_invis_density_contour}. An acceptable shock curvature profile can be observed even on the $150\times150$ grid, which has a grid resolution of about $0.04D$ ($25$ nodes per diameter). In addition, the plane of symmetry is well preserved over all the employed grids. The numerical Schlieren fields are presented in Fig.~\ref{fig:1_cyn_nomv_invis_schlieren}. Compared to the Schlieren photograph measured by \citet{bryson1961diffraction} and the interferometric measurements by \citet{kaca1988interferometric} as well as the numerical results in references \citep{sambasivan2009ghostb, ripley2006numerical, ji2010numerical}, the slip line, reflected shock, and diffracted shock over the immersed boundary are already resolved very well on the $600\times600$ grid, which illustrates the high accuracy of the developed method.
\begin{figure}[!htbp]
    \centering
    \begin{subfigure}[b]{0.24\textwidth}
        \includegraphics[trim = 30mm 0mm 30mm 0mm, clip, width=\textwidth]{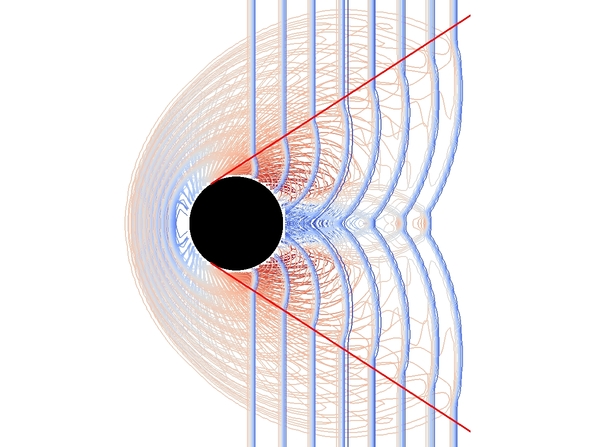}
        \caption{}
        \label{fig:1_cyn_nomv_invis_m0150_overlap}
    \end{subfigure}%
    ~
    \begin{subfigure}[b]{0.24\textwidth}
        \includegraphics[trim = 30mm 0mm 30mm 0mm, clip, width=\textwidth]{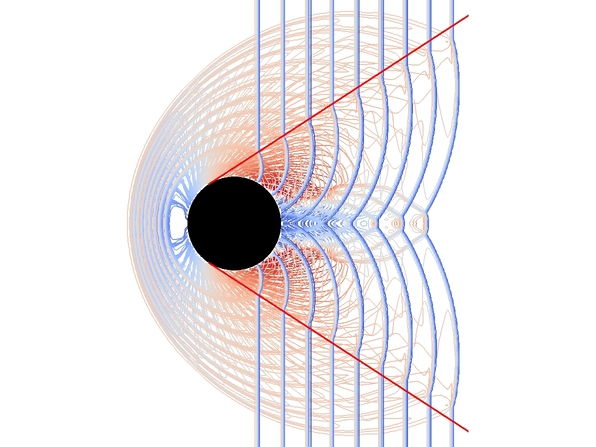}
        \caption{}
        \label{fig:1_cyn_nomv_invis_m0300_overlap}
    \end{subfigure}%
    ~
    \begin{subfigure}[b]{0.24\textwidth}
        \includegraphics[trim = 30mm 0mm 30mm 0mm, clip, width=\textwidth]{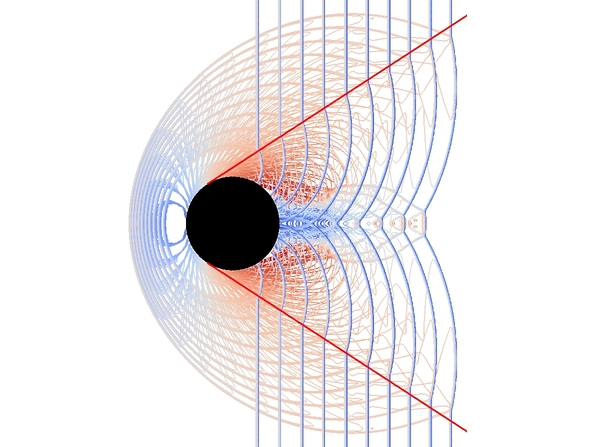}
        \caption{}
        \label{fig:1_cyn_nomv_invis_m0600_overlap}
    \end{subfigure}%
    ~
    \begin{subfigure}[b]{0.24\textwidth}
        \includegraphics[trim = 30mm 0mm 30mm 0mm, clip, width=\textwidth]{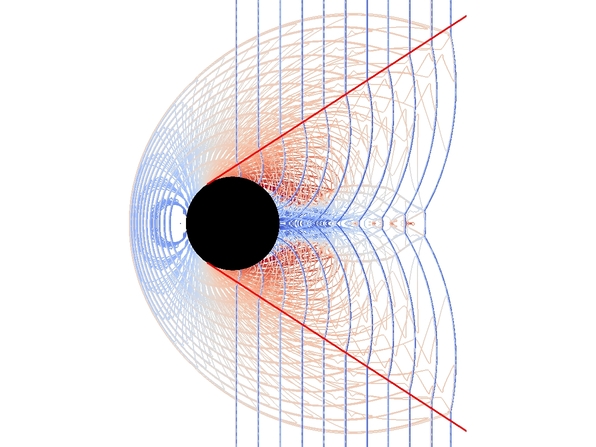}
        \caption{}
        \label{fig:1_cyn_nomv_invis_m1200_overlap}
    \end{subfigure}%
    \caption{Superimposition of density contours showing the predicted propagation path of the triple point (The two straight red lines are the $33^{\circ}$ tangent lines of the cylinder). (a) $150\times150$. (b) $300\times300$. (c) $600\times600$. (d) $1200\times1200$.}
    \label{fig:1_cyn_nomv_invis_overlap}
\end{figure}

As illustrated in Fig.~\ref{fig:shock_cylinder_demo_b}, the intersection of the incident shock, reflected shock, and diffracted shock forms a triple point. During the time-dependent evolution process, this triple point travels in space and produces a triple-point path, as captured in Fig.~\ref{fig:1_cyn_nomv_invis_overlap}. The interferometric measurements of \citet{kaca1988interferometric} predict that this triple-point path is tangent to the cylinder at an angle of $33^{\circ}$ for Mach numbers in the range of $1.42 - 5.96$.
\begin{figure}[!htbp]
    \centering
    \includegraphics[width=0.48\textwidth]{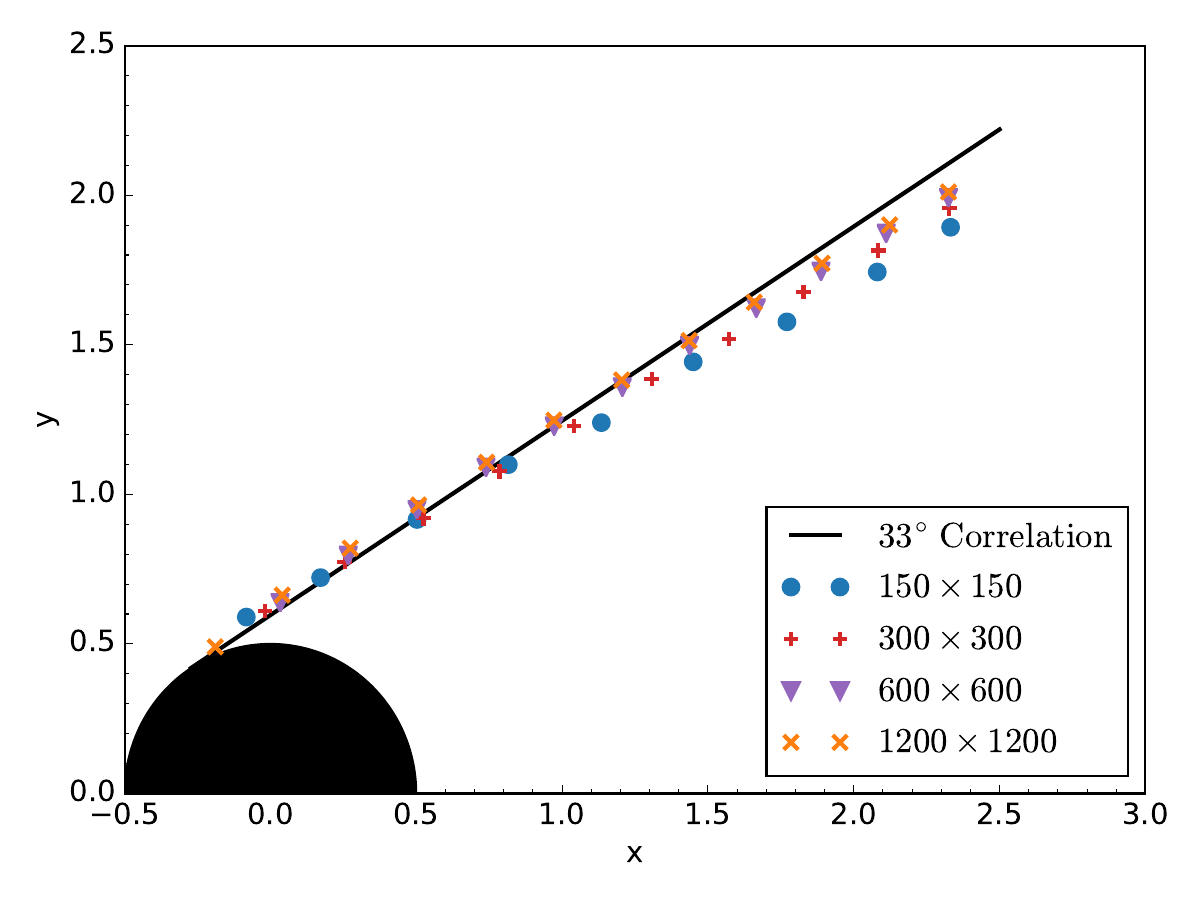}
    \caption{Comparison of the predicted triple-point paths with experimental correlation.}
    \label{fig:1_cyn_nomv_invis_triple_path}
\end{figure}

The predicted triple-point paths are extracted and plotted in Fig.~\ref{fig:1_cyn_nomv_invis_triple_path}. The least square linear regressions of the predicted triple-point paths on the grids of $150\times150$, $300\times300$, $600\times600$, and $1200\times1200$ nodes are about $28.2^{\circ}$, $29.8^{\circ}$, $30.3^{\circ}$, and $30.9^{\circ}$, respectively. These results, which agree well with the experimental correlation of \citet{kaca1988interferometric} and very well with the polynomial reconstruction based results of \citet{sambasivan2009ghostb}, cut-cell based results of \citet{ji2010numerical}, and unstructured mesh based results of \citet{ripley2006numerical}, further demonstrate the validity of the developed method.

\subsection{Mass flux examination}

Due to using non-body conformal Cartesian grids, mass flux over immersed boundary is a fundamental issue in immersed boundary methods \citep{mittal2005immersed}, and efforts such as adopting cut-cell approaches have been devoted in existing studies \citep{ji2008robust, mark2008derivation, seo2011sharp} in an attempt to alleviate this issue. This section examines the mass flux produced by the proposed method herein on generic Cartesian grids with practical grid sizes.

\begin{figure}[!htbp]
    \centering
    \begin{subfigure}[b]{0.32\textwidth}
        \includegraphics[width=\textwidth]{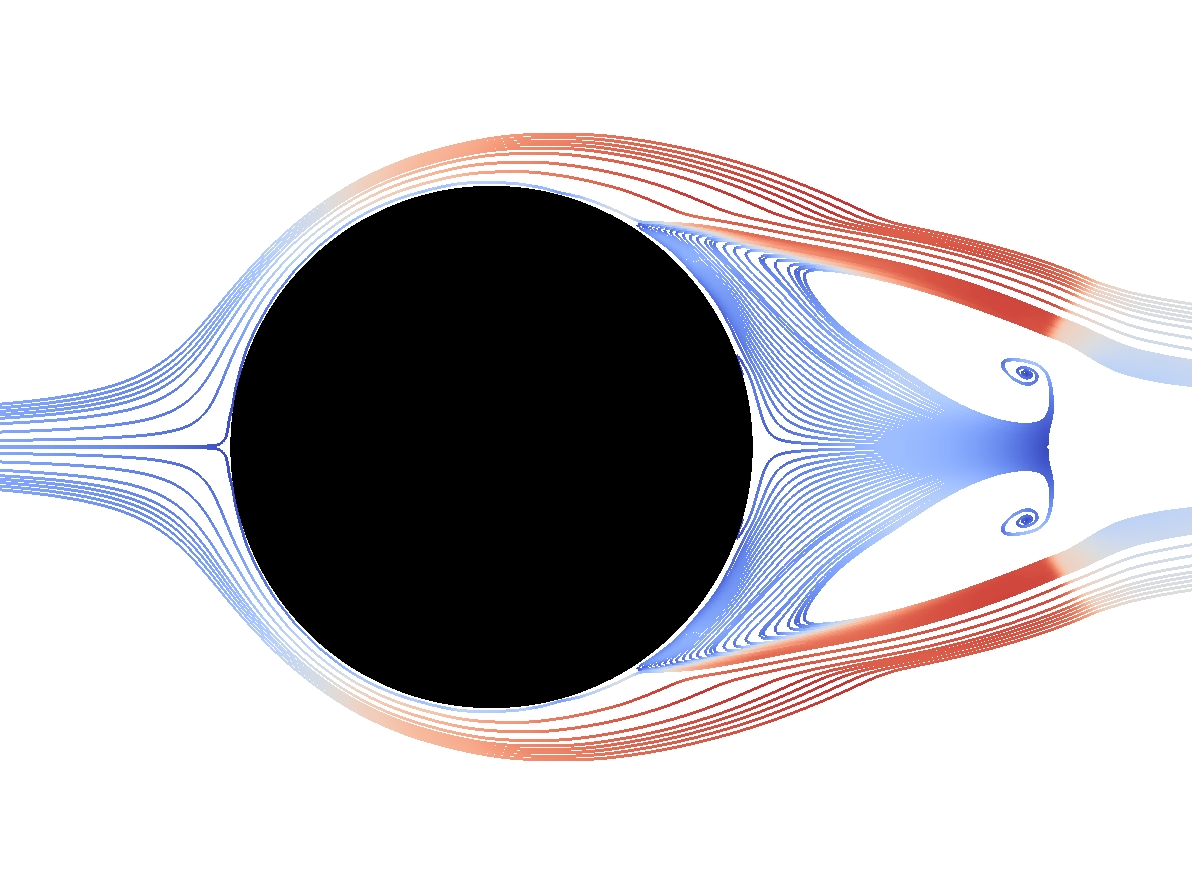}
        \caption{}
        \label{fig:1_cyn_nomv_vis_m0600_streamline}
    \end{subfigure}%
    ~
    \begin{subfigure}[b]{0.32\textwidth}
        \includegraphics[width=\textwidth]{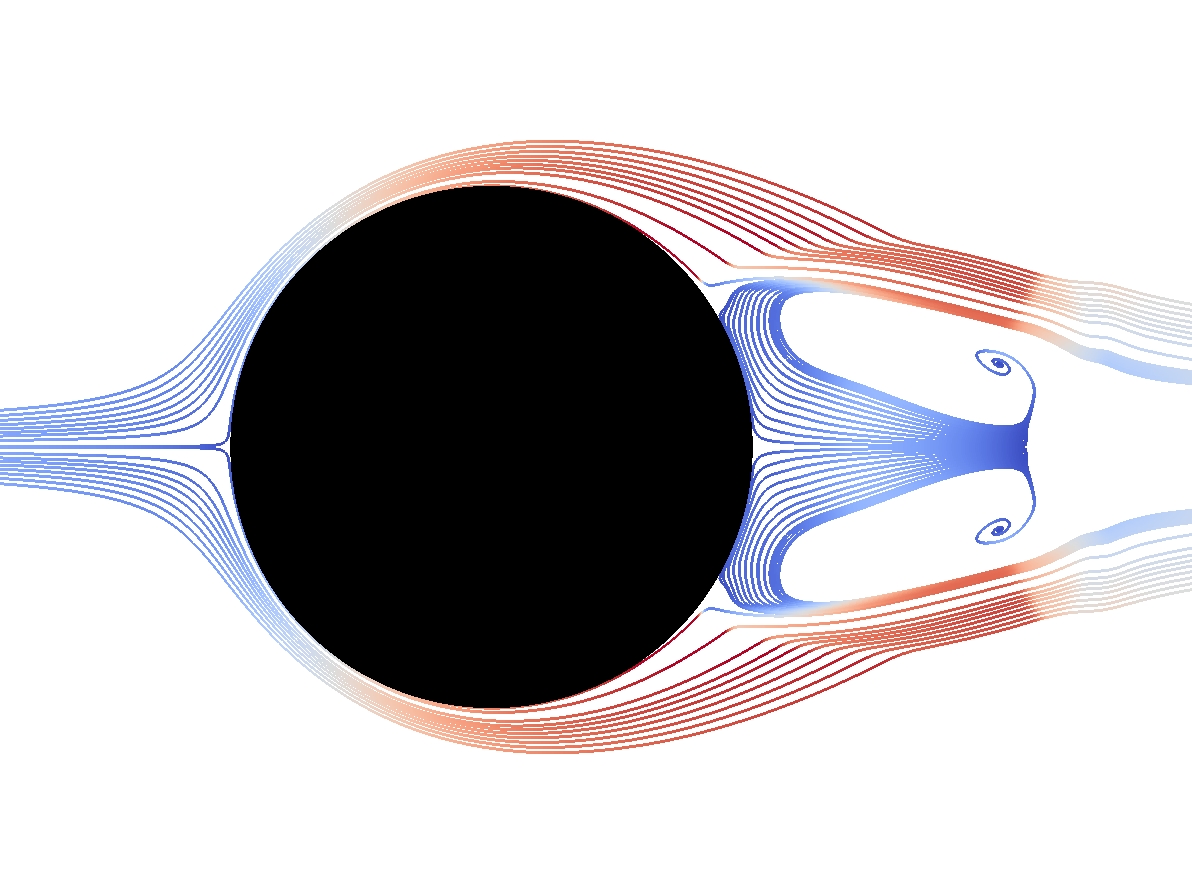}
        \caption{}
        \label{fig:1_cyn_nomv_invis_m0600_streamline}
    \end{subfigure}%
    ~
    \begin{subfigure}[b]{0.32\textwidth}
        \includegraphics[width=\textwidth]{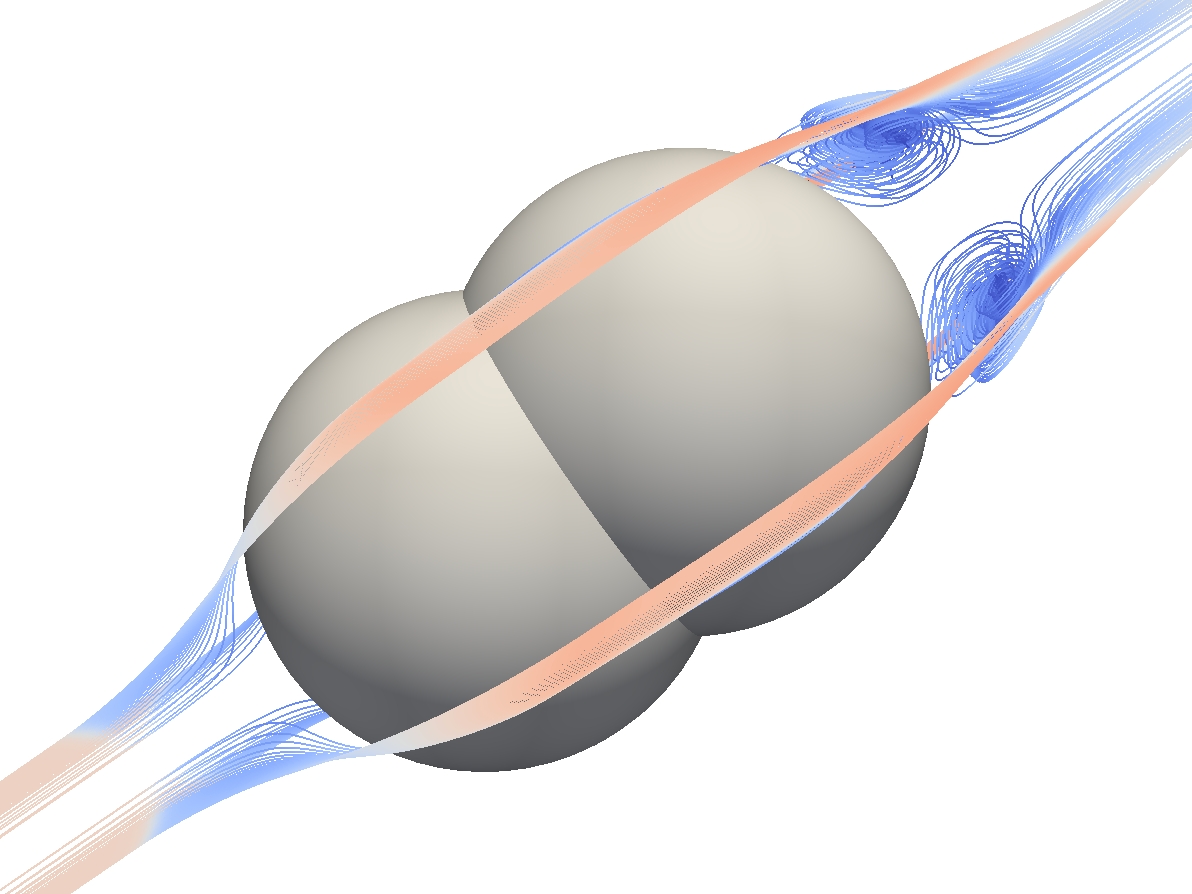}
        \caption{}
        \label{fig:2_sph_nomv_vis_m0250_streamline}
    \end{subfigure}%
    \caption{Streamlines of a Mach $2.81$ shock diffracting over different types of particles. Colored by velocity with corresponding analytical geometry boundaries presented. (a) Shock diffraction over a cylinder solved on a $6D\times6D$ domain discretized by a $600\times600$ grid, no-slip wall. (b) Shock diffraction over a cylinder solved on a $6D\times6D$ domain discretized by a $600\times600$ grid, slip wall. (c) Shock diffraction over two partially overlapped spheres solved on a $6D\times6D\times6D$ domain discretized by a $250\times250\times250$ grid, no-slip wall.}
    \label{fig:stream_trace}
\end{figure}

The streamlines of a Mach $2.81$ shock diffracting over different types of particles are presented in Fig.~\ref{fig:stream_trace}. The solved streamlines by the developed immersed boundary method are closely aligned with the geometry surfaces, even in the three-dimensional problem where a coarse grid is employed.

For the purpose of quantitatively examining the unphysical flux, the surface-normalized absolute flux over the immersed boundary
\begin{equation} \label{eq:absolute_flux}
    f_{\Des{ibm}} = \frac{1}{S}\int\limits_{S} |(\Vector{V} - \Vector{V}_{S})\ \cdot \ \unitVector{n}| \, \mathrm{d}S
\end{equation}
or in a discrete form
\begin{equation}
    f_{\Des{ibm}} = \frac{1}{N}\sum_{n=1}^{N} |(\Vector{V}_n - \Vector{V}_{S})\ \cdot \ \unitVector{n}_n|
\end{equation}
is employed as a quantitative measure, in which, $N$ is the number of the first layer ghost nodes, $\Vector{V}$ is the flow velocity at the ghost node, $\Vector{V}_S$ is the velocity of the geometry, $\unitVector{n}$ is the local unit outward surface normal vector. Since this integration is directly computed on the ghost node layer closest to the geometry without involving interpolation and will overestimate the flux through the geometry, $f_{\Des{ibm}}$ is a reliable measure of the unphysical flux. In addition, the calculation of $f_{\Des{ibm}}$ is on the ghost nodes, whose values are solely determined by the immersed boundary method. Therefore, $f_{\Des{ibm}}$ can serve as a proper global error measure for an immersed boundary method. 
\begin{table}[!htbp]
    \centering
    \caption{The surface-normalized absolute flux $f_{\Des{ibm}}$ in the grid sensitivity study of the supersonic flow over a wedge problem.}
    \label{tab:1_wedge_nomv_flux}
    %\scriptsize
    \begin{tabular}{lccc}
        \hline
        Grid & $f_{\Des{ibm}}$ ($\Unit{m/s}$) & Convergence rate & $f_{\Des{ibm}}/(M_{\infty}a)$ \\
        \hline
        $600\times300$ & $1.825E-1$ & $-$ & $0.456\%$\\
        $900\times450$ & $1.213E-1$ & $1.007$ & $0.303\%$\\
        $1200\times600$ & $9.126E-2$ & $0.990$ & $0.228\%$\\
        $1800\times900$ & $6.099E-2$ & $0.994$ & $0.152\%$\\
        $2400\times1200$ & $4.557E-2$ & $1.013$ & $0.114\%$\\
        \hline
    \end{tabular}
\end{table}

As the supersonic flow over a wedge problem involves a steady supersonic flow passing a relatively strong convex geometry with the intricate slip-wall boundary condition, the surface-normalized absolute flux $f_{\Des{ibm}}$ in the grid sensitivity study of the supersonic flow over a wedge problem is examined and is shown in Table~\ref{tab:1_wedge_nomv_flux}. A higher than first-order global convergence rate is presented over a wide range of grid resolution under the presence of complex shock interactions near the geometry boundary. As either global error measurement or the existence of discontinuities can lead to one order of accuracy degradation for a second-order scheme \citep{roache1998verification}, the presented global convergence behavior on the surface-normalized absolute flux under shocked flow conditions could validate that the proposed method reaches the designed second-order accuracy. In addition, the method generates a very low surface-normalized absolute flux over the immersed boundary even for the coarse grids. For instance, the value of $f_{\Des{ibm}}/(M_{\infty}a_0)$ is about $0.456\%$ for the $600\times300$ grid. According to the discussed qualitative and quantitative results, the developed immersed boundary method herein retains a very sharp interface and is able to effectively alleviate unphysical flux over physical boundaries when grid resolution is improved.

\subsection{Subsonic rotational flow}

\begin{figure}[!htbp]
    \centering
    \begin{subfigure}[b]{0.30\textwidth}
        \includegraphics[trim = 60mm 0mm 60mm 0mm, clip, width=\textwidth]{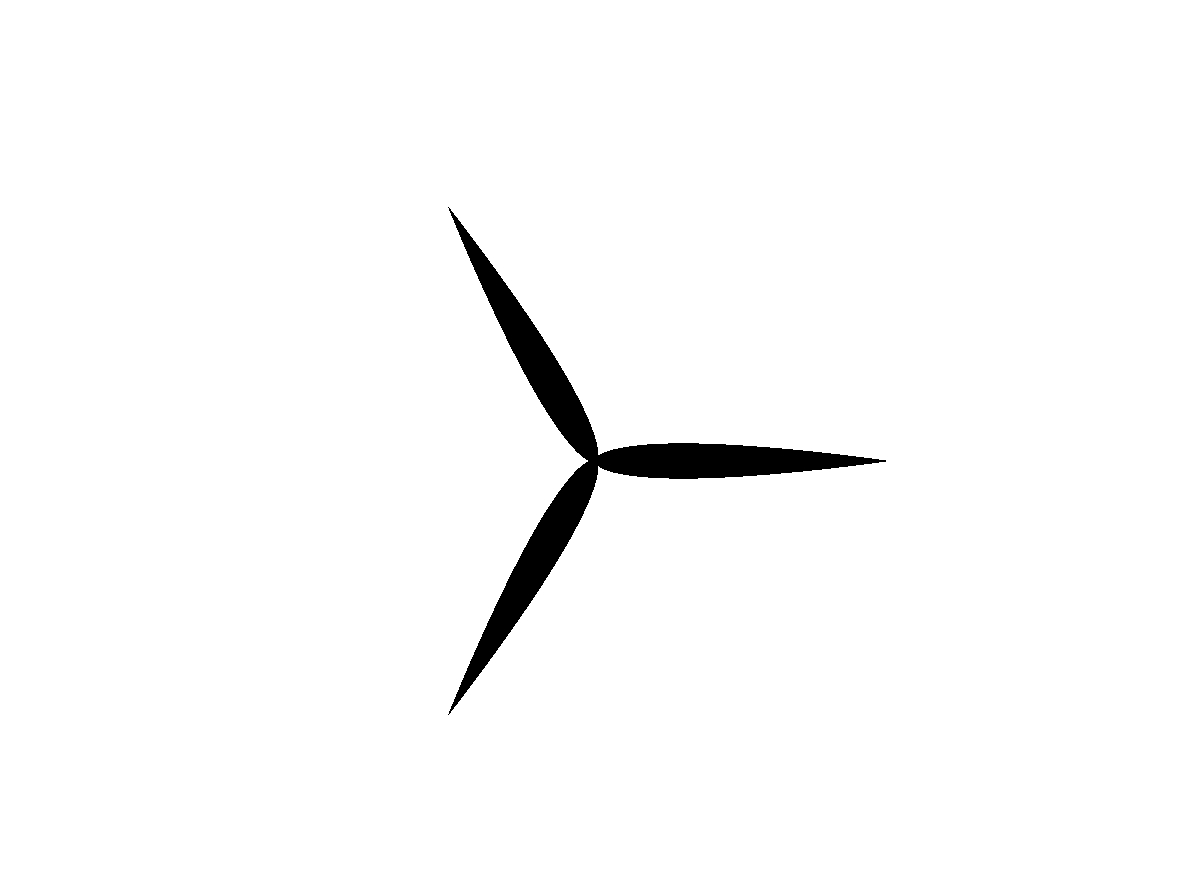}
        \caption{$0.00 \Unit{s}$, $V_{\Des{tip}} = 0 \Unit{m/s}$}
        \label{fig:trans_rot_w5n_vort_000s}
    \end{subfigure}%
    ~
    \begin{subfigure}[b]{0.30\textwidth}
        \includegraphics[trim = 60mm 0mm 60mm 0mm, clip, width=\textwidth]{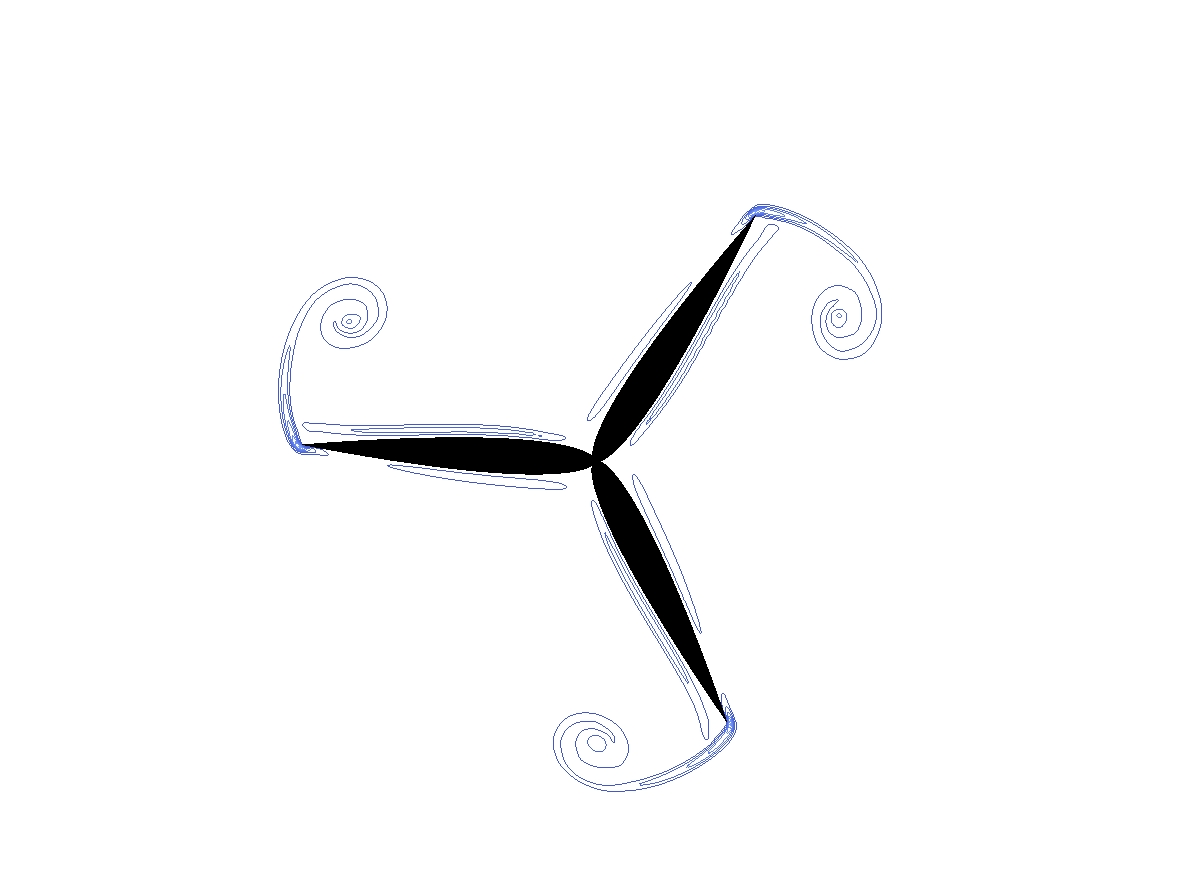}
        \caption{$0.25 \Unit{s}$, $V_{\Des{tip}} = 2.5\pi \Unit{m/s}$}
        \label{fig:trans_rot_w5n_vort_025s}
    \end{subfigure}%
    ~
    \begin{subfigure}[b]{0.30\textwidth}
        \includegraphics[trim = 60mm 0mm 60mm 0mm, clip, width=\textwidth]{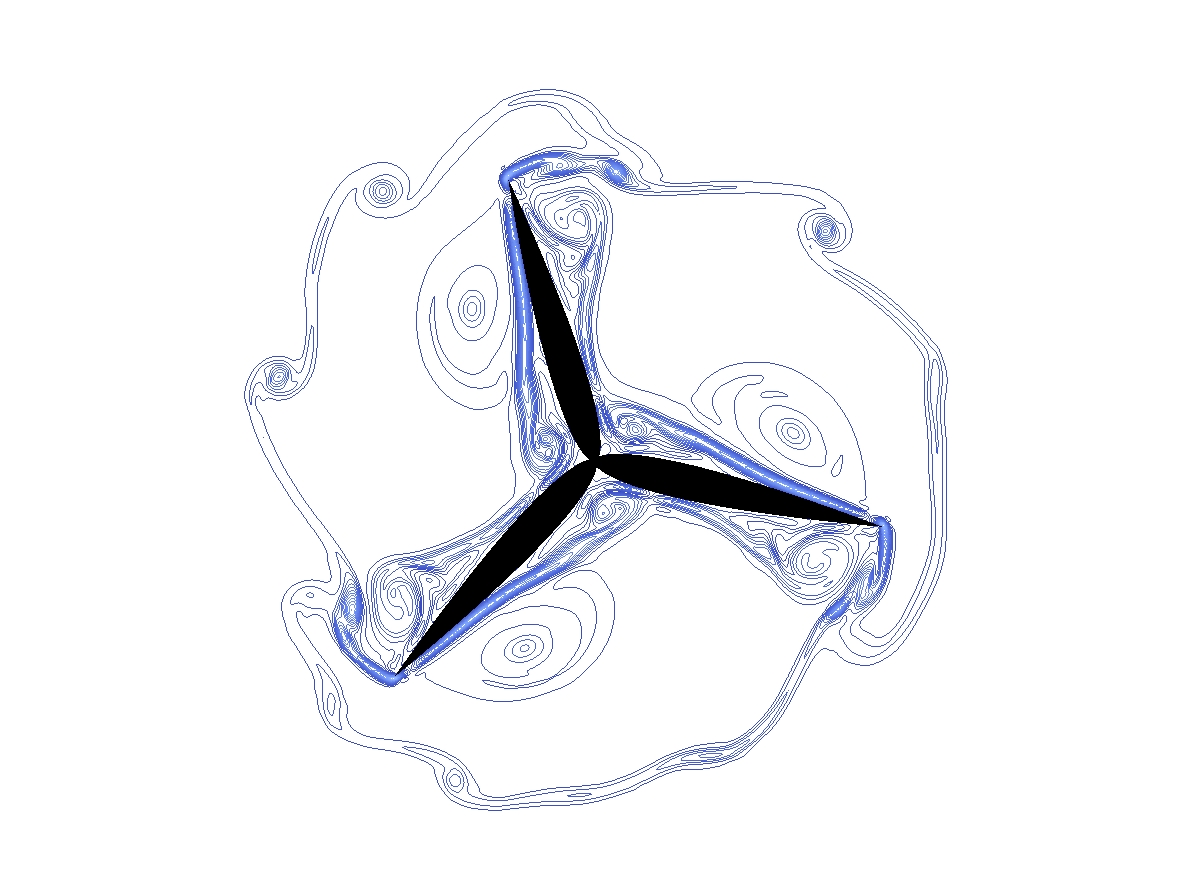}
        \caption{$0.50 \Unit{s}$, $V_{\Des{tip}} = 5.0\pi \Unit{m/s}$}
        \label{fig:trans_rot_w5n_vort_050s}
    \end{subfigure}%
    \\
    \begin{subfigure}[b]{0.30\textwidth}
        \includegraphics[trim = 60mm 0mm 60mm 0mm, clip, width=\textwidth]{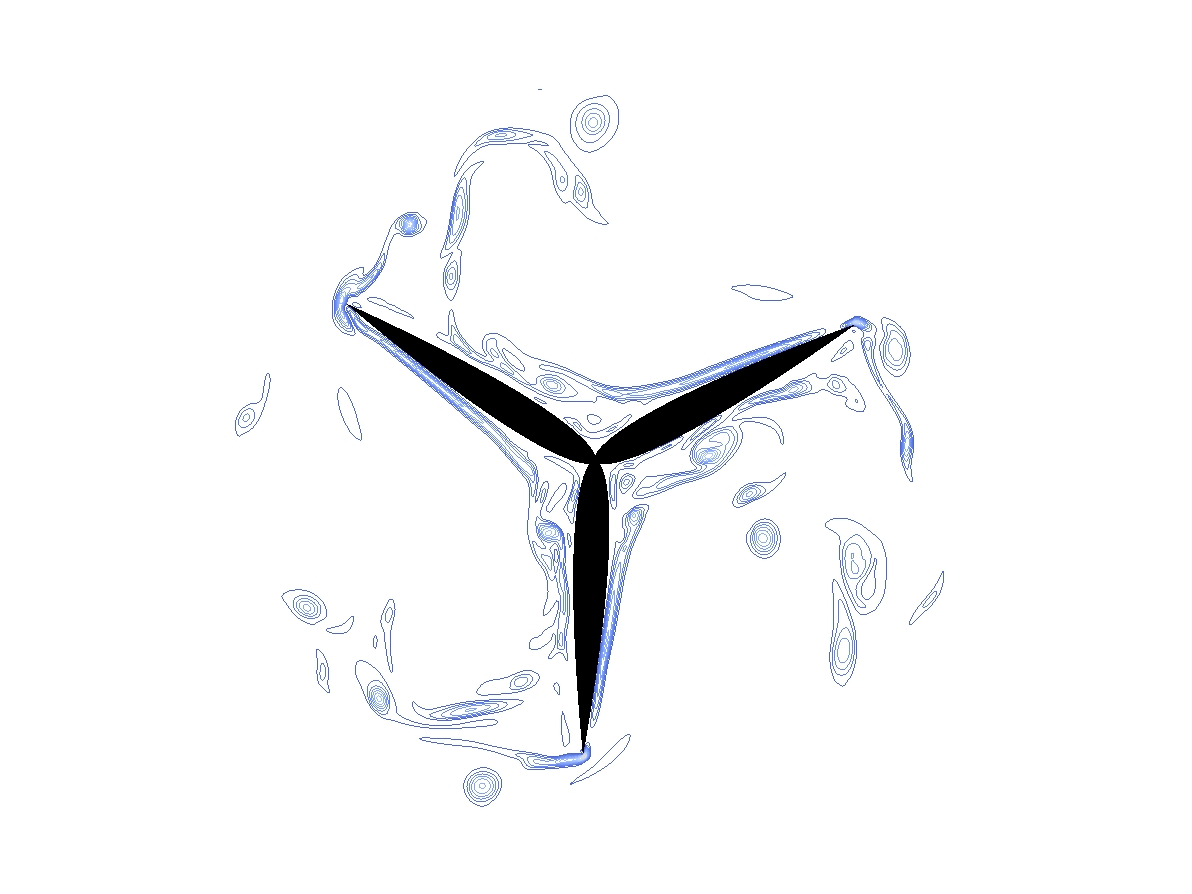}
        \caption{$0.75 \Unit{s}$, $V_{\Des{tip}} = 7.5\pi \Unit{m/s}$}
        \label{fig:trans_rot_w5n_vort_075s}
    \end{subfigure}%
    ~
    \begin{subfigure}[b]{0.30\textwidth}
        \includegraphics[trim = 60mm 0mm 60mm 0mm, clip, width=\textwidth]{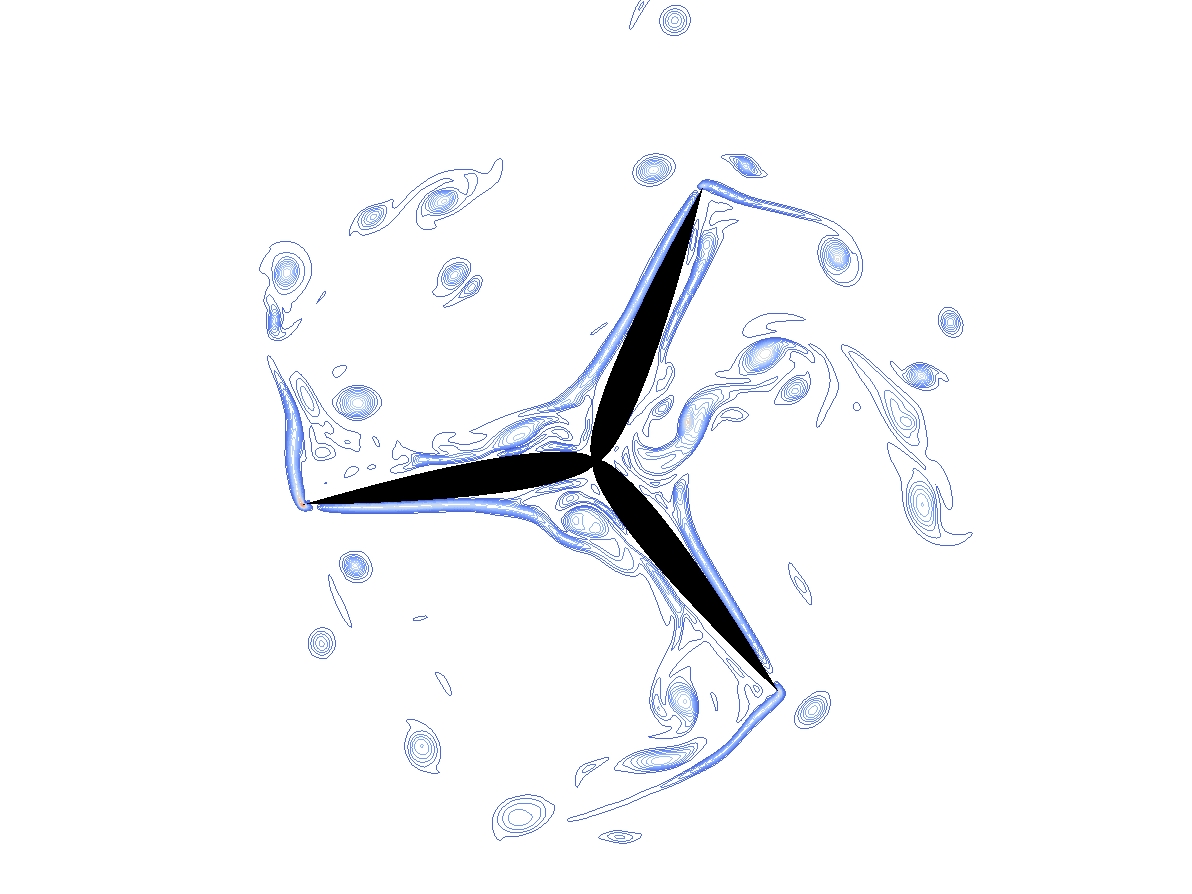}
        \caption{$1.00 \Unit{s}$, $V_{\Des{tip}} = 10.0\pi \Unit{m/s}$}
        \label{fig:trans_rot_w5n_vort_100s}
    \end{subfigure}%
    ~
    \begin{subfigure}[b]{0.30\textwidth}
        \includegraphics[trim = 60mm 0mm 60mm 0mm, clip, width=\textwidth]{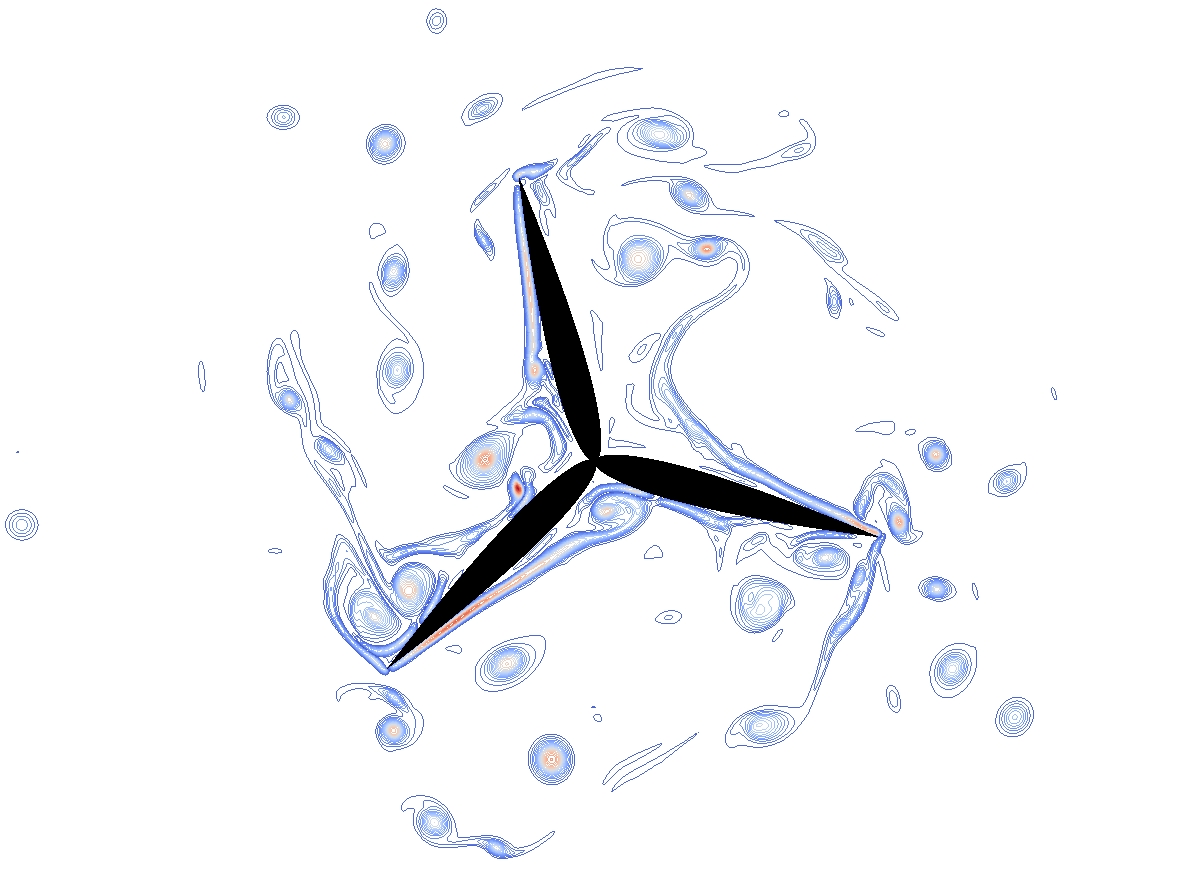}
        \caption{$1.50 \Unit{s}$, $V_{\Des{tip}} = 15.0\pi \Unit{m/s}$}
        \label{fig:trans_rot_w5n_vort_150s}
    \end{subfigure}%
    \\
    \begin{subfigure}[b]{0.58\textwidth}
        \includegraphics[trim = 60mm 145mm 60mm 130mm, clip, width=\textwidth]{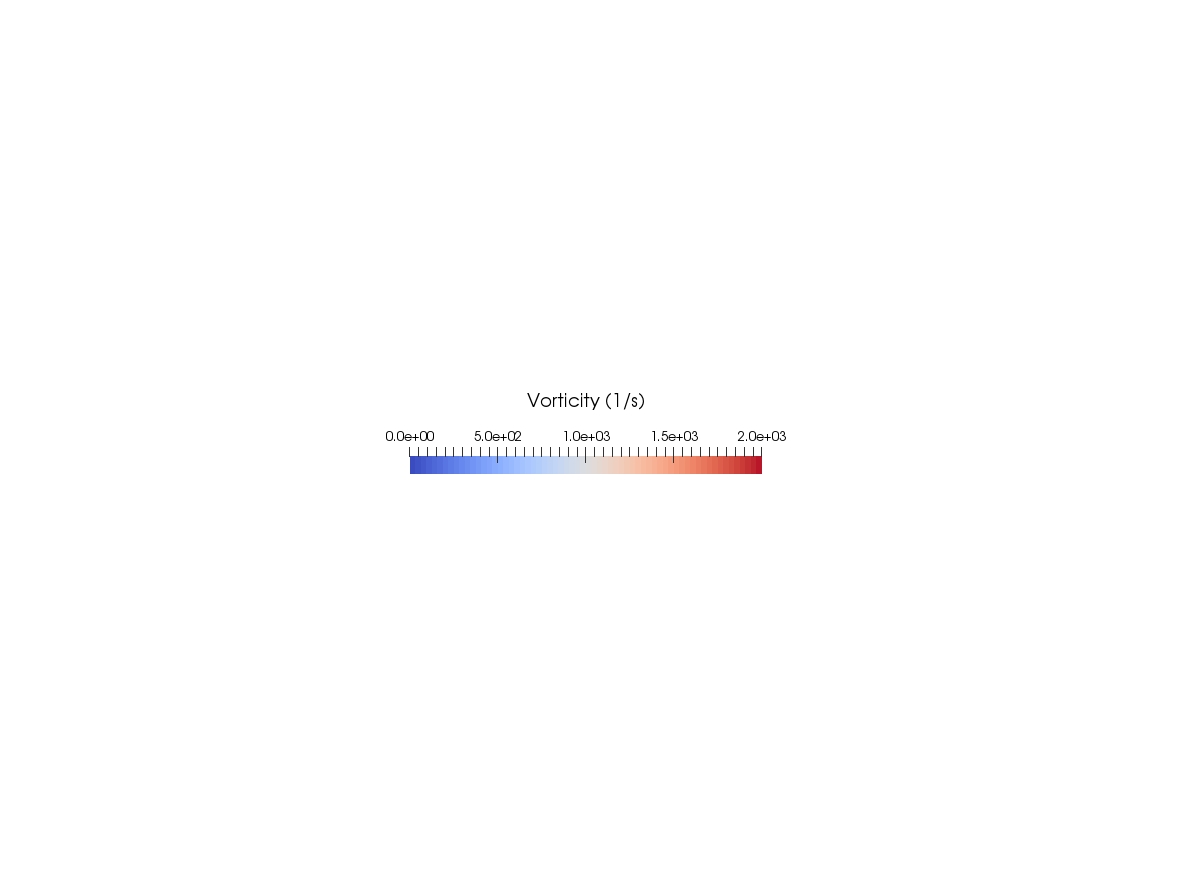}
    \end{subfigure}%
    \caption{Vorticity isocontour at a series of time instants generated by an accelerating rotor. [Nomenclature: $V_{\Des{tip}}$, velocity magnitude of the tip of a blade.]}
    \label{fig:trans_rot_w5n_vort}
\end{figure}

A subsonic rotational flow generated by an accelerating rotor is solved to demonstrate the applicability of the method for fluid-solid systems involving complex geometries. As shown in Fig.~\ref{fig:trans_rot_w5n_vort_000s}, the $2$D rotor consists of three blades with each blade being the shape of NACA $0012$ airfoil, whose chord is $l_c = 1 \Unit{m}$. The rotor is centered in a $6 l_c \times 6 l_c$ domain discretized by a $1200\times1200$ grid. The rotor rotates with an initial angular velocity $\omega(t=0)=0 \Unit{rad/s}$ and a constant angular acceleration $\alpha = 10\pi \Unit{rad/s^2}$. The initial ambient flow state is $(\rho_0, u_0, v_0, p_0)=(1.204 \Unit{kg/m^3}, 0, 0, 101325 \Unit{Pa})$, and the no-slip wall boundary condition is enforced on the blades.

The predicted vorticity isocontour at a series of time instants is captured in Fig.~\ref{fig:trans_rot_w5n_vort}, in which an interesting vortex-induced vortex shedding behavior is observed. As the rotor accelerates, vortices appear at the tips of blades as well as at the sides of the blades. The accelerating tips stretch and deform the generated vortices, causing vortices shedding. The detached vortices soon are entrained by the flow driven by the coming blade, either being merged into the tip vortices of the coming blade or being advected to the front side vortices of the coming blade. The entrained vortices then destabilize the vortex structures they propagate toward, inducing and accelerating the vortex shedding. The above interaction process, combining with the acceleration of the blades, produces a complex and dynamic vortex field showing in the figures.

In the employed case setting, the geometry of the rotor is generated by positioning three independent NACA $0012$ airfoils at a common center. The rotating system is achieved by specifying the same rotational acceleration for the three blades. Therefore, in simulating flow with complex geometries, employing a robust immersed boundary method enables great simplification for the model generation process. For instance, one can obtain an engineering structure by assembling its components via positioning while allowing overlapped surfaces, which avoids the cost of producing a single topology, as required by many mesh generators such as ANSYS ICEM CFD. In addition, the current grid resolution of the domain is $200$ nodes per $l_c$. Meanwhile, the largest width of the blade is only $0.12 l_c$, and the tip of blade is very thin. Consequently, there is only a line of nodes in the region near the tip, whose length is about $5\%$ of the chord. Nonetheless, the proposed method resolves the thin tip region with reasonable sharpness. The overall success of simulating the transient rotational flow illustrates the capability of the presented method for solving general fluid-solid systems. 

\subsection{Explosive dispersal of dense particles}

To demonstrate the robustness of the presented method for solving problems with strongly irregular and moving geometries under challenging flow conditions, a dense particle system dispersed by a high-pressure gas driver is studied.

\begin{figure}[!htbp]
    \centering
    \includegraphics[trim = 00mm 20mm 00mm 20mm, clip, width=0.40\textwidth]{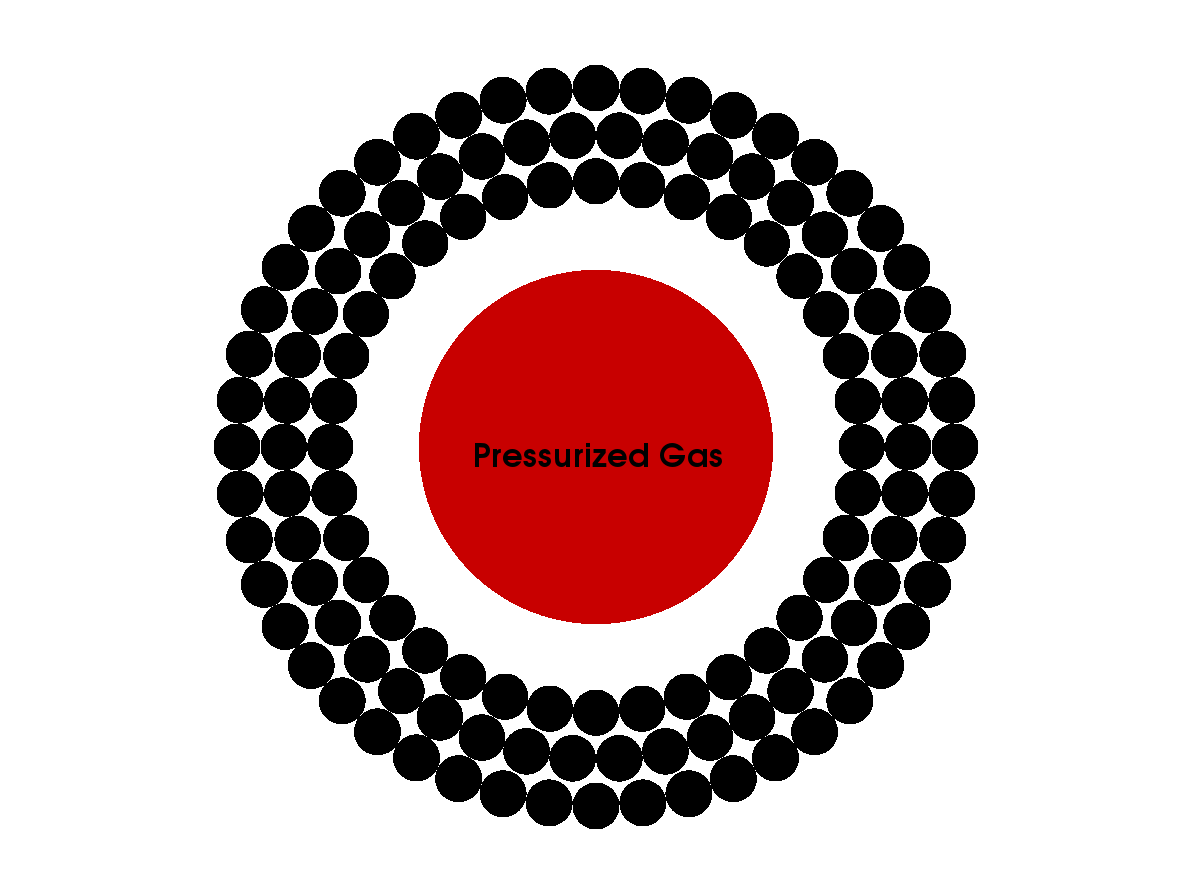}
    \caption{Computational configuration for explosive dispersal of dense particles.}
    \label{fig:2d_n126_ang00_demo}
\end{figure}

As shown in Fig.~\ref{fig:2d_n126_ang00_demo}, in a $1 \Unit{m} \times 1 \Unit{m}$ computational domain, $126$ particles are packed to form a three-layer particulate payload. The number of particles in each layer is $36$, $42$, and $48$, respectively. The centers of particles are evenly distributed on each circular layer. The radius of the first layer is $0.15 \Unit{m}$, and the radius of the other two layers as well as the radius of particles in each layer are determined by ensuring zero gap among neighboring particles. A flow state $(\rho_c, u_c, v_c, p_c)=(1.204 \Unit{kg/m^3}, 0, 0, 1.01325\times10^8 \Unit{Pa})$ is initially positioned at a circular region centered in the domain, whose radius is $0.1 \Unit{m}$. The flow state at the rest of the region is set to $(\rho_0, u_0, v_0, p_0)=(1.204 \Unit{kg/m^3}, 0, 0, 1.01325\times10^5 \Unit{Pa})$. In order to reduce the discrepancy of the timescales between shock propagation and particle acceleration, relatively light particles with a density of $27 \Unit{kg/m^3}$ are used. A $2500 \times 2500$ grid is used for the computational domain, which ensures a grid resolution of $0.03D$ ($32$ nodes per particle diameter) for the smallest particles.
\begin{figure}[!htbp]
    \centering
    \begin{subfigure}[b]{0.30\textwidth}
        \includegraphics[trim = 70mm 0mm 70mm 0mm, clip, width=\textwidth]{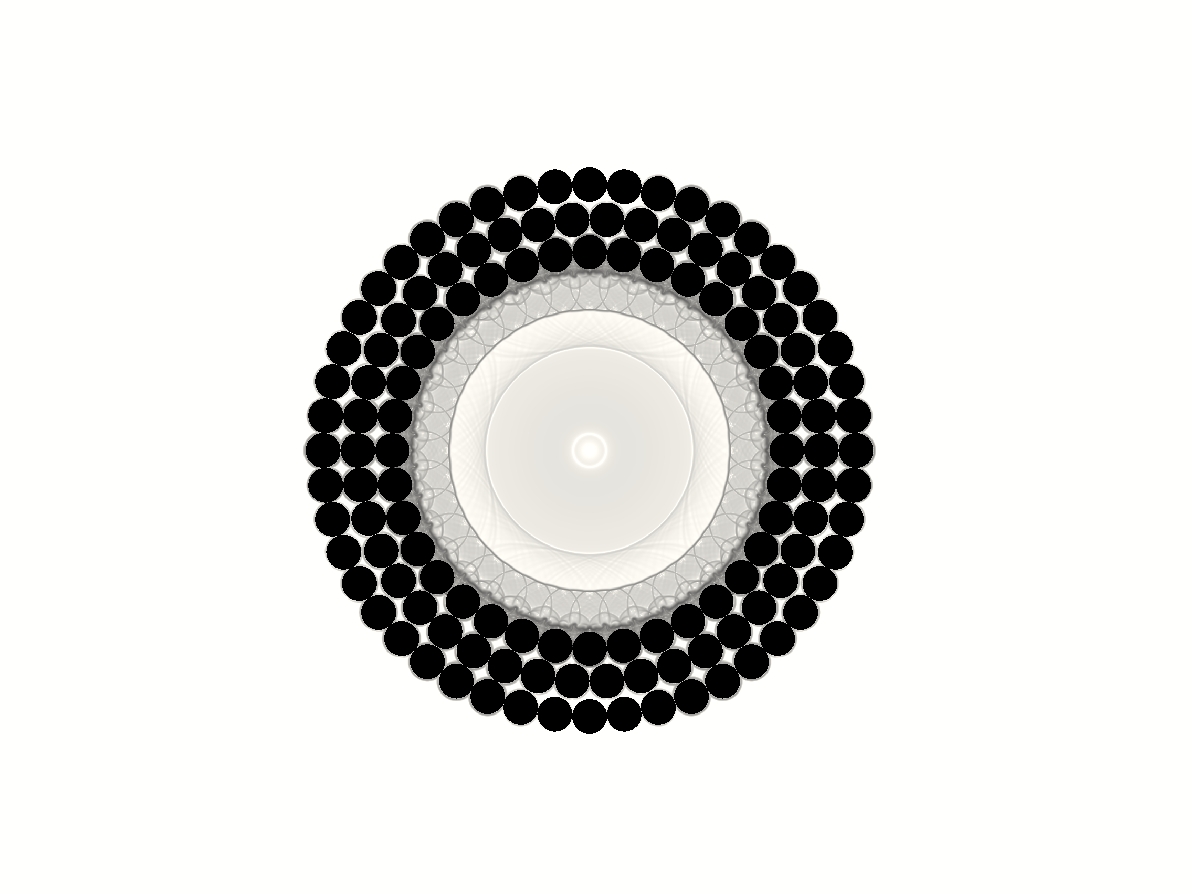}
        \caption{$10 \Unit{\mu s}$}
        \label{fig:2d_n126_ang00_rho27_10p3atm_cr05_010us}
    \end{subfigure}%
    ~
    \begin{subfigure}[b]{0.30\textwidth}
        \includegraphics[trim = 70mm 0mm 70mm 0mm, clip, width=\textwidth]{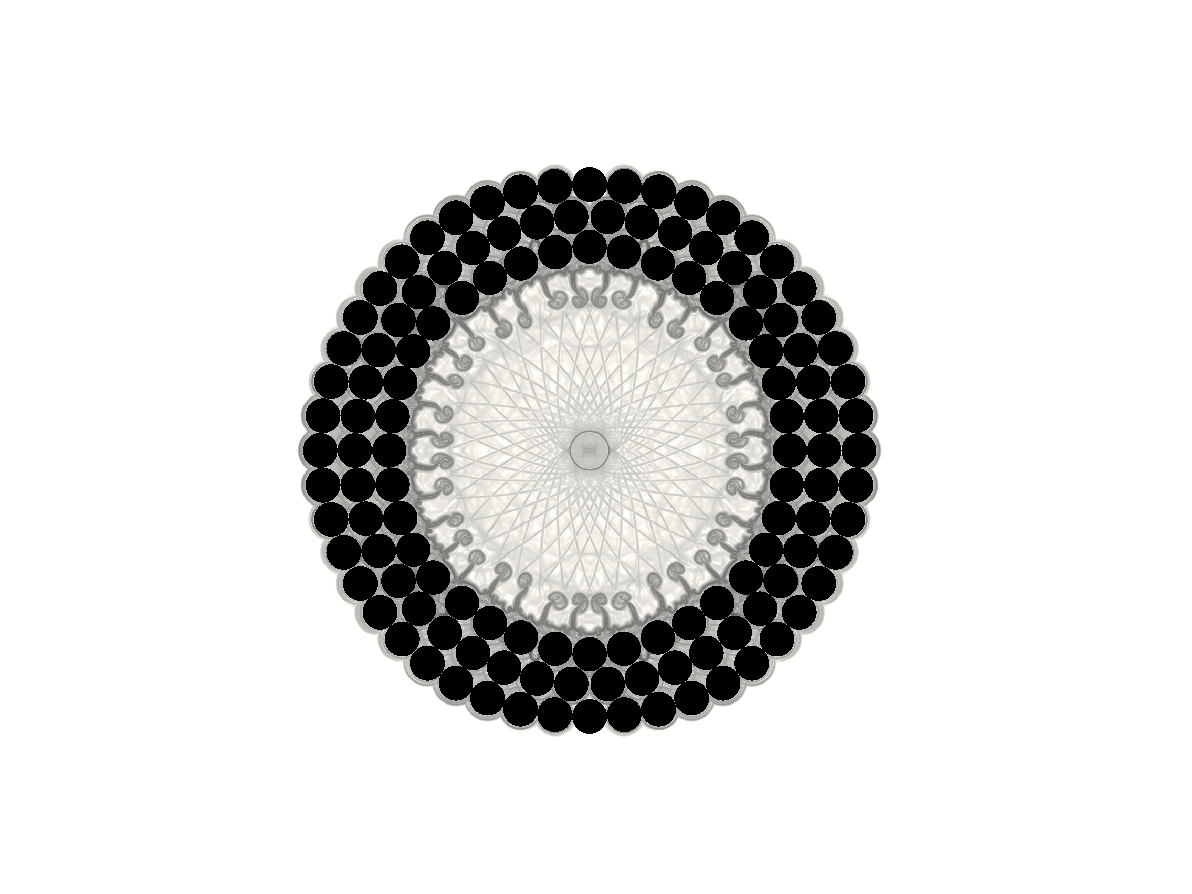}
        \caption{$20 \Unit{\mu s}$}
        \label{fig:2d_n126_ang00_rho27_10p3atm_cr05_020us}
    \end{subfigure}%
    ~
    \begin{subfigure}[b]{0.30\textwidth}
        \includegraphics[trim = 70mm 0mm 70mm 0mm, clip, width=\textwidth]{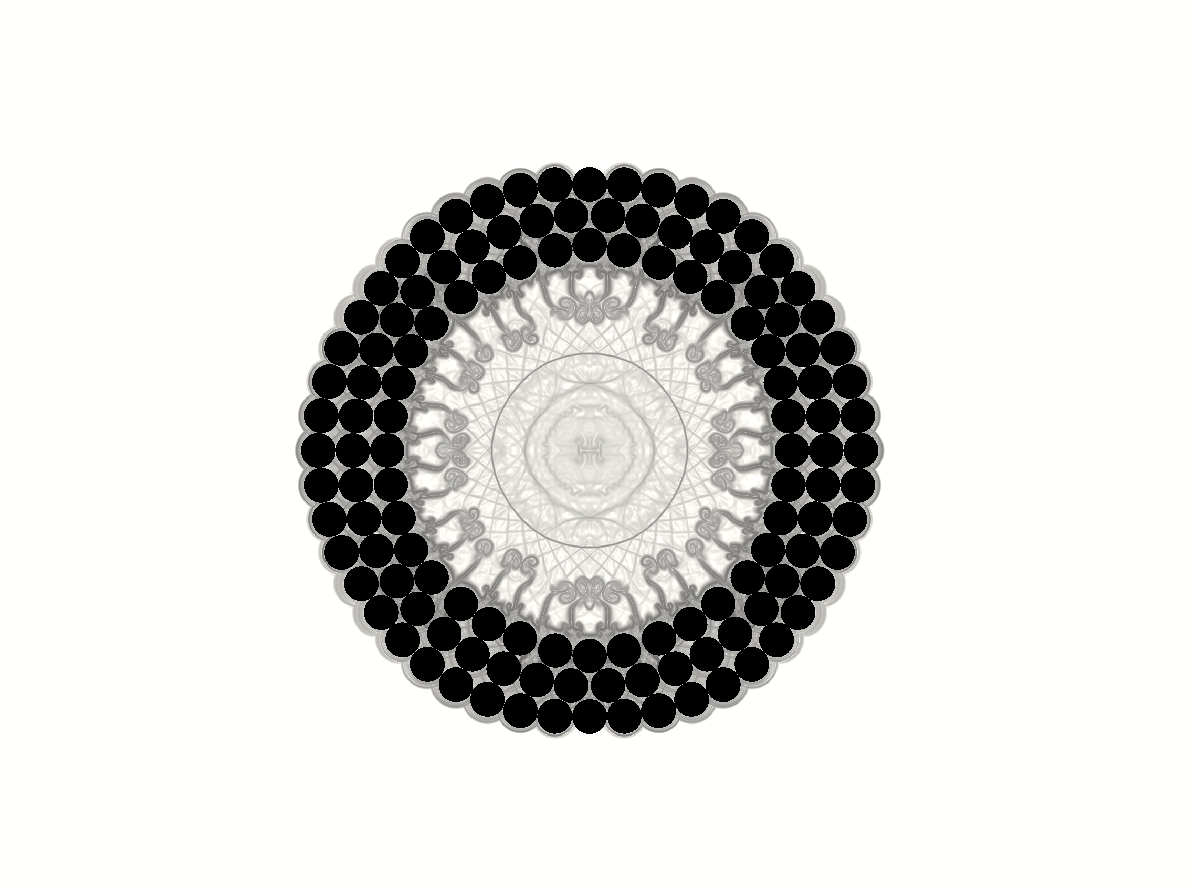}
        \caption{$25 \Unit{\mu s}$}
        \label{fig:2d_n126_ang00_rho27_10p3atm_cr05_025us}
    \end{subfigure}%
    \\
    \begin{subfigure}[b]{0.30\textwidth}
        \includegraphics[trim = 70mm 0mm 70mm 0mm, clip, width=\textwidth]{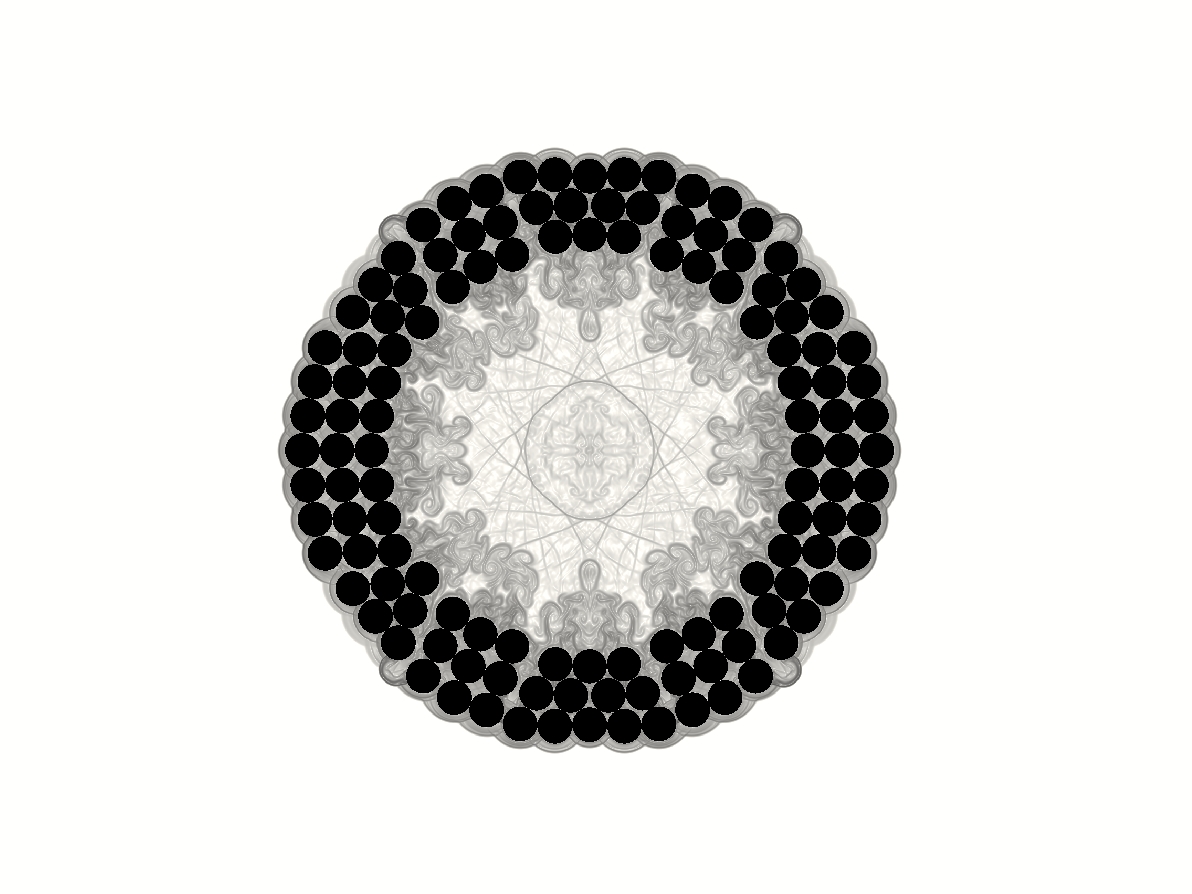}
        \caption{$50 \Unit{\mu s}$}
        \label{fig:2d_n126_ang00_rho27_10p3atm_cr05_050us}
    \end{subfigure}%
    ~
    \begin{subfigure}[b]{0.30\textwidth}
        \includegraphics[trim = 70mm 0mm 70mm 0mm, clip, width=\textwidth]{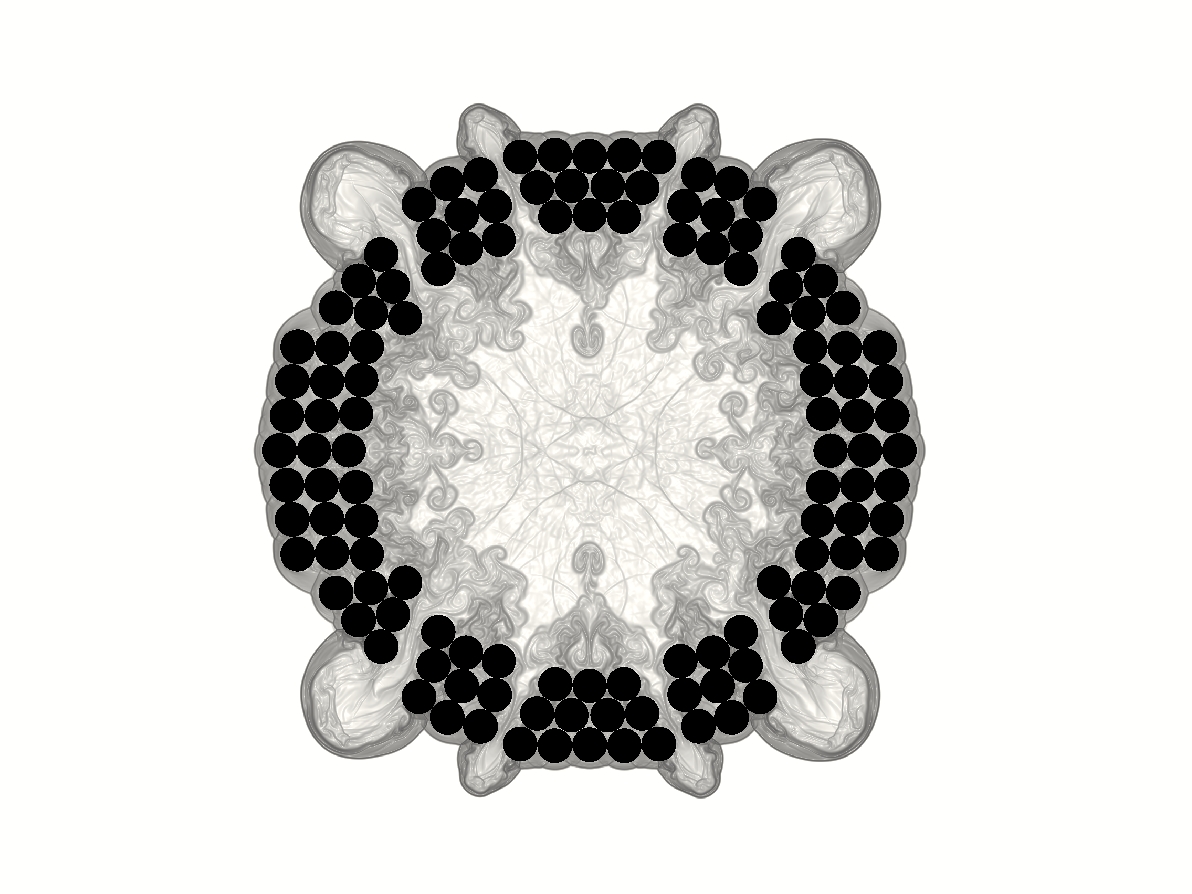}
        \caption{$75 \Unit{\mu s}$}
        \label{fig:2d_n126_ang00_rho27_10p3atm_cr05_075us}
    \end{subfigure}%
    ~
    \begin{subfigure}[b]{0.30\textwidth}
        \includegraphics[trim = 70mm 0mm 70mm 0mm, clip, width=\textwidth]{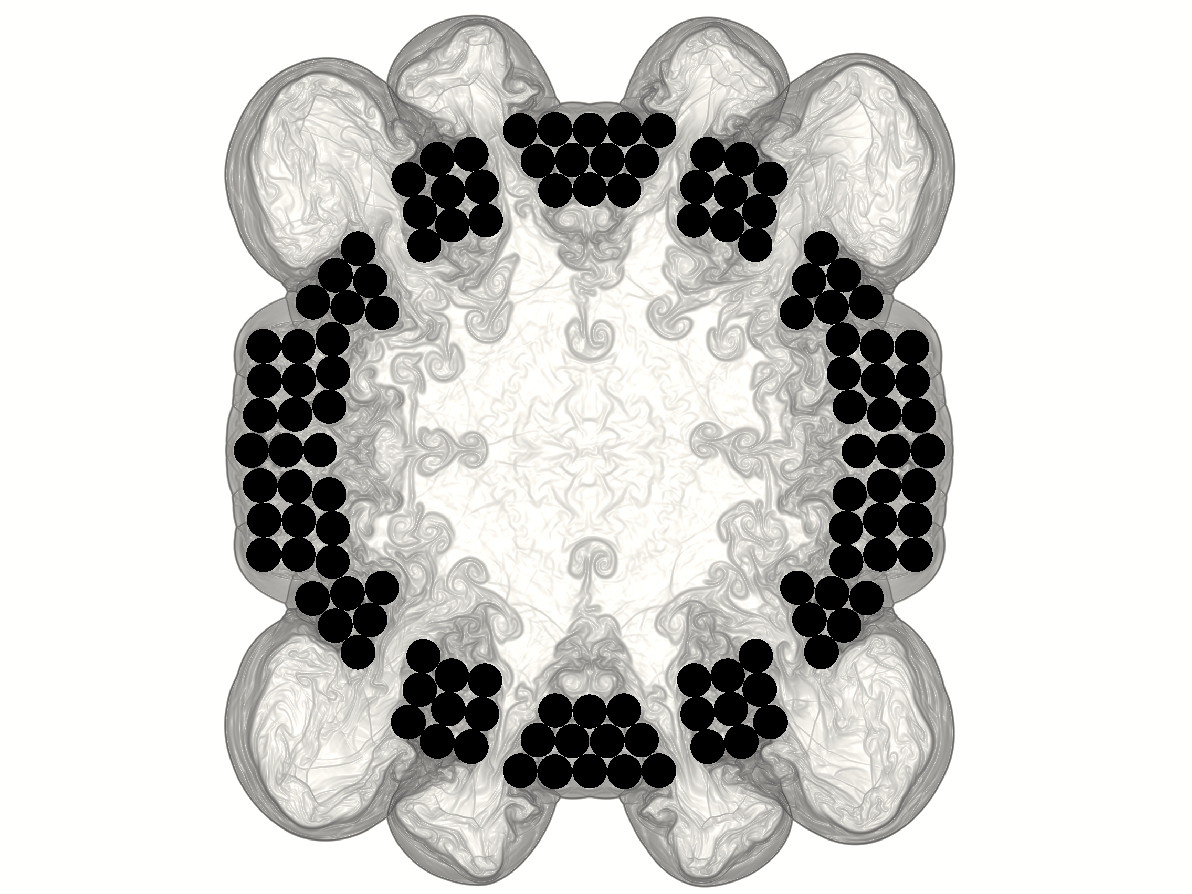}
        \caption{$100 \Unit{\mu s}$}
        \label{fig:2d_n126_ang00_rho27_10p3atm_cr05_100us}
    \end{subfigure}%
    \caption{Numerical Schlieren field during the explosive dispersal process.}
    \label{fig:2d_n126_ang00_rho27_10p3atm_cr05}
\end{figure}

The computed dispersal process of the particle system in the form of numerical Schlieren is presented in Fig.~\ref{fig:2d_n126_ang00_rho27_10p3atm_cr05}. During the particle dispersal, complex shock-shock, shock-particle, and particle-particle interactions are resolved. When reaching the particle payload, shocks diffract and reflect over the particle surfaces. Diffracted shocks propagate along particle surfaces and later collide at the concave regions, forming locally heavy fluids. Richtmyer--Meshkov instability is then induced by the injection of the heavy fluid into ambient light fluid. The reflected shock waves propagate back toward the payload after coalescing at the domain center, which then collide with the growing fluid jets and severely disturb the jet structure. The impacts of repetitive reflected shocks effectively destabilize the particle system, producing gaps among particles. High-speed fluid jets are then generated at those gaps, which further fracture the particle system.

These successful solutions of the explosive dispersal of zero-gap packed dense particles under a high pressure ratio flow condition demonstrate the robustness of the proposed method for solving flow with strongly irregular and moving geometries.

\section{Conclusion} \label{sec:conclusion}

A novel immersed boundary method has been developed, validated, and applied. The effectiveness of the method in solving flow with arbitrarily irregular and moving geometries on Cartesian grids has been illustrated through numerical experiments concerning a variety of flow problems. The main properties of the presented method and the primary conclusions from the numerical experiments are summarized below.

\paragraph{Convergence and accuracy}

The accuracy of the method is established through thorough studies of the supersonic flow over a wedge problem, the supersonic translating wedge problem, and the shock diffraction over a cylinder problem. Employing the analytical $M_{\infty}-\theta-\beta$ relation of oblique shocks, different cases considering the deflection angle ranging from $\theta = 10^{\circ}$ to $\theta = 20^{\circ}$, the Mach number ranging from $M_{\infty} = 2$ to $M_{\infty} = 10$, and stationary as well as moving geometries are tested, and excellent agreement between the numerical and analytical solutions is obtained. For the shock diffraction over a cylinder problem, good agreement between the obtained numerical results and experimental observations as well as other published numerical results is achieved. The successful solutions of these test cases demonstrate the validity and accuracy of the proposed method in solving flow involving irregular and moving geometries under challenging flow conditions. 

The incorporation of physical boundary conditions in the proposed three-step flow reconstruction scheme leads to the property that the constructed $\psi_G$ converges to the exact physical boundary conditions when the ghost node $G$ converges to the boundary point $O$. For non-body conformal Cartesian grids, this property is helpful in alleviating the unphysical flux over immersed boundaries. Two- and three-dimensional streamlines of shock-particle interactions with slip and no-slip boundary conditions have qualitatively illustrated that the developed method maintains a very sharp interface. Through an examination of the surface-normalized absolute flux in the grid sensitivity study of the supersonic flow over a wedge problem, it has been quantitatively shown that, under the presence of a relatively strong convex geometry with the intricate slip-wall boundary condition and complex shock interactions near immersed boundaries, the developed method preserves a higher than first-order global convergence rate on the surface-normalized absolute flux over the immersed boundary and produces surface-normalized absolute flux in very low quantities even on coarse grids. The presented global convergence behavior for flow with discontinuities suggests that the proposed method reaches the designed second-order accuracy.

\paragraph{Uniformity and efficiency}

The proposed method enforces the Dirichlet, Neumann, Robin, and Cauchy boundary conditions in a straightforward and consistent manner and completely avoids the necessity to solve linear systems. As a result, an arbitrary number of field variables that satisfy different types of boundary conditions, such as velocity, pressure, and temperature, can be efficiently and uniformly treated. In addition, the application of the method to different spatial dimensions is also uniform. The uniformity of the method has been illustrated via the solution of two- and three-dimensional flow problems with no-slip and slip wall boundary conditions. Benefited from the use of generic Cartesian grids and the efficiency of the proposed immersed boundary method, all the presented test cases herein were solved using a single processor.

\paragraph{Robustness and stability}

The proposed method employs a three-step flow reconstruction scheme that is scalable to the number of stencil nodes and is uniformly valid under a varying number of stencils, even in the worst situation, in which only one fluid node exists in the domain of dependence of an image point. In addition, as demonstrated by the implementation of the $7$-point-stencil WENO scheme, which requires $3$ ghost node layers, the method can be applied to multiple layers of ghost nodes without imposing extra constraints. This property can greatly facilitate the application of high-order spatial schemes to flow with complex geometries.

Utilizing the three-step reconstruction as well as the convex and extrema-preserving properties of the inverse distance weighting, the proposed method presents strong numerical stability, as demonstrated in the numerical experiments involving challenging flow conditions and dynamic geometries, such as the supersonic translating wedge flow, the subsonic rotational flow, and the explosive dispersal of dense particles. Equipped with suitable discretization schemes, the developed immersed boundary method enables feasible solutions of problems with engineering level of complexity and hence enhances the understanding of physical problems.

On a Cartesian grid, improving the local grid resolution near boundaries inevitably affects the grid globally and significantly increases the grid size. Although the adaptive mesh refinement (AMR) technique can be used to address the grid-size problem \citep{ji2008robust}, it has undesirable effects on the grid topology. In order to increase the local grid resolution with low memory costs but without sacrificing the Cartesian structure, future investigations will focus on using the immersed boundary method to facilitate overset grid methods on generic Cartesian grids.

\section*{Acknowledgements}

Financial support of this work was provided by Natural Sciences and Engineering Research Council of Canada (NSERC) and Defence Research and Development Canada (DRDC). This work was made possible by the facilities of the Shared Hierarchical Academic Research Computing Network (SHARCNET: www.sharcnet.ca) and Compute/Calcul Canada. The first author of this paper is grateful to Prof. Deliang Zhang at the Institute of Mechanics, Chinese Academy of Sciences for introducing Computational Fluid Dynamics and is thankful to Dr. Deyong Wen at Environment Canada for discussions of flow visualization.

\section*{References}

\bibliography{ref}

\end{document}